\documentclass[aps,preprint,preprintnumbers,amsmath,amssymb,floatfix]{revtex4-1}
\usepackage{graphicx}
\usepackage{color}
\usepackage{float}
\usepackage{amsmath}
\usepackage{amssymb}
\usepackage{mathrsfs}
\usepackage{hyperref}
\usepackage{nameref}
\usepackage[normalem]{ulem}
\usepackage{epsfig}
\usepackage{titlesec}
\usepackage{xr}
\newcommand{\noi}{\noindent}
\newcommand{\be}{\begin{equation}}
\newcommand{\ee}{\end{equation}}

\begin{document}
\title{Scalar model for frictional precursors dynamics} 
\author{Alessandro Taloni}
\affiliation{CNR-IENI, Via R. Cozzi 53, 20125
Milano, Italy}
\author{Andrea Benassi}
\affiliation{Empa, Swiss Federal Laboratories for Materials Science and Technology, CH-8600 D\"{u}bendorf, Switzerland.}
\author{Stefan Sandfeld}
\affiliation{Institute of Materials Simulation (WW8), Department of Materials Science, University of Erlangen-N\"urnberg (FAU), Dr.-Mack-Str. 77, 90762 F\"urth, Germany}
\author{Stefano Zapperi}
\affiliation{CNR-IENI, Via R. Cozzi 53, 20125 Milano, Italy}
\affiliation{ISI Foundation, Via Alassio 11/C 10126 Torino, Italy}



\date{}

\begin{abstract} 
Recent experiments indicate that frictional sliding occurs by
the nucleation of detachment fronts at the contact interface that may 
appear well before the onset of global sliding. 
This intriguing precursory activity is not accounted for by traditional friction theories but is
extremely important for friction dominated geophysical phenomena such
as earthquakes, landslides or avalanches. 
Here we simulate the onset of slip of a three dimensional elastic body resting
on a surface and show that experimentally observed frictional precursors depend
in a complex non-universal way on the sample geometry and the loading conditions. 
Our model satisfies  Archard's law and Amontons' first and second laws, reproducing with remarkable precision the real contact area dynamics, the precursors' envelope dynamics prior to the transition to sliding, and the normal and shear internal stress distributions close to the slider-substrate interface. Moreover, it allows to assess which experimental features can be attributed to the elastic equilibrium, 
and which are attributed to the out-of-equilibrium dynamics, suggesting that precursory 
activity is an intrinsically quasi-static physical process.  A direct calculation of the evolution of the 
Coulomb stress before and during precursors nucleation shows large variations across the sample,
which helps to explain why earthquake forecasting methods based only on accumulated slip and 
Coulomb stress monitoring  are often ineffective.
\end{abstract}

\maketitle

\section{Introduction}
The classical laws of friction, due to Amontons and Coulomb, postulate
that a body resting on a surface can be displaced only by applying a shear force
larger than a static friction force, which is proportional to the
normal load and independent of the apparent area of contact.
Recent research has challenged this understanding of friction,
showing that macroscopic slip is due to the formation and propagation
of detachment fronts through the contact interface \cite{rubinstein2004}. The nature of these
fronts and their speed depend on the way shear is applied to the sample and on its geometry \cite{bendavid2010,bendavid2011},
a particularly compelling issue in view of the long held assumption of independence of friction on the sample shape and size.
It is particularly intriguing that, in some cases, localized sliding precursors nucleate long before the applied force reaches the  
static friction force at which the front propagates through the entire contact  interface \cite{rubinstein2004,rubinstein2007,maegawa2010}.
Numerical simulations of  friction models in one \cite{braun2009,bouchbinder2011,amundsen2012,sinai2012,capozza2012} and 
two dimensions \cite{tromborg2011,kammer2012,otsuki2013} allow to study the main features of the 
spatio-temporal dynamics of precursors. These numerical works have mainly focused on the qualitative dynamical aspects of propagation, reproducing the different dynamical regimes observed in experiments \cite{braun2009,bouchbinder2011,amundsen2012,sinai2012}
and the nucleation of the fronts under various loading conditions \cite{tromborg2011}. 

Based on the results of experiments \cite{rubinstein2007}  and numerical simulations \cite{tromborg2011}
it was suggested that frictional precursors evolve according to universal laws:
the sample size and normal load dependences of precursors lengths 
can be rescaled away and different experiments can be collapsed a single master curve. 
Establishing universal forms for slip precursors would be particularly important 
for earthquake forecasting \cite{rubinstein2011}.   Slip or stress accumulation on faults has been often observed 
to accelerate close to large earthquakes \cite{king1994,stein1994,stein1997}, but detailed predictions based on this are considered to be unreliable
\cite{hardebeck2008,grennhough2009}.  It is therefore extremely important to better clarify the
conditions leading to precursors and confirm their universality.  
Another puzzling aspect revealed by experiments is an apparent violation of the Amontons-Coulomb laws:
direct measurement of shear $\tau$ and normal stresses $\sigma$ close to the frictional interface indicated regions where the 
Coulomb stress $\tau_C = \left| \tau\right|-\mu \left|\sigma\right|$ is positive without inducing detachment  \cite{bendavid2010,bendavid2011}.
This result suggests that the friction coefficient $\mu$ might not be a well defined material constant as conventionally assumed.

 Scalar models are commonly used to study the planar crack front propagation in disodered elastic media \cite{barai2013,ramanathan1997}, in quasi-two dimensional geometries \cite{zapperi2000} and under antiplane shearing conditions \cite{deangelis1985}. On the other hand, recent experiments have been provided the evidence  that classical shear cracks singular solutions, originally devised to account for brittle fractures, offer a  quantitative excellent description of the static-to-dynamic friction transition \cite{svetlizky2014}. Here, combining  the solution of three dimensional scalar elastic equations in finite geometries
with simple contact mechanical rules for local slip at the frictional interface,  we reproduce accurately the experimentally
observed evolution of the contact area as the sample is loaded. In this way, we obtain a complete picture of the
role played by sample geometry and loading conditions on the precursors nucleation. Moreover we  
show that precursors originate from stress gradients on the contact interface and are therefore 
absent when loading is applied uniformly through the top of the slider.  Disorder induced precursors nucleation
of the kind predicted for sliding thin films and monolayers \cite{reguzzoni2010} should be strongly suppressed in three dimensions
due to long-range elastic interactions which make the coherence length extremely large
\cite{lorenz2012}.  When shear is applied on the slider side, however, we observe precursors whose evolution depends in a non-universal
way on the sample geometry. The occurrence of
universal profiles is explained by the symmetries of the interfacial shear stress obtained analytically.

\noi Despite its quasi-static nature, our model incorporates the Achard's law and the Amontons' first and second laws, reproducing several key features observed in
experiments including the discrete stress drops observed in correspondence of the slip
precursors. Most importantly, our model can help to assess which experimental feature can be attributed to the static elastic equilibrium, and which instead is a pure dynamical out-of-equilibrium aspect. 
Our calculations reproduce the experimental interfacial stress profiles detected at the frictional interface, 
before slip. We discuss the large fluctuations of the internal stresses during precursors activity in the bulk of the material, and we provide the numerical and analytical evidence of this large heterogeinity. This observation, substantiated also by finite element model simulations, suggests that drawing firm conclusions
based on the value of the Coulomb stress measured away from the contact interface, both in
laboratory experiments and in earthquake faults, could be problematic.

\section{Results \label{sec:evolution}}

\subsection{Simulations for different sample sizes and loading conditions}
Following Ref. \cite{rubinstein2007}, we first study how precursors  depend on  $L_x$ and on the normal load
$F_N$ when $F_S^{lat}$ is applied through a rod placed on the trailing edge, at height $h=6$mm. 
Experimental evidence suggests that the precursor size $\ell$ obtained for different values of $L_x$ and $F_N$ can be collapsed into a 
single master curve when normalized by $L_x$ and plotted versus $F_S^{lat}/\mu F_N$. Our numerical results reproduce quantitatively
the experimental findings as shown in Fig. \ref{fig:5}(a). In our model, however, we are able to change  $L_x$ over
a wider range than  in the experiments, revealing that  data collapse is in fact only approximate (see inset of Fig. \ref{fig:5}(a)). 
Similar behavior is obtained when we vary $L_y$ (Fig. \ref{fig:S4}) or $L_z$ (Fig. \ref{fig:S5}) keeping constant the other parameters: in all cases front precursors exhibit a dependence on the sample dimensions. We have also  changed $L_x$ and $L_y$ holding their ratio $L_x/L_y$ unchanged 
(Fig. \ref{fig:S6}),  $L_x$ and $L_z$ with $L_x/L_z$ constant (Fig. \ref{fig:S7}), or $L_y$ and $L_z$ with $L_y/L_z$ constant (Fig. \ref{fig:S8}).  Again, data collapse is not obtained, indicating that for this loading condition the precursor lengt $\ell$ depends in a non-trivial way on the sample dimensions $(L_x,L_y,L_z)$.  The general trend however is that precursory activity tends to decrease as the varying dimension is increased: for a larger sample we typically need a larger shear force to observe a precursor of a given length.

Experimental results in Ref.  \cite{rubinstein2007} also suggest that the height $h$ at which the lateral force is applied to the sample trailing edge has no influence on the evolution of the front precursors. While  this is true  for the range of $h$ used in experiments (see Fig \ref{fig:5}(b)), when we increase $h$ further the curves  no longer collapse. In particular, we find that the lateral force needed to nucleate the first precursor increases considerably with $h$ (see the inset of Fig \ref{fig:5}(b)). Remarkably, this effect persists when we increase both $h$ and $L_z$ leaving their ratio constant (Fig. \ref{fig:S9}). Yet, experiments have provided the evidence that the precursors length $\ell$ advances by periodical  discrete leaps of roughly equal size, which take place at nearly constant increments of  $F_S$. Moreover, this periodicity exhibits an apparent scaling with $h$, becoming larger with increasing $h$  \cite{rubinstein2009}. While the envelope of the curves reproduced by our model do show periodicity in the increments of the precursors sizes, the size of these discrete jumps and the corresponding increments of $F_S$ seem to remain unaffected by varying $h$, at least within the range of heights used in the experiments.

We explore further the dependence of precursors on the sample geometry by considering a different loading condition
in which the lateral force is applied uniformly on the sample side surface ($2\Delta h=L_z$). In this case, we find that the 
precursors are size independent when we vary $L_x$ and $L_y$  keeping their ratio $L_x/L_y$ constant (Fig. \ref{fig:6}(a)) or
$L_x$ and $L_z$ with constant $L_x/L_z$ (Fig. \ref{fig:6}(b)). When we vary instead $L_y$ and $L_z$, keeping constant $L_y/L_z$, no universality is found and
precursors again tend to disappear for large sample sizes  (Fig. \ref{fig:6}(c)). A similar effect is obtained using mixed mode loading
as in Refs. \cite{bendavid2010,bendavid2011}, applying simultaneously a shear force on the top surface 
and on the trailing edge. As the ratio between both forces $n \equiv F_S^{top}/F_S^{lat}$ increases, the length of the  precursor shrinks 
(Fig. \ref{fig:6}(d)) and disappear when loading is only applied  on the sample top plate.

Precursors are  defined by detecting the  decay of the real area of  contact. This feature is perfectly reproduced by our model, and, roughly speaking, it is ultimately  due to 
the  detachment of regions of the frictional interface satisfying the local static friction condition (\ref{eq:local_rule}).  Thus, the  dependence of the precursor envelope profile on the sample geometry should reflect the properties of the Coulomb stress across the entire contact interface. In the Supplementary Information (sec. \ref{geometrical}), we discuss how some general aspects of the precursor shape can be deduced from the symmetry of the Green function.

\subsection{Normal and shear stresses at the interface and the Amontons law \label{sec:coulomb}}

Direct experimental measurements of shear and normal stress profiles close to the contact interface show that the Coulomb stress can exceed zero locally, without inducing any detachment front, precursor or local slip \cite{bendavid2010,bendavid2011}. This is puzzling since it would
represent a local violation of the Amontons-Coulomb law, suggesting that the friction coefficient might not be a material constant. 
In our model, however, the local and global friction coefficient $\mu$ is fixed across the whole interface, and  local detachment
occurs if   $\tau_C(x,y,0) > 0$ by construction (Eqs.(\ref{eq:local_rule}) and (\ref{eq:local_rule_dscrt})). Yet, this apparent contradiction can be resolved by noting that local stresses in Refs.  \cite{bendavid2010,bendavid2011} are measured on a reference plane located at a height of $z_P=2$mm above the frictional interface. 

\noi Thanks to the analytical solvability of our model we can compute the shear and normal stresses at any points $(x,y,z)$ of the slider bulk: this is perfomed in the Suppl. Mat. sec. \ref{sec:calculation} (see Eqs.(\ref{tau_P_def}), (\ref{sigma_P_def}) or Eqs.(\ref{tau_P_def_dcrt}),(\ref{sigma_P_def_dcrt})). Calculating the stresses on the plane $z_P=2$mm yields a good quantitative agreement with experiments (Fig \ref{fig:7}(a), \ref{fig:S10}). In particular, the curves shown in Fig. \ref{fig:7} represent the shear and normal stresses averaged over the $y$ direction ($\tau(x,z_P)=\int_0^{L_y}dy\,\tau(x,y,z_P)/L_y$ and $\sigma(x,z_P)=\int_0^{L_y}dy\,\sigma(x,y,z_P)/L_y$), along the entire sample $0<x<L_x$, just before the onset of the first precursor, i.e. when no detachment is yet present at the contact interface. This can be seen also from  Fig.\ref{fig:S10} where the full quasi-static dynamics of $\tau(x,z_P)$ and $\sigma(x,z_P)$ are plotted.  The corresponding Coulomb stress on the same plane ($\tau_C(x,z_P)=\left|\tau(x,z_P)\right|-\mu\left|\sigma(x,z_P)\right|$ or $\tau^C_{z_P}[i]=\left|\tau_{z_P}[i]\right|-\mu\left|\sigma_{z_P}[i]\right|$ from Eq.(\ref{coulomb_stress_dscrt})) prior to the nucleation of the first precursor event is reported in Fig.\ref{fig:7}(b) (solid green line) showing again a good agreement with the experimental data.

Our result may suggest  the observed apparent violation of the Amontons first law \cite{bendavid2010,bendavid2011} could be due to the fluctuation undergone by the internal stresses in the material bulk, even in the vicinity (but not on) the slider frictional interface. Defining a local friction coefficient as the ratio $\mu(x,y,z)=\frac{\left|\tau(x,y,z)\right|}{\left|\sigma(x,y,z)\right|}$, is not an eligible procedure if the point $(x,y,z)$ does not lie on the frictional plane $(x,y,0)$. To substantiate this statement, in Fig.\ref{fig:7}(b) we compare the $y$-averaged Coulomb stress on the plane $z_P=2mm$  above the slider-bottom surface ($\tau_C(x,z_P)$, solid green curve) with the corresponding quantity at frictional the interface $\tau_C(x)=\int_0^{L_y}dy\,\tau_C(x,y)/L_y$ ( solid magenta curve). As it can be clearly seen, the Coulomb stress value may suffer large fluctuations according to the sample position where it is measured. Although the authors of the experiments in \cite{bendavid2010} were careful to perform the measurements at locations $x$ to ``avoid the effects of large stress gradients'', the agreement shown in Fig.\ref{fig:7} and the analytical calculations in Suppl. Mat. sec. \ref{sec:calculation} demonstrate that large fluctuations in the (positive) Coulomb stress appear in the bulk of the material, and are mainly due to the internal shear stress gradient resulting from the lateral shear applied.  A pictorial intuitive illustration of this argument is reported in Fig. \ref{fig:8}, where the $y$-averaged Coulomb stress $\tau_C(x,z)$ is shown for several values of the adiabatic force $F_S$, see also  Movie S2. Regions where $\tau_C(x,z)$ exceeds zero occur even before precursors nucleation (Fig. \ref{fig:8}(a)) and even far from the trailing edge. This may lead one to believe incorrectly that the friction coefficient is not a material constant.  In the same Fig.\ref{fig:8} (bottom panels) we report the full Coulomb stress $\tau_C(x,y)$ at the plane ($z=z_P$), showing that the average over $y$ is indeed a correct protocol to average out the noise-induced fluctuations.

\noi The bulk fluctuations of Coulomb stress are also present while considering a fully tensorial elasticity model. As a matter of fact, in the Suppl. Mat. sec. \ref{sec:FEM} we report shear and normal stress calculated by means of a finite element model (FEM). In Fig.\ref{fig:S12} we plot $\tau_C(x,y=L_y/2,z)$ arising from FEM simulations: large gradients of  Coulomb stress make any claim about the local value of $\mu$ highly questionable. This is seen before any detachment occurred at the interface (panel (a)), and also when a portion of the contact area is disconnected from the rough surface (panels (b)-(d)). We notice that in this case no analytical calculations can be carried out in the fashion of Suppl. Mat. sec. \ref{sec:calculation}, but the shear and normal stresses are numerically obtained by means of the FEM software. In particular we stress once again that the the normal component of the positive Coulomb stress  $\sigma(x,y,z)$ is largely influenced by the shear force $F_S$ as opposite to Eq.\ref{sigma_P_def}.

\subsection{Role of disorder in front nucleation}

The role of the substrate roughness heterogeneity on the nucleation of front appears, from Eqs.(\ref{eq:tau_surf_cont}),(\ref{eq:sigma_surf_cont}), rather complex. In general one can say that heterogeneity amplifies and modulate the different contributions to normal and shear interface stresses $\sigma_{surf}$ and $\tau_{surf}$. For narrow roughness distributions (in most of the experiments the roughness appears to be a very well-controlled parameter), we do not expect that the substrate disorder may play a major role promoting or suppressing the precursors dynamics, at least not comparable to the role expressed by the force-induced stress gradients. This is what clearly appears from our analysis, in step with the simplified 1D model \cite{capozza2012} and with the experimental outcomes. It is possible, however, that large roughness fluctuations may induce internal stress gradients leading to precursors nucleation, even in regions far from the trailing edge. This is a particularly interesting issue, since the precursor could originate as a stable detachment droplet, irrespective of the type of loading exerted (whether lateral or top shearing). The key question is therefore to determine the conditions for which a detachment droplet constitutes a meta-stable state.  The question could be addressed by defining the  contact interface energy density as $\varepsilon(x,y)=\tau_{surf}(x,y)u_x(x,y)+\sigma_{surf}(x,y)\left[u_z(x,y)-u_z^0(x,y)\right]$ \cite{persson2001}, and the energy change associated to the transition from an initial stable configuration to a second on which the droplet has formed: $\Delta E (r) =  \int_{\Sigma(r)}dx dy \, \tau_{surf}(x,y)u_x(x,y)+\sigma_{surf}(x,y)\left[u_z^{(R)}(x,y)-u_z^{0(R)}(x,y)\right]-\int_0^{L_x}dx\int_0^{L_y} dy \, \tau_{surf}(x,y)u_x(x,y)+\sigma_{surf}(x,y)\left[u_z(x,y)-u_z^0(x,y)\right]$, where $\Sigma(r)$ represents the contact surface configuration including the nucleated droplet of average size $r$. A detachment droplet configuration will be stable if the energy penalty $\Delta E (r)$ has a positive maximum for some $r$. Now, asking which physical conditions allow the formation of a stable droplet, means  which sample dimensions $L_x$, $L_y$, $L_z$, roughness $w^2$ and force $F_S(<\mu F_N)$ 
give a  $\Delta E (r)$  with a positive maximum (with $r<L_x,L_y,L_z$). Unfortunately, due to the intricacy of expressions (\ref{eq:tau_surf_cont}) and (\ref{eq:sigma_surf_cont}),  we could obtain the answer only by numerical simulations. However, albeit one cannot completely exclude that detachment regions appear on length-scales which are well below our and experimental resolution ($\sim 1$mm), the set of parameters used in our simulations and in the experiments does not allow for a stable disorder-induced droplet, therefore sliding occurs either by precursors nucleation from the trailing edge or as first-order phase transition  for top shearing. On the other hand, it is expected that in thin films  ($L_y/L_z\to\infty$, $L_x/L_z\to\infty$), interfacial disorder may induce a droplet nucleation
of the kind predicted in Ref.\cite{reguzzoni2010}. However, since the calculation of $\Delta E (r)$ involves two equilibrium configurations, a quasi-static model is the right candidate to tackle it.   

\section{Discussion \label{sec:discussion}}

In this paper we have introduced a scalar  model for the onset of frictional sliding of a three dimensional elastic object resting on a rough surface. We have devised a scalar elasticity model which allows an analytical treatment of the relevant quantities, and the straightforward implementation of the  quasi-static dynamics. This model incorporates, for the first time, mesoscopic laws of contact mechanics at the frictional interface,  reproducing with remarkable precision Archard's law and Amonton's first and second laws. Most importantly the scalar model is capable of reproducing with good accuracy the real contact area dynamics, the precursors' envelope dynamics prior to the transition to sliding, and the normal and shear internal stress distributions close to the slider-substrate interface. The model stems from a strong Ansatz, namely that  the  components of displacements $u_x$ and $u_z$ are decoupled and $u_y\simeq 0$. However, the solution of the model is exact: if one accepts the initial Ansatz, one has at hand the analytical expression for any physical observable in static equilibrium. The numerical implementation of the model is required to take into account the statistical heterogeneity inherent to  the asperity disorder of the underlying rough substrate.


 The first limitation of our model has been discussed previously, and consists in neglecting the Poisson expansion and the sample torque, which can have strong implications only in the case of top uniform shearing conditions, although for very large samples the scalar models conclusions should be respected. Hence no firm general conclusions nor predictions on the occurrence of frictional sliding and precursor dynamics can be drawn based on these observations. 
Earthquakes faults are mostly driven uniformly from a distance implying that, in average,  precursory activity should not be present. Stress gradients and hence precursor activity could,  however, arise either due to local heterogeneities or
because the fault plane is tilted with respect to the earth crust \cite{doglioni2011}. The lack of universal scaling forms dictating the precursors evolution, however, makes any forecasting of catastrophic events extremely difficult, especially when we do not know precisely the loading conditions.

The second limitation certainly lies in the quasi-static nature of the  model. But what at first glance may  seem like a strong simplification is in fact a point of strength. The reasons are the following:

\noi First, the equilibrium problem requires a very limited number of adjustable parameters to set up the model. Our derivation indeed requires the fine tuning of only two parameters, connected to the normal and transverse spring stiffness inspired by the theory of contact mechanics, and whose physical meaning and interpretation are straightforward (see Suppl. Mat. sec. \ref{sec:calibration}). To estimate these parameters, we only need a direct comparison with simple experiments, such as the validation of the Archard's law \cite{archard1953,rubinstein2006}  to tune the normal stiffness (see Fig.\ref{fig:2}(a)), and an experiment like the one reported in Ref. \cite{berthoud1998} for the transverse stiffness. This eliminates from the picture
a host of dynamical quantities that are often difficult to quantify, or even to justify from the physical point of view. This is the case, for instance, of phenomenolgical friction coefficients interpolating between statics and dynamics employed in 1D \cite{amundsen2012} and 2D models \cite{tromborg2011,otsuki2013,kammer2012}, or the bulk damping coefficient $\gamma$, whose numerical value is usually put in by hand \cite{braun2009,bouchbinder2011,amundsen2012,sinai2012,capozza2012,tromborg2011,otsuki2013,kammer2012}, or of the actual value of the reattachment delay time $\tau$ responsible for the frictional interface contacts rejuvenation \cite{braun2009,bouchbinder2011,capozza2012}. While it is out of doubt that, upon cessation of motion, the contacts at the interface reform and strengthen \cite{rubinstein2006, maegawa2010}, it is very hard to infer the rejuvenation characteristic time $\tau$ from experimental data. In our model, we did not include the contacts reformation process within our quasi-static protocol, showing that it is not a necessary ingredient to recover the precursors overall profile.
 
\noi Second, our model may assess which of the observed experimental features are due to the out-of-equilibrium dynamics and which are mostly due to equilibrium properties.  For instance, our model is able to recover the precursors steps and the shape of their envelope,  but fails to reproduce the increase in the precursors 
waiting times when the shear is applied at higher and higher $h$. Thus, we can conclude that this intriguing aspect is probably due to inertial effects present when shear is applied through an external spring (see Eq.(\ref{eq:spring_shear_force})). To check this experimentally, it would be sufficient to change the spring displacement $U_S$ rate  and detect any possible change in the leaps phenomenology. To the contrary, our model allows to establish that the occurrence of precursors is in fact a quasi-static physical process. Any of the equilibrium states reached by the slider during the adiabatic evolution, is just one of the meta-stable configurations in which the system can dwell. This large number of meta-stable states is mainly due to the disorder heterogeneity of the roughness at the interface, and to a much minor degree to the rules adopted to detach the contacts when they satisfy the condition $\tau_C(x,y,0)>0$. 

\noi Because of its quasi-static nature, our model cannot reproduce the detachment front dynamics. According to the definition provided in Refs.\cite{rubinstein2004,rubinstein2006bis,rubinstein2007,rubinstein2009,bendavid2010} a detachment front indicates a drastic reduction of the real area of contact which takes place on time scales which are roughly in the millisecond range. The entire precursor experiment occurs instead over a few minutes \cite{rubinstein2007,rubinstein2009}.  Experiments have revealed three different types of crack-like rupture fronts, slow, sub-Rayleigh, and intersonic (or supershear), according to their propagation velocity through the frictional interface. Precursors advance by arrested front propagation: discrete increments, indeed, occur by rupturing the contact interface at a velocity which corresponds to sub-Rayleigh fronts at the begining, and to slow fronts close to  the sliding transition \cite{rubinstein2009}. In particular a final slow front is responsible of the static-to-dynamics frictional sliding. Our model does not capture the crack-like propagation of fronts, since the fronts are a dynamical out-of-equilibrium processes in between two equilibrium states, namely between precursors. Nevertheless, our model might substantiate the experimental observation on the relation between precursors appearance and slow fronts  triggering the frictional sliding. As a matter of fact, in Fig.\ref{fig:7} we were able to reproduce quantitatively the shear and normal stress profiles before any precursor nucleation occurred. In Ref.\cite{bendavid2010} these stress distributions were related to the ensuing slow rupture front (see Fig.2A in \cite{bendavid2010}). Thus it is possible to argue that whenever we observe a precursor activity, the transition to sliding is triggered by slow fronts. 

To summarize the central finding of our work, three dimensional finite body scalar Green's function makes it possible to investigate the dependence of many physical observables on any sample parameter. Our results show that the evolution of the fronts depends in a non-universal way on the loading conditions and the sample dimensions and shape. Only for some loading condition, the precursors follow a curve  that allows for a simple universal rescaling in terms of the sample dimension: this prediction can be experimentally checked. Moreover we have shown that large stress gradients take place not only at the frictional interface but also within the material bulk. These gradients are mainly due to the way the external shear is applied and to the sample geometry, on top of frustrated Poisson expansion and elastic torque. Hence no firm general conclusions nor predictions on the occurrence of frictional sliding and precursor dynamics can be drawn based on these observations. 

Earthquakes faults are mostly driven uniformly from a distance implying that, in average,  precursory activity should not be present. Stress gradients and hence precursor activity could,  however, arise either due to local heterogeneities or
because the fault plane is tilted with respect to the earth crust \cite{doglioni2011}. 
Measurements of local variations of the Coulomb stress  around earthquake faults
have been used to assess the correlation between stress accumulation and earthquake triggering
\cite{king1994,stein1994,stein1997}. Predicting earthquakes based on slip or stress accumulation has been so far an elusive task  \cite{hardebeck2008,grennhough2009},  and the reason behind this failure can be addressed in the scenario pictured in our analysis.  Indeed, as illustrated,  we find that for loading condition leading to large stress gradients, the evolution of the Coulomb stress measured above the contact interface provides only a rough indicator of the ensuing  detachment front dynamics, which instead appears to be very well characterized by the real contact area variation. Furthermore, the lack of universal scaling forms dictating the precursors evolution, makes any forecasting of catastrophic events extremely difficult, especially when we do not know precisely the loading conditions. 

\section{Methods}

\subsection{The scalar model \label{sec:model}}
We consider an elastic PMMA macroscopic body resting in equilibrium on a rough surface.  We derive  the equations for the displacements of the elastic body subject to external forces, within the scalar elasticity approximation. In particular, we are interested in the solutions for the displacement fields at the frictional interface. There is no direct connection of the model theory to actual elasticity. The latter involves three displacement components and a system of coupled equilibrium equations enforced with boundary conditions.

\begin{figure}[ht]
\centering
\includegraphics[width=0.5\textwidth]{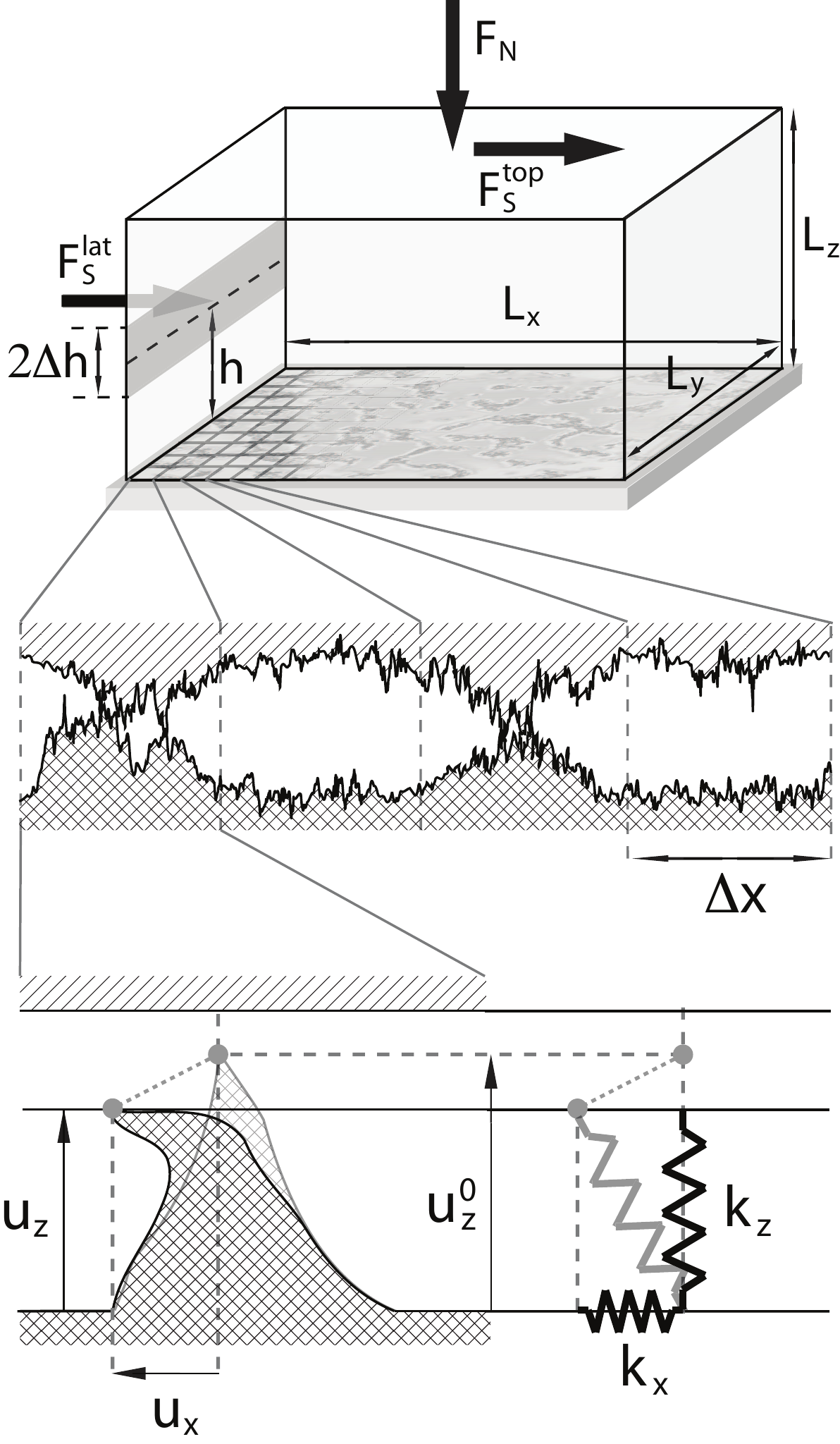} 
\caption{{\bf Graphical illustration of the model}.  We consider a block of dimensions $L_x$, $L_y$ and $L_z$ in contact with a rough surface (sketched in the middle panel).  A normal force $F_N$ and a shear force $F^{top}_S$ are applied uniformly on the top surface along $z$ and $x$ respectively,  a lateral force $F^{lat}_S$ is applied on the sample trailing edge over a rectangular region of width $2\Delta h$ at height $h$. The bottom surface of the block is discretized on a grid of size $\Delta x\times \Delta y$ (we invariably chose $\Delta y=\Delta x=1$mm, see also Fig.\ref{fig:S1}). Each grid element central point  may form an elastic contact with the rough surface, that is modelled by a set of elastic asperities
of height $u_z^0$, and effective transverse and normal stiffness $k_x$ and $k_z$, respectively.
\label{fig:1}}
\end{figure}

Let us first illustrate the physical system that our model aims at reproducing, which coincides with the experimental setup described in Refs.\cite{rubinstein2004,bendavid2010,bendavid2011,rubinstein2007,rubinstein2011,svetlizky2014},  see Fig.\ref{fig:1} . The experiments were conducted using  two PMMA blocks in contact, one  on top of the other. The top block, of dimensions 140,150,200mm $\times$ 6mm $\times$ 75,100mm (according to the different experiments performed), was pushed against a bottom block of  dimensions 250mm $\times$ 30mm $\times$ 28mm in the $\hat{x}, \hat{y}$ and $\hat{z}$ directions respectively. In general, the condition $L_y\ll L_x, L_z$ was always satisfied. The two blocks were pushed together by a normal load  $F_N$ and, while the bottom block was fixed, the top block was subject to a shearing process by means of the lateral force $F_S$ applied solely on the $\hat{x}$ direction. This experimental system was usually adopted, with the only exception of Ref.\cite{bendavid2010}, where the experiments were also conducted clamping the top block at the top edge and applying the shear $F_S$ to the bottom block. However, the relative blocks movement is constrained by the frictional resistance at the interface offered by roughness-induced surface forces. In our model, we consider for simplicity the bottom block to be infinite (\emph{the substrate}) and only the top block shearing. The surface stresses at the interfaces,  $\varphi_{surf}(x,y,0)$, are formally distributions accounting for the spatial hetereogeneity of the PMMA roughness.

Since no external force is acting on the $\hat{y}$ direction, and because the sample geometry fulfills the condition $L_y\ll L_x, L_z$, we have assumed 

\be
u_y\simeq 0.
\label{eq:approximation_1}
\ee

\noi Moreover, the scalar elasticity yields that  the stress tensor satisfies $\sigma_{yk}=0$ (where $k=x$, $y$, or $z$). Thus the scalar equations for the decoupled displacement fields take the following form

\be
\begin{array}{c}
        \frac{E}{(1+\nu)}\left[\frac{\partial^2u_{x}}{\partial x^2}+\frac{\partial^2u_{x}}{\partial y^2}+\frac{\partial^2 u_{x}}{\partial z^2}\right]\simeq 0\\
        \\
        \frac{E}{(1+\nu)}\left[\frac{\partial^2u_{z}}{\partial x^2}+\frac{\partial^2u_{z}}{\partial y^2}+\frac{\partial^2u_{z}}{\partial z^2}\right]\simeq 0\\
    \end{array}
\label{eq:equilibrium}
\ee

\noi where  $E$ represents the  Young's modulus and 
$\nu$  the  Poisson's ratio. The scalar elasticity Eqs.(\ref{eq:equilibrium}) can be analyically treatable, once one specifies the proper boundary conditions. At the  equilibrium,  internal stresses at the surface must counterbalance the external forces acting on the sample. Since we consider a slider of dimensions $\left[L_x, L_y, L_z\right]$ the boundary conditions for Eqs.(\ref{eq:equilibrium})  are

\be
\begin{array}{ccccc}
      \sigma_{xx}(0,y,z)&=&\frac{E}{(1+\nu)}\left.\frac{\partial u_{x}}{\partial x}\right|_{0,y,z}&=&-\frac{F_S^{lat}}{L_y2\Delta h}\theta\left(\zeta-h+\Delta h\right)\left[1-\theta\left(\zeta-h-\Delta h\right)\right]\\
      \sigma_{xz}(x,y,L_z)&\simeq&\frac{E}{(1+\nu)}\left.\frac{\partial u_{x}}{\partial z}\right|_{x,y,L_z}&=&-\frac{F_S^{top}}{L_xL_y}\\
      \sigma_{xz}(x,y,0)&\simeq&\frac{E}{(1+\nu)}\left.\frac{\partial u_{x}}{\partial z}\right|_{x,y,0}&=&-\varphi_{surf}^x(x,y,0)\\
      \sigma_{zz}(x,y,L_z)&=&\frac{E}{(1+\nu)}\left.\frac{\partial u_{z}}{\partial z}\right|_{x,y,L_z}&=&-\frac{F_N}{L_xL_y}\\
      \sigma_{zz}(x,y,0)&=&\frac{E}{(1+\nu)}\left.\frac{\partial u_{z}}{\partial z}\right|_{x,y,0}&=&-\varphi_{surf}^z(x,y,0).\\
    \end{array}
\label{eq:boundary}
\ee

\noi   As shown in Fig.\ref{fig:1}(a), $F_S^{lat}$ corresponds to a shear force in the $x$ direction, applied to the elastic slider on a portion of the plane $(0,y,z)$ of size $L_y\times 2\Delta h$ centered around  $z=h$ and $\theta(x)$ stands for the Heavyside step function;  $F_s^{top}$ is a shear force (also pointing to the $x$ direction) uniformly applied on  top  of the slider; $F_N$ is the normal force, i.e. a force applied on the entire top plane and pointing toward $-\hat{z}$; the surface stresses $\varphi_{surf}^{x,z}(x,y,0)$ represent the interaction between the elastic body and the rough underlying surface at the plane $(x,y,0)$, in the $x$ and $z$ direction respectively (see Fig.\ref{fig:S1}). With the boundary conditions (\ref{eq:boundary}), we can solve the equilibrium Eqs. (\ref{eq:equilibrium}) for the displacement fields on the slider  bottom plane. In technical term, we have to solve two independent Laplace equations with von Neuman boundary conditions.  The solutions of Eqs.(\ref{eq:equilibrium}) are obtained by  generalizing to three dimensions the corresponding solution for the von Neuman problem  in two dimensions \cite{roach1970}.
The result reads

\be
\begin{array}{l}
u_x(x,y,0)=\langle u_x\rangle + \frac{(1+\nu)}{E}\left\{\int_0^{L_x}d\xi\int_0^{L_y}d\eta
G(x,\xi;y,\eta;0,0)\varphi_{surf}^x(\xi,\eta,0)-\right.\\
\left.-\frac{F_s^{top}}{L_xL_y}\int_0^{L_x}d\xi \int_0^{L_y}d\eta\,G(x,\xi;y,\eta;0,L_z)+\frac{F_s^{lat}}{L_y2\Delta h}
\int_0^{L_y}d\eta\int_{h-\Delta h}^{h+\Delta h}d\zeta G(x,0;y,\eta;0,\zeta)\right\}
\end{array}
\label{eq:x_solution}
\ee

\be
\begin{array}{l}
u_z(x,y,0)=\langle u_z\rangle + \frac{(1+\nu)}{E}\left\{\int_0^{L_x}d\xi\int_0^{L_y}d\eta\, G(x,\xi;y,\eta;0,0)\varphi_{surf}^z(\xi,\eta,0)-\right.\\
\left.-\frac{F_N}{L_xL_y}\int_0^{L_x}d\xi\int_0^{L_y}d\eta\, G(x,\xi;y,\eta;0,L_z)\right\}
\end{array}
\label{eq:z_solution}
\ee

\noi where   $G(x,\xi;y,\eta;z,\zeta)$ is the Green function:

\be
\begin{array}{l}
G(x,\xi;y,\eta;z,\zeta)=\\
\frac{8}{L_xL_yL_z}\sum\limits_{n=0}\sum\limits_{m=0}\sum\limits_{p=0}\gamma_{nmp}\frac{\cos\left(\frac{n\pi z}{L_z}\right)\cos\left(\frac{n\pi \zeta}{L_z}\right)\cos\left(\frac{m\pi y}{L_y}\right)\cos\left(\frac{m\pi \eta}{L_y}\right)\cos\left(\frac{p\pi x}{L_x}\right)\cos\left(\frac{p\pi \xi}{L_x}\right)}{\left(\frac{n\pi}{L_z}\right)^2+\left(\frac{m\pi}{L_y}\right)^2+\left(\frac{p\pi}{L_x}\right)^2}
\end{array}
\label{eq:green}
\ee

\noi with  $\gamma_{000}=0,\, \gamma_{0mp}=\gamma_{n0p}=\gamma_{nm0}=1/2,\, \gamma_{00p}=\gamma_{0m0}=\gamma_{n00}=1/4,$ and  $\gamma_{nmp}=1$ otherwise.
In the Suppl. Mat. sec. \ref{sec:rapidly} we furnish the analytical procedure for the fast convergence  of the sum in Eq.(\ref{eq:green}).

 \noi The quantitites $\langle u_x\rangle$ and $\langle u_z\rangle$ represent two arbitrary constants corresponding to the displacement fields averaged over the entire  sample volume. They must be chosen imposing two additive equilibrium  constraints, namely that each component of the the surface forces, over the whole contact surface, must be equal and opposite to the external forces:

\be
\begin{array}{l}
\int_0^{L_x}d\xi\int_0^{L_y}d\eta\,\varphi_{surf}^x(\xi,\eta,0) =-F_S^{top}-F_S^{lat}\\
\int_0^{L_x}d\xi\int_0^{L_y}d\eta\,\varphi_{surf}^z(\xi,\eta,0) =-F_N.
\end{array}
\label{eq:additional_constraints}
\ee

 The last step is to provide an adequate analytical expression accounting for the effects of the interface on the slider mechanics, this is done by introducing the surface forces $\varphi_{surf}^x$ and $\varphi_{surf}^z$. In  Suppl. Mat. sec. \ref{sec:surface} we derive the expression of these forces, according to the contact mechanics theories. In  first approximation they are both linear in the displacements $u_z$ and $u_x$, i.e., for the force acting on the $\hat{z}$ direction, we have

\be
\varphi_{surf}^z(x,y)=\left\{
\begin{array}{ccc}
k_z(x,y)\left[u^0_z(x,y)-u_z(x,y,0)\right] & & u_z<u^0_z\\
0                                         & & u_z>u^0_z
\end{array}
\right.
\label{eq:surface_z}
\ee

\noi with $k_z(x,y)= \frac{c_z}{u^0_z(x,y)}$;  $u^0_z(x,y)$ is a random displacement that models the height fluctuations of the rough substrate (see Fig.\ref{fig:S1}). It is formally an uncorrelated noise extracted from a Gaussian distribution with a variance $\sigma_\xi^2=1 \mu m$, being $\sigma_\xi^2$ the roughness of the underlying surface \cite{bendavid2011,rubinstein2004,rubinstein2006,rubinstein2007,bendavid2010,bendavid2010b} (see Suppl. Mat. sec. \ref{sec:calibration}).
The interfacial interactions along $\hat{x}$ are given by 

\be
\varphi_{surf}^x(x,y)=-k_x(x,y)u_x(x,y,0)
\label{eq:surface_x}
\ee

\noi where $k_x(x,y)= c_xe^{-\frac{u_z(x,y,0)}{u^0_z(x,y)}}$ and $u_x(x,y,0)\gtreqless 0$. The constant $c_z$ and $c_x$ appearing in the expression of $k_z$ and $k_x$ are the only two adjustable parameters that our model encompasses (see the next subsection and Suppl. Mat. sec. \ref{sec:calibration}).   The expressions for $k_x$ and $k_z$ respect the laws of contact mechanics \cite{persson2001,persson2006,persson2007} and are entirely motivated by experiments:  the transverse (or tangential) stiffness $k_x$
of PMMA is indeed proportional to the normal load \cite{berthoud1998}, which in general decreases exponentially with
the vertical elastic displacement  \cite{benz2006}; the normal stiffness is  $k_z\sim -dP/du_z$, where $P\sim \exp(-u_z/u_z^0)$ is the squeezing pressure \cite{campana2011,benz2006}. Therefore, internal stresses are not decoupled at the interface, as they are connected via the local normal pressure entering the definition of the tangential stiffness $k_x$.  

\noi  Finally, introducing the linear expressions (\ref{eq:surface_z}) and (\ref{eq:surface_x}) into the Eqs.(\ref{eq:z_solution}) and  (\ref{eq:x_solution}) respectively, we obtain closed equations for the displacements  at the contact plane:

\be
\begin{array}{l}
u_x(x,y,0)=\langle u_x\rangle-\frac{(1+\nu)}{E}\left\{\int_0^{L_x}d\xi\int_0^{L_y}d\eta\, G(x,\xi;y,\eta;0,0)k_x(\xi,\eta)u_x(\xi,\eta,0)+\right.\\
+\frac{F_s^{top}}{L_xL_y}\int_0^{L_x}d\xi\int_0^{L_y}d\eta\, G(x,\xi;y,\eta;0,L_z)
-\\
\left.-\frac{F_s^{lat}}{L_y2\Delta h}\int_0^{L_y}d\eta\int_{h-\Delta h}^{h+\Delta h}d\zeta\, G(x,0;y,\eta;0,\zeta)\right\}
\end{array}
\label{eq:x_solution_linear}
\ee

\be
\begin{array}{l}
u_z(x,y,0)=\langle u_z\rangle+\\
+\frac{(1+\nu)}{E}\left\{\int_0^{L_x}d\xi\int_0^{L_y}d\eta\, G(x,\xi;y,\eta;0,0)k_z(\xi,\eta)\left[u^0_z(\xi,\eta)-u_z(\xi,\eta,0)\right]-\right.\\
\left.-\frac{F_N}{L_xL_y}\int_0^{L_x}d\xi\int_0^{L_y}d\eta\, G(x,\xi;y,\eta;0,L_z)\right\}.
\end{array}
\label{eq:z_solution_linear}
\ee

\noi In the former expressions  for the interfacial displacements, terms involving the contributions arising from the external shear or normal forces $F_S$ and $F_N$ can be calculated analytically. This is, indeed, one of the novelties that our model introduces: the complete expression of the Green function  (see Eq.(\ref{eq:green})) allows the determination of any of the force-induced components in the interfcial displacements equations. It will be clear in the next sections that this property entails the critical interpretation of the experimental and numerical results and, in particular, it furnishes precise predictions on the precursors' appeareance and dynamics and their dependence on the slider dimensions.  In the Suppl. Mat. sec. \ref{sec:external} we give the full analytical derivation of the terms proportional to $F_s^{top}$, $F_s^{lat}$ and $F_N$ appearing in Eqs.(\ref{eq:x_solution_linear}) and (\ref{eq:z_solution_linear}). Hereby, to simplify the displacements expressions, we introduce the following shorthand notations:
 
\be
u_S^{top}(x,y,0)=-\frac{(1+\nu)}{E}\frac{F_s^{top}}{L_xL_y}\int_0^{L_x}d\xi\int_0^{L_y}d\eta\, G(x,\xi;y,\eta;0,L_z) 
\label{eq:u_s_top}
\ee

\be
u_S^{lat}(x,y,0)=\frac{(1+\nu)}{E}\frac{F_s^{lat}}{L_y2\Delta h}\int_0^{L_y}d\eta\int_{h-\Delta h}^{h+\Delta h}d\zeta\, G(x,0;y,\eta;0,\zeta).
\label{eq:u_s_lat}
\ee

\be
u_N(x,y,0)= -\frac{(1+\nu)}{E}\frac{F_N}{L_xL_y}\int_0^{L_x}d\xi\int_0^{L_y}d\eta\, G(x,\xi;y,\eta;0,L_z),
\label{eq:u_n}
\ee

\noi thanks to which the Eqs.(\ref{eq:x_solution_linear}) and (\ref{eq:z_solution_linear}) take the form

\be
u_x(x,y,0)=\langle u_x\rangle-\frac{(1+\nu)}{E}\int_0^{L_x}d\xi\int_0^{L_y}d\eta\, G(x,\xi;y,\eta;0,0)k_x(\xi,\eta)u_x(\xi,\eta,0)+u_S^{top}(x,y,0)+u_S^{lat}(x,y,0)
\label{eq:x_solution_linear_simplified}
\ee

\noi and 

\be
u_z(x,y,0)=\langle u_z\rangle
+\frac{(1+\nu)}{E}\int_0^{L_x}d\xi\int_0^{L_y}d\eta\, G(x,\xi;y,\eta;0,0)k_z(\xi,\eta)\left[u^0_z(\xi,\eta)-u_z(\xi,\eta,0)\right]+u_N(x,y,0).
\label{eq:z_solution_linear_simplified}
\ee

\subsection{Discretization, numerical solution and quasi-static dynamics}

The slider interfacial displacements, i.e. the solutions of the Eqs. (\ref{eq:x_solution_linear_simplified}) and (\ref{eq:z_solution_linear_simplified}), are  achieved by discretization of the slider  bottom plane. As a matter of fact, we  place a square grid on the contact plane, composed by  elements having an area  $\Delta x\times\Delta y$, 
so that  $L_x=N_x\Delta x$ and $L_y=N_y\Delta y$ with $\Delta x=\Delta_y=1mm$ (see Fig.\ref{fig:S1}).  Albeit the terms in Eqs. (\ref{eq:x_solution_linear_simplified}) and (\ref{eq:z_solution_linear_simplified}) are defined on the 
entire contact plane $(x,y,0)$, we calculate them only on each single point $(x,y)$, which is the center of  the grid element. This is the case, by instance, of the surface forces $\varphi_x$ and $\varphi_z$ in Eqs.(\ref{eq:surface_x}) and (\ref{eq:surface_z}) respectively, which are formally distributions: we interpret them as acting effectively only on the grid center point,  representative of the enclosed area $\Delta x\times\Delta y$, as shown in Fig.\ref{fig:S1}.

\noi  In the Suppl. Mat. sec. \ref{sec:discretization} we report the formal derivation of the discretization technique, which leads to the following expressions for the linear inversion Eqs.(\ref{eq:x_solution_linear_simplified}) and (\ref{eq:z_solution_linear_simplified}):

\be
u_x[i]=\sum\limits_{j=1}^{N_xN_y}A^x_{ij}\left\{u_S^{top}[j]+u_S^{lat}[j]+\langle u_x\rangle\right\}
\label{eq:x_solution_vec_i}
\ee

\noi and 

\be
u_z[i]=\sum\limits_{j=1}^{N_xN_y}A^z_{ij}\left\{v_z^0[j]+u_N[j]+\langle u_z\rangle\right\},
\label{eq:z_solution_vec_i}
\ee

\noi where the matrices $A^x_{ij}$ and $A^z_{ij}$ are defined in Eqs.(\ref{Ax}) and (\ref{Az}) respectively and the vector $\vec{v}_z^0$ in Eq.(\ref{v_0}). With the former expressions at hand, we can calculate the equilibrium displacements along $\hat{x}$ and $\hat{z}$, compatible with given values of the external normal and shear forces: we hereby recall that the two constants $\langle u_x\rangle$  and $\langle u_z\rangle$ are set to ensure that the surface forces counterbalance the external loads (see Eqs.(\ref{average_u_x}) and (\ref{average_u_z})).

 In a typical simulation, external shear forces are increased quasi-statically
and the actual values of the local interfacial displacements are calculated numerically at the discretized bottom interface thanks to Eqs.(\ref{eq:z_solution_vec_i}) and (\ref{eq:x_solution_vec_i}) respectively (see Fig.\ref{fig:S1}). Indeed,  we first check for the equilibrium along  $\hat{z}$, and secondly along $\hat{x}$.
Contact springs are disconnected, i.e. irreversibly broken, whenever the local Coulomb stress satisfies

\be
\tau_C(x,y,0) \equiv |\tau_{surf}(x,y)|-\mu|\sigma_{surf}(x,y)|=|\varphi_{x}(x,y)|-\mu|\varphi_{z}(x,y)|>0
\label{eq:local_rule}
\ee

\noi where $\mu$ represents the static local friction coefficient set to $\mu=0.5$.  When the condition (\ref{eq:local_rule}) is fulfilled, the corresponding interface portion is detached from the underlying surface resulting in a local slip event. Every time a spring is broken, the equilibrium Eqs.(\ref{eq:z_solution_vec_i}) and (\ref{eq:x_solution_vec_i}) are recalculated with the new boundary condition, i.e. setting to 0 the interfacial forces corresponding to the broken spring.

\noi The overall sliding occurs when none of the interfacial contacts has survived, i.e. when  $\tau_C(x,y,0)>0$ across  the entire bottom plane (see the flowchart in Fig.\ref{fig:S11}).  This happens when the static friction force is equal to $\mu F_N$, satisfying the Amonton's first law. Therefore $\mu$ represents both the local and global friction coefficient. The detailed protocol of the quasi-static protocol enforced is 
reported in Suppl. Mat. sec. \ref{sec:quasi}.


\subsection{Model calibration}

One of the key features of our scalar quasi-static model is that Archard's principle is imposed at the mesoscopic scale \cite{archard1953} since the area of real contact of a grid element is function of the vertical load acting on it, i.e. $A_R(x,y)\sim e^{-\frac{u_z(x,y,0)}{u_z^0(x,y)}}$, see Eq.(\ref{rubinstein_local_final}). This is in  agreement with the mechanics of the interfacial asperities, whose transverse stiffnesses are $k_x\sim e^{-\frac{u_z(x,y,0)}{u_z^0(x,y)}}\sim A_R(x,y)$, and with the elastic-response picture emerging from the experiment of Berthoud and Baumberger \cite{berthoud1998}. The definition of the on-site real contact area offers the advantage of controlling the fluctuations of the total contact area  during the entire quasi-static evolution. As a matter of fact, it is possible to define the  the total real area of contact as

\be
A_R=\int _0^{L_x}dx\int _0^{L_y}dy A_R(x,y)
\label{eq:real_area_of_contact_ct}
\ee

\begin{figure}[ht]
 \centering
 \includegraphics[width=0.8\textwidth]{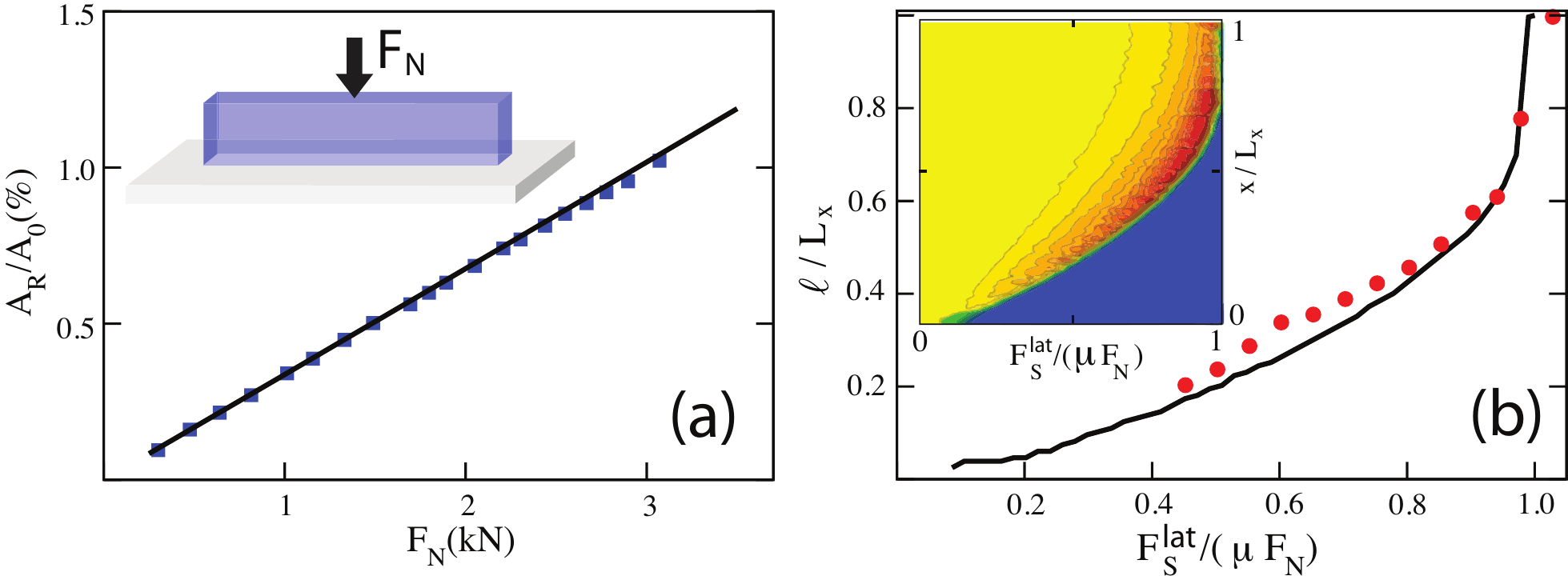} 
\caption{ {\bf Scalar model numerical calibration}. (a) The value of the normal stiffness $k_z$ is set by measuring the change of the real area of contact (in $\%$ of the nominal contact area $A_0$) as a function of the normal load $F_N$,  and comparing the numerical results (squares) with the experimental data from Ref. \protect\cite{rubinstein2006} (solid line). $L_x=30$mm, $L_y=6$mm, $L_z=150$mm.  (b) The transverse stiffness $k_x$ is obtained by matching 
the quasi-static evolution of the precursor position $\ell$  (solid black line) with the corresponding  experimental data reported in Ref. \protect\cite{rubinstein2007} (red circles).  $L_x=140$mm, $L_y=7$mm, $L_z=75$mm, $F_N=3.5$kN.   The inset shows the front propagation by imaging the real contact area $A_R(F_S)$ averaged over the $y$ direction and normalized to its initial value $A_R(0)$ (the image is the top view of the histogram shown in  Fig.\ref{fig:3}(b)).  Color code: blue reflects a decrease of the real area of contact with respect to the initial value, while red highlights a prominent increase. Precursor fronts correspond to the edge of the blu part of the plot.
\label{fig:2}}
\end{figure}

\begin{figure}[ht]
\centering
\includegraphics[width=0.8\textwidth]{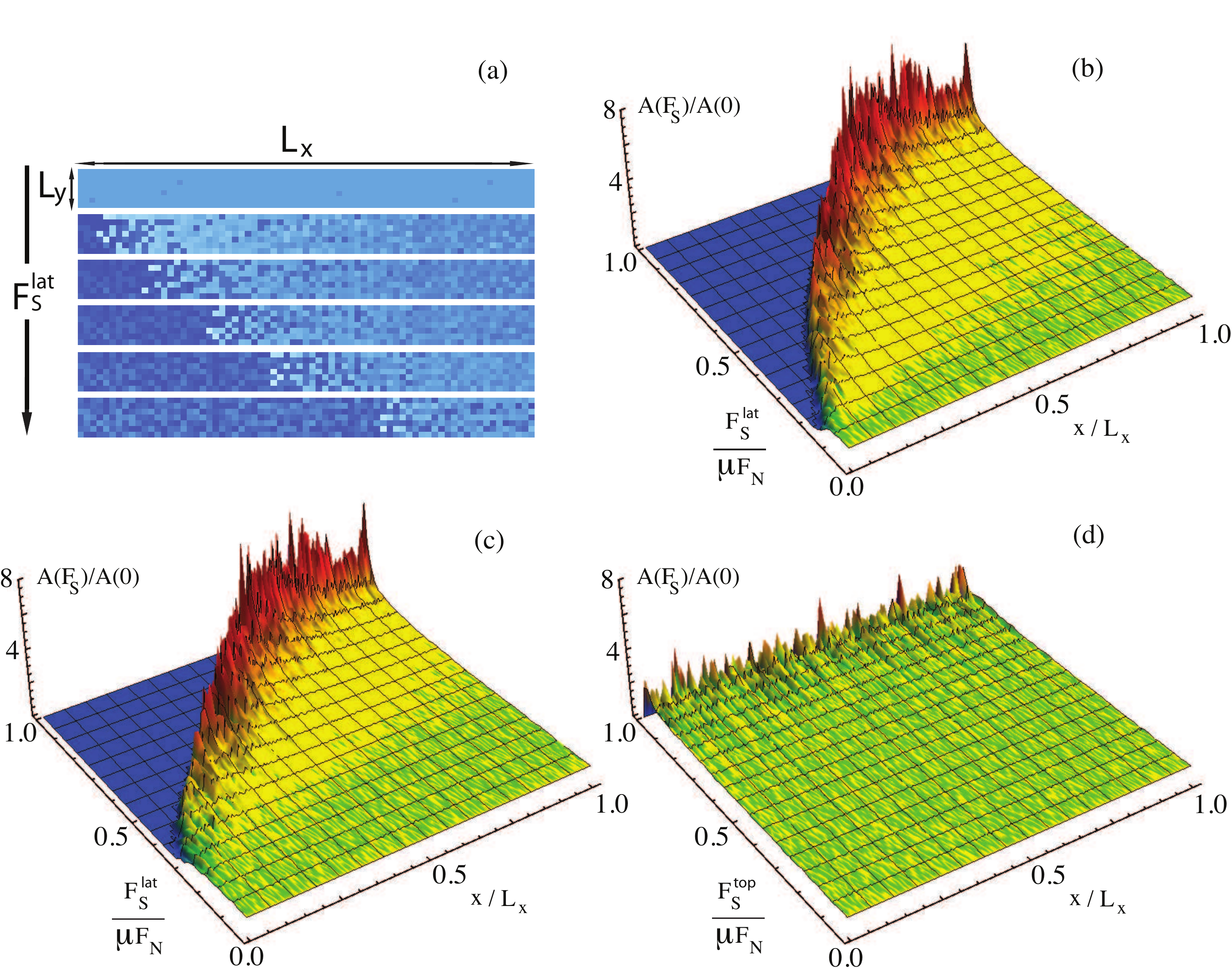} 
\caption{ {\bf Real area of contact  and determination of precursors size}. (a) Successive snapshots of the contact area, normalized to its initial value, show the advancement of the slip precursors as $F^{lat}_S$ is increased. Dark (pale) blue indicates a decrease (increase) in the contact area. (a)-(c) Quasi-static evolution of the average real area of contact $A_R(F_S)=\frac{\int_0^{L_y} dy\,A_R(x,y)}{L_y}$  (normalized to the zero-shear  value $A_R(0)$) for three type of loading conditions: with a rod ($h=6mm, \Delta_h=2$mm) (b), uniformly from a side (c), and uniformly from top (d).  $F_N=2.7$kN, $L_x=200$mm, $L_y=7$mm, $L_z=75$mm. Color map goes from blue ($A_R(F_S)/A_R(0)\ll 1$) to red ($A_R(F_S)/A_R(0)\gg 1$): for any value of $F_S$,  blu region corresponds to the precursor size, and the boundary between blu and red/yellow regions represents the precursor size $\ell$.  Regions close to the trailing edge experience a decay of the real contact area as $F_S$ is adiabatically increased, whereas the real area of contact area considerably grows on the opposite side ((a) and (b)). When the sample is loaded uniformly from top,  the sliding takes place without precursors appearence (c). 
\label{fig:3}}
\end{figure}

\noi and monitor its change as a function of the applied loads $F_N$ and $F_S$. This is a notable progress provided by our model  when compared with previous 1D \cite{braun2009,bouchbinder2011,rubinstein2011,capozza2012,amundsen2012} and 2D
models \cite{tromborg2011,otsuki2013,radiguet2013}, allowing for a direct quantitative comparison with experiments whose main analysis tool is, indeed, the  observation of the real contact area evolution \cite{rubinstein2004,rubinstein2006,rubinstein2007,bendavid2010}. We demonstrate the validity of the Archard's principle at macroscopic scale  (see Figs.\ref{fig:2}(a)), i.e. $A_R$ increases linearly with the applied vertical force $F_N$. Moreover, our model provides the remarkable result that $A_R$ only depends on the load $F_N$ and not on the nominal area $L_x\times L_y$, what is commonly known as the Amonton's second law (Fig.\ref{fig:S2},\ref{fig:S3}).   
As the shear force $F_S$ is adiabatically increased on the other hand, we detect the variation of the real contact area as illustrated in Movie S1 and Fig.\ref{fig:3} for three typical loading conditions used in experiments \cite{rubinstein2004,rubinstein2006,rubinstein2007,bendavid2010}.

 In Suppl. Mat. sec. \ref{sec:calibration} we discuss the calibration of the parameters appearing in our model. The only parameters that have to be adjusted are the two constants $c_x$ and $c_z$ which define the local stiffnesses. To do so, we have to compare the variation of $A_R$ as a function of $F_N$ and $F_S$ provided by our numerical simulations, with the corresponding variations observed in the experiments. 

\noi The normal stiffness is obtained by measuring the real contact area as a function of the normal load when no shear is applied, and tuning $c_z$ until the resulting area matches that reported in  Ref. \cite{rubinstein2006} (see Fig. \ref{fig:2}(a)). Indeed, as detailed in the Suppl. Mat. sec. \ref{sec:calibration}, we define the total real area of contact (\ref{eq:real_area_of_contact_ct}) in the discrete form as

\be
A_R=\sum\limits_{i=1}^{N_xN_y}\frac{\Delta x\Delta y}{ \left[e-1\right]}\left[e^{-\frac{u_z[i]-u_z^0[i]}{u_z^0[i]}}-1\right]\theta(u_z^0[i]-u_z[i]).
\label{eq:real_area_of_contact}
\ee

\noi Changing the value of the constant $c_z$ corresponds to change the equilibrium set of $u_z[i]$ given by Eq.(\ref{eq:z_solution_vec_i}): higher is $c_z$, stiffer are the interfacial springs, smaller will get the coresponding  real contact area.  The best value for $c_z= 1.65\times  10^8\frac{N}{m^2}$ yields the curve reported in Fig. \ref{fig:2}(a), showing a remarkable agreement with the experimental observation.

\noi  To determine the transverse stiffness, we compare the quasi-static evolution of $A_R$ detected in experiments with the corresponding  one obtained from simulations (Fig. \ref{fig:3}(a)). In particular, we consider a block of dimensions $L_x=140$mm $L_y=6$mm  and $L_z=75$mm under a normal load $F_N=3.3$kN and increase adiabatically the lateral force $F_S^{lat}$ applied at height $h=6$mm as in Ref. \cite{rubinstein2007}. As shown in Fig. \ref{fig:2}(b), as the lateral shear force is increased, the portion of contact area close to the trailing edge decreases drastically.  According to the definition furnished in  Ref.\cite{rubinstein2007}, precursors correspond to the regions of the frictional interface which undergo a reduction of the area of real contact, for values of the applied shear well before the static frictional force. A pictorial view of the adiabatic precursor evolution  is reported in the inset of Fig. \ref{fig:2}(b), where the color code represents the variation of the average local contact area with respect to its value at $F_S=0$. The boundary between the portion of contact surface which has decreased and that which has increased during the shearing process, determines the precursor size  $\ell$. This yieldis a curve that we compare with experiments to estimate the best value of  $c_x=1.65\times  10^{12}\frac{N}{m^3}$  (see the caption of Fig. \ref{fig:2}(b) and Suppl. Mat. sec. \ref{sec:calibration} for more details). 

\subsection{Loading mode and stick-slip events}

Throughout the paper, we will consider the conceptually simple case in which the sample
is loaded by imposing a constant shear force on the appropriate boundaries. This means that we will adopt $F_S$ as the adiabatic variable (quasi-static parameter), and calculate the equilibrium interfacial displacement from  Eqs.(\ref{eq:x_solution_vec_i}) and (\ref{eq:z_solution_vec_i}) each time that $F_S$ is slowly increased. This leads to discrete ``leapfroggy'' precursors
for which we study the continuum envelope as reported for instance in Fig. \ref{fig:2}(b). 
  However, the discrete
nature of the precursors dynamics is more apparent if we drive the system as in the experiments reported in Ref.\cite{rubinstein2007}, where the lateral force is applied through a spring with elastic constant $K_S=4\times 10^6$N/m. 
To model this case, we replace the external force appearing in  Eq.(\ref{eq:x_solution}) with the expression

\be
F_s^{lat}= K_S\left(U_s-\langle u_x\rangle\right)
\label{eq:spring_shear_force}
\ee

\begin{figure}[ht]
\centering
\includegraphics[width=0.8\textwidth]{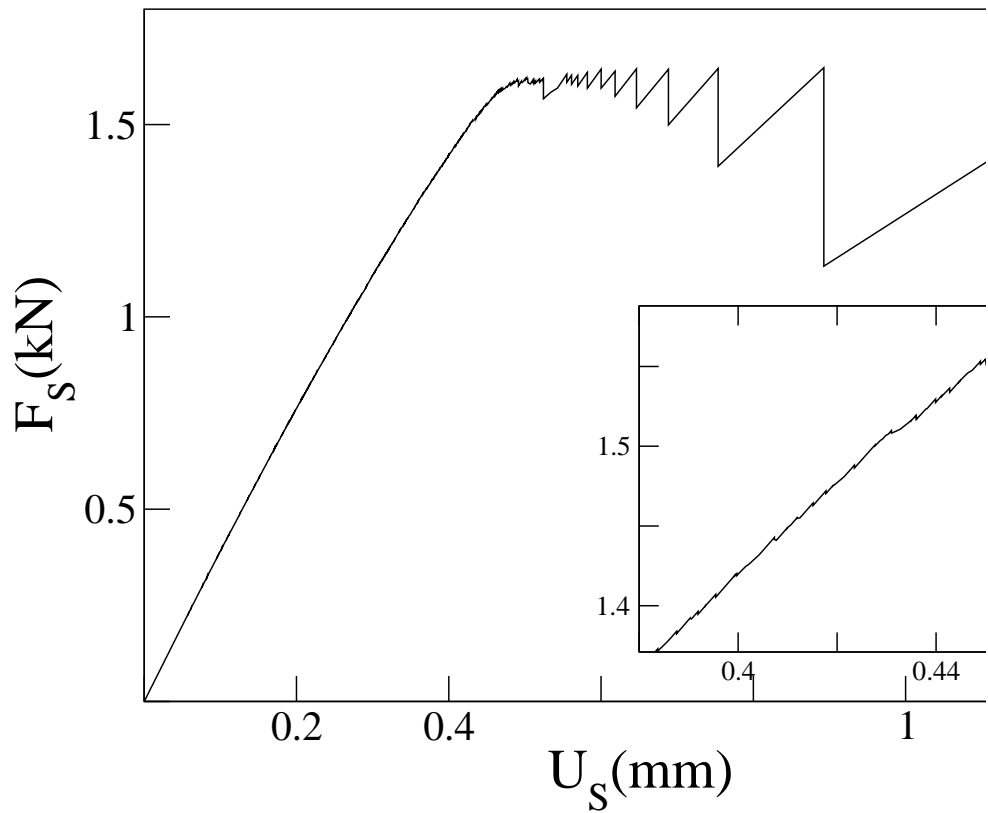} 
\caption{{\bf Stick-slip under displacement controlled driving.} We report the force on the driving spring as a function of the imposed displacement for the following same sample  parameters:  $L_x=140$mm, $L_y=7$mm, $L_z=75$mm, $F_N=3.5$kN. 
A magnification of the curve is reported in the inset showing the precursory stick slip events. 
\label{fig:4}}
\end{figure}

\begin{figure}[ht]
\centering
\includegraphics[width=0.5\textwidth]{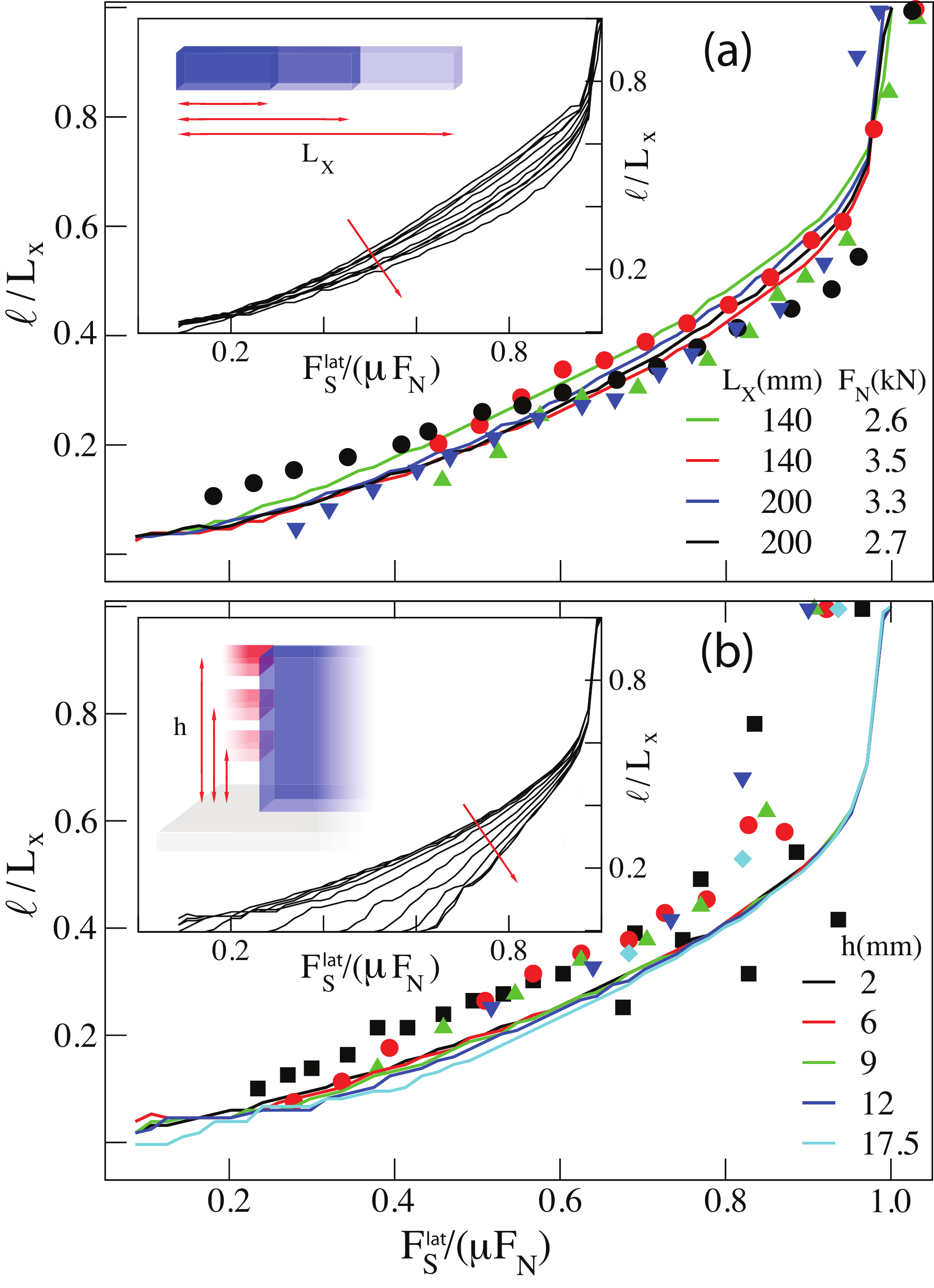} 
\caption{{\bf The dependence of slip  precursors on the sample size and the point of application of the lateral force}. (a) Experimental data (symbols) from Ref. \protect\cite{rubinstein2007} obtained for different $L_x$ and $F_N$ show an approximate data collapse when the rescaled front position $\ell/L_x$ is plotted against the rescaled  lateral shearing  force $F^{lat}_S/\mu F^N$ ($L_y=7$mm, $L_z=75$mm, $h=6$mm). This result is accurately reproduced by our model (solid lines). Inset shows that when the range of $L_x$ is increased further, collapse is lost ($100mm<L_x<350$mm). (b) Similarly, experimental data (symbols)  from Ref. \protect\cite{rubinstein2007} indicate that the rescaled precursors profiles are approximately independent of the height $h$ (the point of application of the lateral force $F_S^{lat}$), in perfect agreement with the  numerical outcomes (solid lines). $L_x=140$mm, $L_y=7$mm, $L_z=75$mm, $F_N=3$kN.  In the inset, we show that no collapse arises if $h$ is increased beyond the experimental values ($2mm<h<73$mm).
 \label{fig:5}}
\end{figure}

\begin{figure}[ht]
\centering
\includegraphics[width=0.8\textwidth]{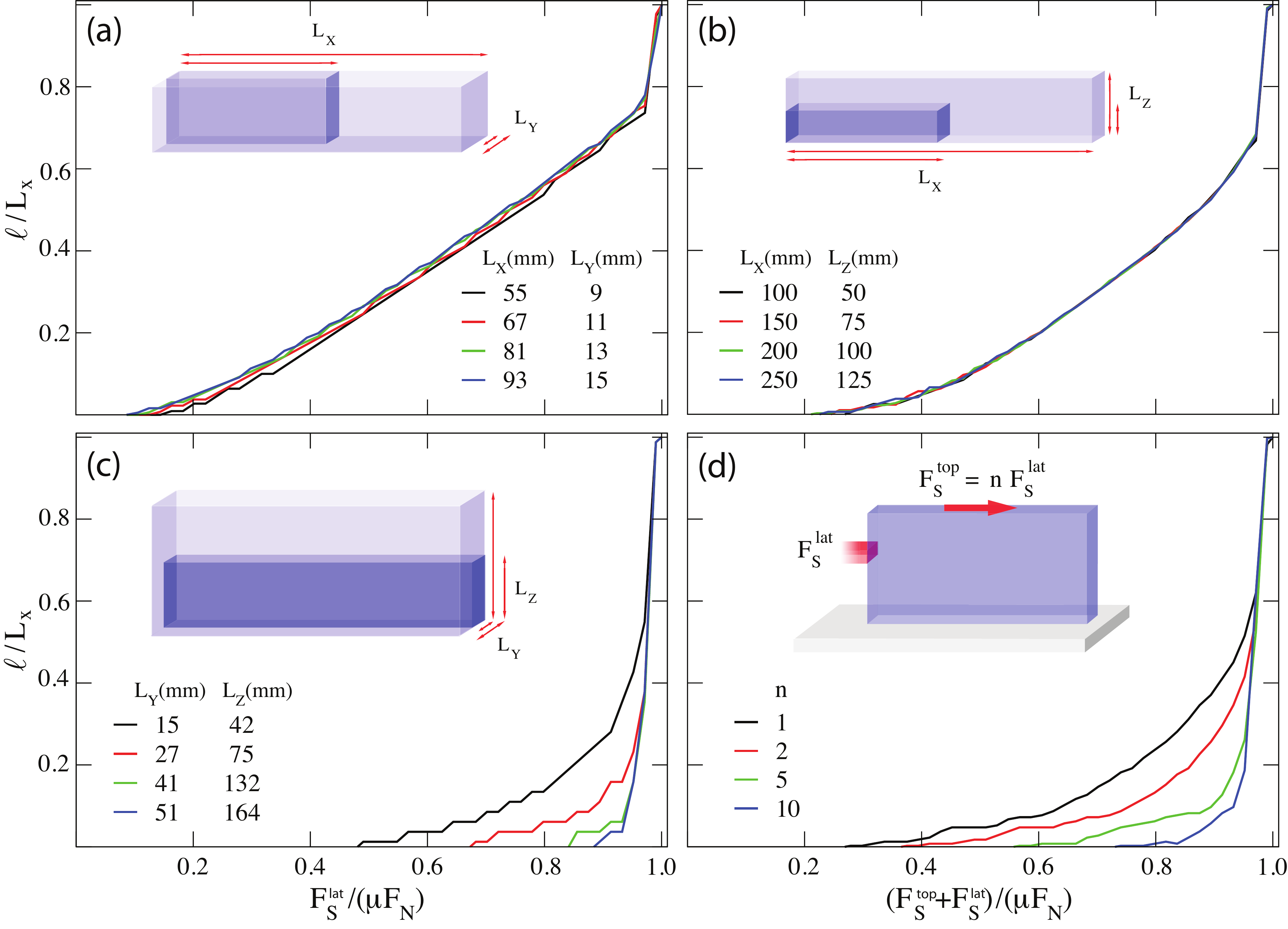}
\caption{{\bf The dependence of slip precursors on the sample aspect ratios and on the loading conditions} . 
Rescaled precursors quasi-static evolution  obtained when an uniform side shear $F_S^{lat}$ is applied. (a) Curves exhibit the same universal behavior  for different  $L_x$ and $L_y$ but same aspect ratio $\frac{L_x}{L_y}\simeq 6.1$, with $F_N=4$kN and $L_z=10$mm. (b) Perfect collapse of the curves is obtained when the aspect ratio $L_x/L_z$ and $L_y$ are kept constant. $L_x/L_z\simeq 2$,  $L_y=7$mm, $F_N=4$kN. (c)
Precursors progressively disappear when  $L_y$ and $L_z$ are increased by leaving unchanged the ratio $L_y/L_z$, and $L_x$ is constant. $L_y/L_z\simeq 0.36$, $L_x=40$mm, $F_N=4$kN. This findings are consistent with the assumption that precursors evolution profiles reflect the same simmetries appearing in the shear stress at the frictional interface \ref{u_x_lat_uniform},  which is function of the quantity $R=\frac{L_x}{L_yL_z}$. (d) While simultaneously loading the sample from top and from the edge with a rod  ($h=6$mm  and $\Delta_h=2$mm), $F_S^{top}=nF_S^{lat}$,  precursors dynamics is suppressed for large $n$. Sample parameters are $F_N=2.7$kN, $L_x=201$mm, $L_y=7$mm $L_z=132$mm.
\label{fig:6}}
\end{figure}

\begin{figure}[ht]
\centering
\includegraphics[width=0.8\textwidth]{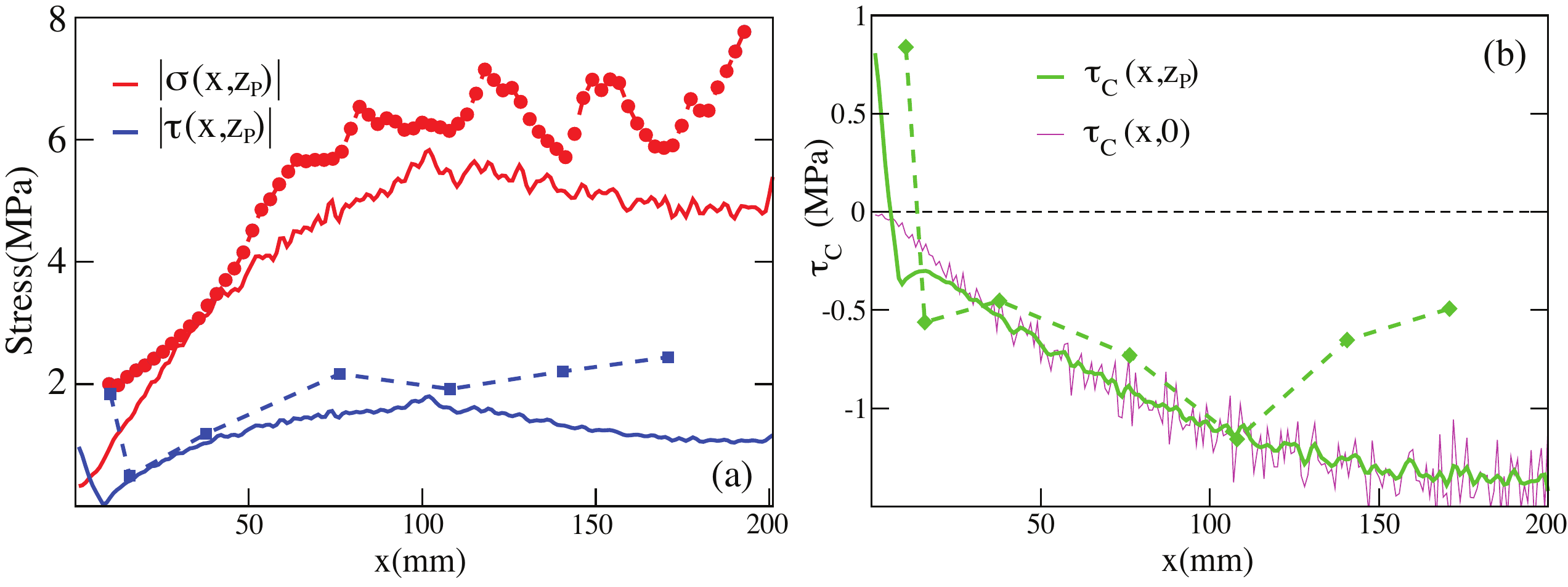}
\caption{{\bf Coulomb stress as slip precursor}. (a) Normal (red) and shear stress (blu) averaged over the $y$ direction. Profiles of $\left|\sigma(x,z_P)\right|=\left|\frac{\int_0^{L_y}dy\, \sigma(x,y,z_P)}{L_y}\right|$ and $\left|\tau(x,z_P)\right|=\left|\frac{\int_0^{L_y}dy\, \tau(x,y,z_P)}{L_y}\right|$ across the sample length are calculated on a reference plane $z_P=2$mm above the frictional interface (solid lines) and compared with the experimental data from Ref.\cite{bendavid2010} (Fig.2A, slow front) (symbols-dashed lines). Shear force was applied on top as well as on the sample trailing edge according to the experimental setup (Ref.\cite{bendavid2010}). Stress calculation was performed at a value of the shear force $F_S$ right before the nucleation of the precursor (see Fig.\ref{fig:S10} and Fig.\ref{fig:8}(a)). $L_x=200$mm, $L_y=7$mm, $L_z=100$mm, $F_N=6.25$kN. (b) Coulomb stress calculated on the reference plane above the contact interface and averaged over the $y$ direction (solid  green line): $\tau_C(x,z_P)=\left|\sigma(x,z_P)\right|-\mu\left|\tau(x,z_P)\right|$. Comparison with data inferred from Ref.\cite{bendavid2010} is excellent (green symbols-dashed line). The magenta line represents the $y$-averaged Coulomb stress at the frictional interface ($\tau_C(x,0)=\int_0^{L_y}dy\, \tau(x,y,0)$): although no detachment front is yet present at the contact plane ($\tau_C(x,0)<0$ throughout the surface), the value of the Coulomb stress at the reference plane $z_P=2$mm can exceed locally the threshold, leading to the erroneous conclusion that Amonton-Coulomb law might be violated.
 \label{fig:7}}
\end{figure}

\begin{figure}[ht]
\centering
\includegraphics[width=0.6\textwidth]{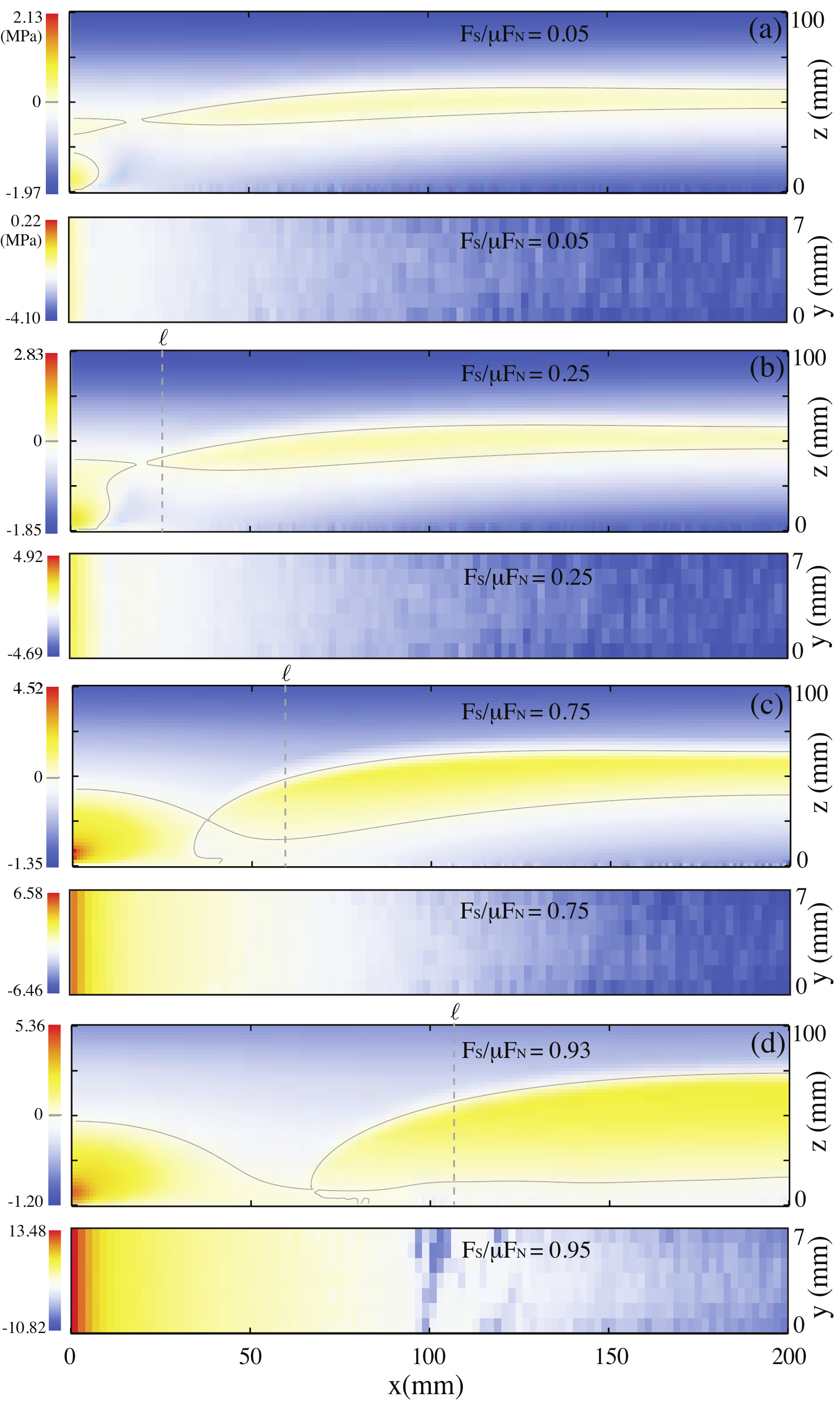}
\caption{{\bf Full Coulomb stress at the frictional interface}. (a)-(d) Quasi-static evolution of the Coulomb stress (averaged over $y$) along the slab $x-z$ plane (on top), and on the plane $z=z_P$ (bottom panels). Color code indicates regions where $\tau_C>0$ (yellow-red) from those for which $\tau_C<0$  (blu), grey solid lines correspond to the set of points fulfilling $\tau_C=0$. Panel (a) refers to the slider sitution before the first precursor event nucleates, the plane $z=z_P=2$mm (bottom panels) is where quantities in Fig.\ref{fig:7} are calculated (see also Fig.\ref{fig:S10} dashed black lines). Grey dashed lines represent the precursor envelope $\ell$ at the frictional interface, obtained from the real contact  area  decay (see Fig.\ref{fig:2}(b)(inset) and  Fig.\ref{fig:3}).
 \label{fig:8}}  
\end{figure}

\noi where $U_S$ is the externally applied displacement, which now  corresponds to the adiabatic adjustable parameter. $\langle u_x\rangle$, on the other hand, has the same meaning as the force-controlled protocol. In Fig. \ref{fig:4}, we report the evolution of the spring force (Eq.(\ref{eq:spring_shear_force})) as a function of the applied displacement for a typical simulation. Small stick-slip events, corresponding to discrete precursors leaps, are shown in
the inset of Fig. \ref{fig:4}, closely resembling the  experimental observations. Increasing the displacement further leads to larger stick-slip events
that in the constant-force case correspond to the last system size spanning event, that leads to the
slip of the entire block. 
More details on the
solution of the elastic equations for this particular case can be found in Suppl. Mat. sec. \ref{sec:displacement}.

\noi  Our model does not encompasses the rejuvenation of the real area of contact once the precursor has passed, because once a spring is broken no rehealing is allowed. However, in previous models such those in Ref.s \cite{capozza2012,tromborg2011,otsuki2013,radiguet2013}, once a precursor has detached a portion of interface, the corresponding interfacial contacts always reform, and the whole previous precursor path is broken again by each new precursor. This clearly contradicts the experimental evidence \cite{rubinstein2007}, where no rehealing of the real area of contact can be appreciated between subsequent precursor jumps, and  the discrete jumps in the precursors dynamics can be observed only by displaying the derivative $\frac{dA_R}{dt}$ (see by instance Fig.2(a) of Ref.\cite{rubinstein2007} or Fig.14 of Ref.\cite{rubinstein2009}).

\section*{Acknowledgments}
This work is supported by CNR through the ERANET ComplexityNet pilot project LOCAT and by ERC AdG-2011 SIZEFFECTS. AB has bee partly supported by SINERGIA Project CRSII2\,136287/1 from
the Swiss National Science Foundation. We are thankful to Dr. M. Maldini,  Dr. R. Donnini, Dr.  D. Ripamonti, Dr. D. Vilone and Dr. A. Sellerio for illuminating discussions. 



%
%
%
%








\appendix

\section{Rapidly converging representation of the scalar Green's function \label{sec:rapidly}}

Albeit in principle Eqs.(\ref{eq:x_solution}), (\ref{eq:z_solution}) provide the required displacements on the contact plane for any value of the external forces, in practice they are of very limited usefulness, owing to the extremely  poor convergence properties of the Green's function in Eq.(\ref{eq:green}) \cite{marshall1999}.
In this section we derive an expression for  Eq.(\ref{eq:green}) which is rapidly converging. Firstly, we expand it  by assigning to the coefficients $\gamma_{mnp}$ their explicit values:

\be
\begin{array}{l}
G(x,\xi;y,\eta;z,\zeta)=\frac{8}{L_xL_yL_z}\left\{\frac{1}{4}\sum\limits_{p=1}\frac{\cos\left(\frac{p\pi x}{L_x}\right)\cos\left(\frac{p\pi \xi}{L_x}\right)}{\left(\frac{p\pi}{L_x}\right)^2}+\right.\\
+\frac{1}{4}\sum\limits_{m=1}\cos\left(\frac{m\pi y}{L_y}\right)\cos\left(\frac{m\pi \eta}{L_y}\right)\left[\frac{1}{\left(\frac{m\pi}{L_y}\right)^2}+
  2\sum\limits_{p=1}\frac{\cos\left(\frac{p\pi x}{L_x}\right)\cos\left(\frac{p\pi \xi}{L_x}\right)}{\left(\frac{m\pi}{L_y}\right)^2+\left(\frac{p\pi}{L_x}\right)^2}\right]+\\
+\frac{1}{4}\sum\limits_{n=1}\cos\left(\frac{n\pi z}{L_z}\right)\cos\left(\frac{n\pi \zeta}{L_z}\right)\left[\frac{1}{\left(\frac{n\pi}{L_z}\right)^2}+2\sum\limits_{m=1}
\frac{\cos\left(\frac{m\pi y}{L_y}\right)\cos\left(\frac{m\pi \eta}{L_y}\right)}{\left(\frac{n\pi}{L_z}\right)^2+\left(\frac{m\pi}{L_y}\right)^2}\right]+\\
+\frac{1}{2}\sum\limits_{p=1}\sum\limits_{n=1}\cos\left(\frac{p\pi x}{L_x}\right)\cos\left(\frac{p\pi \xi}{L_x}\right)\cos\left(\frac{n\pi z}{L_z}\right)\cos\left(\frac{n\pi \zeta}{L_z}\right)\times\\
\times\left[\frac{1}{\left(\frac{n\pi}{L_z}\right)^2+\left(\frac{p\pi}{L_x}\right)^2}+\left.2\sum\limits_{m=1}\frac{\cos\left(\frac{m\pi y}{L_y}\right)\cos\left(\frac{m\pi \eta}{L_y}\right)}{\left(\frac{m\pi}{L_y}\right)^2+\left(\frac{n\pi}{L_z}\right)^2+\left(\frac{p\pi}{L_x}\right)^2}\right]\right\}.
\end{array}
\label{green_2}
\ee

\noi We can then study the different terms appearing in the former expression. The first sum is $\frac{1}{4}\sum\limits_{p=1}\frac{\cos\left(\frac{p\pi x}{L_x}\right)\cos\left(\frac{p\pi \xi}{L_x}\right)}{\left(\frac{p\pi}{L_x}\right)^2}=\mathscr{F}(x,\xi;L_x)$, which can be straightforwardly evaluated \cite{prudnikov1981} 

\be
\mathscr{F}(x,\xi;L_x)=
\frac{1}{48}\left\{
\begin{array}{lll}
3\left(x^2+\xi^2\right)+2L_x^2-6\xi L_x & & 0\leq x\leq \xi \\
3\left(x^2+\xi^2\right)+2L_x^2-6x L_x   & & \xi\leq x\leq  L_x .
\end{array}
\right.
\label{green_1term}
\ee

\noi The second term can be recast as \cite{prudnikov1981} 

\be
\begin{array}{l}
\frac{1}{4}\sum\limits_{m=1}\cos\left(\frac{m\pi y}{L_y}\right)\cos\left(\frac{m\pi \eta}{L_y}\right)\left[\frac{1}{\left(\frac{m\pi}{L_y}\right)^2}+
  2\sum\limits_{p=1}\frac{\cos\left(\frac{p\pi x}{L_x}\right)\cos\left(\frac{p\pi \xi}{L_x}\right)}{\left(\frac{m\pi}{L_y}\right)^2+\left(\frac{p\pi}{L_x}\right)^2}\right]=\\
\frac{L_xL_y}{16\pi}\sum\limits_{m=1}\frac{\cos\left(\frac{m\pi |y-\eta|}{L_y}\right)+\cos\left(\frac{m\pi (y+\eta)}{L_y}\right)}{m}\left\{\frac{\cosh\left[\frac{m\pi\left(L_x -\left|x-\xi\right|\right)}{L_y}\right]}{\sinh\left[\frac{m\pi L_x}{L_y}\right]}+\frac{\cosh\left[\frac{m\pi\left(L_x-x-\xi\right)}{L_y}\right]}{\sinh\left[\frac{m\pi L_x}{L_y}\right]}\right\},
\end{array}
\label{green_2term}
\ee

\noi and we can get a compact representation  introducing  the function 

\be
\mathscr{L}(a,b; L_a,L_b)=\frac{L_aL_b}{16\pi}\sum\limits_{m=1}\frac{\cos\left(\frac{m\pi b}{L_b}\right)}{m}\frac{\cosh\left[\frac{m\pi\left(L_a -a\right)}{L_b}\right]}{\sinh\left[\frac{m\pi L_a}{L_b}\right]}
\label{elle}.
\ee

\noi Hence Eq.(\ref{green_2term}) acquires the final form

\be
\begin{array}{l}
\frac{1}{4}\sum\limits_{m=1}\cos\left(\frac{m\pi y}{L_y}\right)\cos\left(\frac{m\pi \eta}{L_y}\right)\left[\frac{1}{\left(\frac{m\pi}{L_y}\right)^2}+
  2\sum\limits_{p=1}\frac{\cos\left(\frac{p\pi x}{L_x}\right)\cos\left(\frac{p\pi \xi}{L_x}\right)}{\left(\frac{m\pi}{L_y}\right)^2+\left(\frac{p\pi}{L_x}\right)^2}\right]=\\
\mathscr{L}\left(\left|x-\xi\right|,\left|y-\eta\right|; L_x,L_y\right)+\mathscr{L}\left(x+\xi,\left|y-\eta\right|; L_x,L_y\right)+\\
+\mathscr{L}\left(\left|x-\xi\right|,y+\eta; L_x,L_y\right)+\mathscr{L}\left(x+\xi,y+\eta; L_x,L_y\right).
\end{array}
\label{green_2term_final}
\ee

\noi The same can be done for the third term in the righthand side of Eq.(\ref{green_2}).

\noi We now address the last term, that we partially sum as \cite{prudnikov1981}

\be
\begin{array}{l}
\sum\limits_{p=1}\sum\limits_{n=1}\frac{\cos\left(\frac{p\pi x}{L_x}\right)\cos\left(\frac{p\pi \xi}{L_x}\right)\cos\left(\frac{n\pi z}{L_z}\right)\cos\left(\frac{n\pi \zeta}{L_z}\right)}{2}\times\\
\times\left[\frac{1}{\left(\frac{n\pi}{L_z}\right)^2+\left(\frac{p\pi}{L_x}\right)^2}+2\sum\limits_{m=1}\frac{\cos\left(\frac{m\pi y}{L_y}\right)\cos\left(\frac{m\pi \eta}{L_y}\right)}{\left(\frac{m\pi}{L_y}\right)^2+\left(\frac{n\pi}{L_z}\right)^2+\left(\frac{p\pi}{L_x}\right)^2}\right]=\\
\frac{L_y}{8\pi}
\sum\limits_{p=1}\left[\cos\left(\frac{p\pi \left|x-\xi\right|}{L_x}\right)+\cos\left(\frac{p\pi \left(x+\xi\right)}{L_x}\right) \right]\sum\limits_{n=1}\frac{\cos\left(\frac{n\pi \left|z-\zeta\right|}{L_z}\right)+\cos\left(\frac{n\pi \left(z+\zeta\right)}{L_z}\right)}{\sqrt{\left(\frac{p}{L_x}\right)^2+\left(\frac{n}{L_z}\right)^2}}\times\\
\times\left\{\frac{\cosh\left[\pi\sqrt{\left(\frac{p}{L_x}\right)^2+\left(\frac{n}{L_z}\right)^2}\left(L_y -\left|y-\eta\right|\right)\right]+\cosh\left[\pi\sqrt{\left(\frac{p}{L_x}\right)^2+\left(\frac{n}{L_z}\right)^2}\left(L_y -y-\eta\right)\right]}{\sinh\left[\pi\sqrt{\left(\frac{p}{L_x}\right)^2+\left(\frac{n}{L_z}\right)^2} L_y\right]}\right\}.
\end{array}
\label{green_3term}
\ee

\noi Defining the function 

\be
\begin{array}{l}
\mathscr{H}\left(a,b,c; L_a,L_b,L_c\right)=\\
\frac{L_b}{8\pi}\sum\limits_{p=1}\sum\limits_{n=1}\frac{\cos\left(\frac{p\pi a}{L_a}\right)\cos\left(\frac{n\pi c}{L_c}\right)\cosh\left[\pi\sqrt{\left(\frac{p}{L_a}\right)^2+\left(\frac{n}{L_c}\right)^2}\left(L_b -b\right)\right]}{\sqrt{\left(\frac{p}{L_a}\right)^2+\left(\frac{n}{L_c}\right)^2}\sinh\left[\pi\sqrt{\left(\frac{p}{L_a}\right)^2+\left(\frac{n}{L_c}\right)^2} L_b\right]}
\end{array}
\label{acca}
\ee

\noi we express Eq.(\ref{green_3term}) as

\be
\begin{array}{l}
\sum\limits_{p=1}\sum\limits_{n=1}\frac{\cos\left(\frac{p\pi x}{L_x}\right)\cos\left(\frac{p\pi \xi}{L_x}\right)\cos\left(\frac{n\pi z}{L_z}\right)\cos\left(\frac{n\pi \zeta}{L_z}\right)}{2}\times\\
\times\left[\frac{1}{\left(\frac{n\pi}{L_z}\right)^2+\left(\frac{p\pi}{L_x}\right)^2}+2\sum\limits_{m=1}\frac{\cos\left(\frac{m\pi y}{L_y}\right)\cos\left(\frac{m\pi \eta}{L_y}\right)}{\left(\frac{m\pi}{L_y}\right)^2+\left(\frac{n\pi}{L_z}\right)^2+\left(\frac{p\pi}{L_x}\right)^2}\right]=\\
\mathscr{H}\left(\left|x-\xi\right|,\left|y-\eta\right|,\left|z-\zeta\right|; L_x,L_y,L_z\right)
+\\
+\mathscr{H}\left(x+\xi,\left|y-\eta\right|,\left|z-\zeta\right|; L_x,L_y,L_z\right)+\\
+\mathscr{H}\left(\left|x-\xi\right|,y+\eta,\left|z-\zeta\right|; L_x,L_y,L_z\right)+\\
+\mathscr{H}\left(\left|x-\xi\right|,\left|y-\eta\right|,z+\zeta; L_x,L_y,L_z\right)+\\
+\mathscr{H}\left(x+\xi,y+\eta,\left|z-\zeta\right|; L_x,L_y,L_z\right)+\\
+\mathscr{H}\left(x+\xi,\left|y-\eta\right|,z+\zeta; L_x,L_y,L_z\right)+\\
+\mathscr{H}\left(\left|x-\xi\right|,y+\eta,z+\zeta; L_x,L_y,L_z\right)+
\mathscr{H}\left(x+\xi,y+\eta,z+\zeta; L_x,L_y,L_z\right).
\end{array}
\label{green_3term_1}
\ee

\noi By inserting the expressions (\ref{green_1term}), (\ref{green_2term_final}) and (\ref{green_3term_1}) in (\ref{green_2}) we obtain the desidered result. However a close analysis reveals that the ensuing formula does provide a fast convergence for any set of points $(x,\xi; y,\eta;z,\zeta)$, but for those who lie on the hypersurface $x\neq\xi, y=\eta, z=\zeta$. In this case the Green function can be recast in the following form:

\be
\begin{array}{l}
G(x,\xi;y,\eta;z,\zeta)=\frac{8}{L_xL_yL_z}\left\{\mathscr{F}(y,\eta;L_y)+\right.\\
\mathscr{L}\left(\left|y-\eta\right|,\left|x-\xi\right|; L_y, L_x\right)+\mathscr{L}\left(\left|y-\eta\right|,x+\xi; L_y,L_x\right)+\\
+\mathscr{L}\left(y+\eta,\left|x-\xi\right|; L_y,L_x\right)+\mathscr{L}\left(y+\eta,x+\xi; L_y,L_x\right)+\\
+\mathscr{L}\left(\left|x-\xi\right|,\left|z-\zeta\right|; L_x,L_z\right)+\mathscr{L}\left(x+\xi,\left|z-\zeta\right|; L_x,L_z\right)+\\
+\mathscr{L}\left(\left|x-\xi\right|,z+\zeta; L_x,L_z\right)+\mathscr{L}\left(x+\xi,z+\zeta; L_x,L_z\right)+\\
+\mathscr{H}\left(\left|y-\eta\right|,\left|x-\xi\right|,\left|z-\zeta\right|; L_y,L_x,L_z\right)+\\
+\mathscr{H}\left(\left|y-\eta\right|,x+\xi,\left|z-\zeta\right|; L_y,L_x,L_z\right)+\\
+\mathscr{H}\left(y+\eta,\left|x-\xi\right|,\left|z-\zeta\right|; L_y,L_x,L_z\right)+\\
+\mathscr{H}\left(\left|y-\eta\right|,\left|x-\xi\right|,z+\zeta; L_y,L_x,L_z\right)+\\
+\mathscr{H}\left(y+\eta,x+\xi,\left|z-\zeta\right|; L_y,L_x,L_z\right)+\\
+\mathscr{H}\left(\left|y-\eta\right|,x+\xi,z+\zeta; L_y,L_x,L_z\right)+\\
+\left.\mathscr{H}\left(y+\eta,\left|x-\xi\right|,z+\zeta; L_y,L_x,L_z\right)+
\mathscr{H}\left(y+\eta,x+\xi,z+\zeta; L_y,L_x,L_z\right)\right\}.
\end{array}
\label{green_3}
\ee

\section{Surface forces \label{sec:surface}}

To complete the solutions in Eqs.(\ref{eq:x_solution}), (\ref{eq:z_solution}) for the displacements fields at the contact plane, we
have to introduce a functional form for the surface stresses $\varphi_{surf}^{x,z}$. This expression represents the slider-surface interaction, embodying the microsopic details into a coarse-grained description. Albeit the surface forces $\varphi_{surf}^{x,z}(x,y)$ are formally  distributions defined  on the contact plane $(x,y,0)$, 
they must be interpreted as effectively  acting only on a single point $(x,y)$ representative  of the  surface  portion $\Delta x\times\Delta y$ that  surrounds it (Fig.\ref{fig:S1}). In the next section we will provide a clear description of the slider-surface interface discretization scheme. Here  we only clarify the assumptions  which $\varphi_{surf}^{x,z}(x,y)$ are built from:

\begin{itemize}  

\item the  surface element $\Delta x\times\Delta y$ behaves as a macroscopic object, fulfilling  the macroscopic laws  of friction;

\item the surface forces are purely elastic, i.e. linear in the displacement fields.

\end{itemize}

\noi We consider first the surface force along the $z$ direction $\varphi_{surf}^z$.  Experimental evidence 
\cite{benz2006}
shows that an elastic body, under a squeezing pressure $P$, exhibits an average surface separation $\langle u_z\rangle$ which decreases with $P$ as

\be
P\sim e^{-\frac{\langle u_z\rangle}{u^0_z}};
\label{Benz_formula}
\ee

\noi here $u^0_z$ depends on the nature of the surface rougheness but is independent on $P$. Thus,  the first assumption requires that the internal stress $\sigma_{zz}(x,y,0)$ on the microscopical surface $\Delta x\times\Delta y$  centered around $(x,y)$, can be written as

\be
\sigma_{zz}(x,y,0)\sim -e^{-\frac{u_z(x,y,0)}{u^0_z(x,y)}}.
\label{Benz_formula_revisited}
\ee

\noi The minus sign relates to  the tensorial nature of $\sigma_{zz}(x,y,0)$, always pointing along the $-\hat{z}$ direction. Now, since at the equilibrium $\sigma_{zz}(x,y,0)=-\varphi_{surf}^z(x,y)$, as noticed in Ref. \cite{persson2007}
the local interfacial stiffness is given by

\be
k_z(x,y)=-\frac{d\varphi_{surf}^z(x,y)}{du_z(x,y,0)}=\frac{\varphi_{surf}^z(x,y)}{u^0_z(x,y)}\sim \frac{e^{-\frac{u_z(x,y,0)}{u^0_z(x,y)}}}{u^0_z(x,y)},
\label{Perrson_interfacial_stiffness}
\ee
 
\noi which, for $u_z(x,y,0)\simeq u^0_z(x,y)$ can be recast as

\be
k_z(x,y)= \frac{c_z}{u^0_z(x,y)},
\label{k_z}
\ee

\noi where $c_z$ is a constant to be set. Thus, the surface stress can be written in the following linear form

\be
\varphi_{surf}^z(x,y)=k_z(x,y)\left[u^0_z(x,y)-u_z(x,y,0)\right]= \frac{c_z}{u^0_z(x,y)}\left[u^0_z(x,y)-u_z(x,y,0)\right]
\label{stress_z}
\ee

\noi fulfilling the second assumption. We notice that the previous form holds only for $u^0_z(x,y)>u_z(x,y,0)$, and $\varphi_{surf}^z(x,y)=0$ when $u^0_z(x,y)<u_z(x,y,0)$ ($u_z(x,y,0)\geq 0$), see Fig.\ref{fig:1}(a).

\noi We now consider the interfacial stress along $x$ $\varphi_{surf}^x(x,y)$. For small displacements $u_x(x,y,0)$, the second assumption requires that it can be expanded to first order as 

\be
\varphi_{surf}^x(x,y)=-k_x(x,y)u_x(x,y,0)
\label{stress_x_ansatz}
\ee

\noi where $u_x(x,y,0)\gtreqless 0$. We have to find an expression for $k_x(x,y)$.   From Eq.(\ref{Benz_formula_revisited})  the normal load acting on the surface element $\Delta x\Delta y$ is given by

\be
F_N(x,y)=\Delta x\Delta y\left|\sigma_{zz}(x,y,0)\right|\sim\Delta x\Delta y e^{-\frac{u_z(x,y,0)}{u^0_z(x,y)}}.
\label{microscopical_load}
\ee

\noi Now, the experiment in \citep{berthoud1998}
 shows that an elastic body  behaves like an harmonic spring (with constant $K_x$), if subject to a small shear force: the spring constant is linear in the applied load, i.e.

\be
K_x(F_N)\sim F_N.
\label{baumberger_relation}
\ee

\noi Hence, fulfilling the first hypothesis, we can assume that the surface element $\Delta x\times\Delta y$ is characterized by a microscopic transverse spring stiffness

\be
k_x(x,y)= c_xe^{-\frac{u_z(x,y,0)}{u^0_z(x,y)}},
\label{k_x}
\ee

\noi with $c_x$ constant to be set. Thanks to (\ref{stress_x_ansatz}) and (\ref{k_x}) we can finally write the interfacial stress along the $x$ direction as

\be
\varphi_{surf}^x(x,y)=-k_x(x,y)u_x(x,y,0)=-c_x e^{-\frac{u_z(x,y,0)}{u^0_z(x,y)}}u_x(x,y,0).
\label{stress_x}
\ee

\noi Finally, introducing the linear expressions  Eqs.(\ref{stress_x}) and (\ref{stress_z}) in Eqs.(\ref{eq:x_solution}) and (\ref{eq:z_solution}) respectively, we obtain the closed Eqs. (\ref{eq:x_solution_linear}), (\ref{eq:z_solution_linear}).

\section{External force contributions to the interfacial displacements \label{sec:external}}

In the expressions (\ref{eq:x_solution_linear}) and (\ref{eq:z_solution_linear}) for the interfacial displacements, terms involving the contributions arising from the external shear or normal forces $F_S$ and $F_N$ can be calculted analytically.

\noi  In Eq.(\ref{eq:x_solution_linear}) we can evaluate the term proportional to the shear force applied uniformly at the top surface $F_s^{top}$. According  to \cite{prudnikov1981}, 
  plugging Eq.(\ref{eq:green}) in Eq.(\ref{eq:x_solution}), we obtain

\be
\frac{F_s^{top}}{L_xL_y}\int_0^{L_x}d\xi\int_0^{L_y}d\eta\, G(x,\xi;y,\eta;0,L_z)=-\frac{L_z}{6L_xL_y}F_s^{top}.
\label{G_x_top}
\ee

\noi Hence, the contribution  arising from applied shear on the top surface  is constant for any point $(x,y)$ on the contact surface. We can thus define a component of the displacement at the interface

\be
u_S^{top}(x,y,0)=-\frac{(1+\nu)}{E}\frac{F_s^{top}}{L_xL_y}\int_0^{L_x}d\xi\int_0^{L_y}d\eta\, G(x,\xi;y,\eta;0,L_z) 
\label{u_s_top}
\ee

\noi which does not display any dependence on $(x,y)$. On the other hand, the contribution due to the lateral shear $F_s^{lat}$ can be evaluated as

\be
\begin{array}{l}
\frac{F_s^{lat}}{L_y2\Delta h}\int_0^{L_y}d\eta\int_{h-\Delta h}^{h+\Delta h}d\zeta\, G(x,0;y,\eta;0,\zeta)=\frac{L_x}{L_yL_z}F_s^{lat}\left[\frac{x^2}{2L_x^2}-\frac{x}{L_x}+\frac{1}{3}\right]+\\
+\frac{F_s^{lat}L_z}{\pi^2  L_xL_y\Delta h}\left\{\frac{(h+\Delta h)\pi^2}{12}\left[\frac{(h+\Delta h)^2}{L_z^2}-3\frac{(h+\Delta h)}{L_z}+2\right]-\right.\\
-\frac{(h-\Delta h)\pi^2}{12}\left[\frac{(h-\Delta h)^2}{L_z^2}-3\frac{(h-\Delta h)}{L_z}+2\right]+\\
+\left.L_x\sum\limits_{n=1}\frac{\cosh\left[n\pi\frac{(L_x-x)}{L_z}\right]}{n^2\sinh\left[n\pi\frac{L_x}{L_z}\right]}\left[\sin\left(n\pi\frac{(h+\Delta h)}{L_z}\right)-\sin\left(n\pi\frac{(h-\Delta h)}{L_z}\right)\right]\right\}.
\end{array}
\label{G_x_lat}
\ee

\noi In the case of a shear force applied uniformly on the slider trailing edge ($x=0$, $h=\Delta h=\frac{L_z}{2}$), the former expression simplifies to

\be
\frac{F_s^{lat}}{L_yL_z}\int_0^{L_y}d\eta\int_{0}^{L_z}d\zeta\, G(x,0;y,\eta;0,\zeta)=\frac{L_x}{L_yL_z}F_s^{lat}\left[\frac{x^2}{2L_x^2}-\frac{x}{L_x}+\frac{1}{3}\right].
\label{G_x_lat_simple}
\ee

\noi Again we can define the contribution to the local displacement due to a lateral shear force as 

\be
u_S^{lat}(x,y,0)=\frac{(1+\nu)}{E}\frac{F_s^{lat}}{L_y2\Delta h}\int_0^{L_y}d\eta\int_{h-\Delta h}^{h+\Delta h}d\zeta\, G(x,0;y,\eta;0,\zeta).
\label{u_s_lat}
\ee

\noi We can perform the same analysis for normal displacements, obtaining again a constant
contribution due to the normal force

\be
u_N(x,y,0)= -\frac{(1+\nu)}{E}\frac{F_N}{L_xL_y}\int_0^{L_x}d\xi\int_0^{L_y}d\eta\, G(x,\xi;y,\eta;0,L_z) = \frac{(1+\nu)}{E}\frac{L_z}{6L_xL_y}F_N.
\label{u_n}
\ee

\section{Model discretization \label{sec:discretization}}

We hereby show the procedure to discretize the contact surface between the slider and the underlying interface. Therefore we take a grid of the slider  bottom plane, with an individual element having an area  $\Delta x\times\Delta y$, i.e.

\be
\begin{array}{l}
L_x=N_x\Delta x\\
L_y=N_y\Delta y
\end{array}
\label{L_dscrt}
\ee

\noi as it is shown in Fig.\ref{fig:S1}. The central point $(x,y)$  of each grid  element takes the following discrete form

\be
\begin{array}{ccc}
x=\left(n_x-\frac{1}{2}\right)\Delta x &  & n_x\in[1, N_x]\\
y=\left(n_y-\frac{1}{2}\right)\Delta y &  & n_x\in[1, N_y].
\end{array}
\label{x_dscrt}
\ee

\noi The solutions Eqs.\ref{eq:x_solution_linear} and  \ref{eq:z_solution_linear} at any point $(x,y)$ require the integration over the whole surface, i.e. the sum over the discrete set of the points $(\xi,\eta)$

\be
\begin{array}{ccc}
\xi=\left(n_\xi-\frac{1}{2}\right)\Delta x &  & n_\xi\in[1, N_x]\\
\eta=\left(n_\eta-\frac{1}{2}\right)\Delta y &  & n_\eta\in[1, N_y].
\end{array}
\label{xi_dscrt}
\ee

\noi Therefore any grid  center point $(x,y)$, as well as any point $(\xi,\eta)$ can be represented by a single index $i$ and $j$ respectively,

\be
\begin{array}{ccc}
i=(n_y-1)N_x+n_x=\left(\frac{y}{\Delta y}-\frac{1}{2}\right)\frac{L_x}{\Delta x}+\left(\frac{x}{\Delta x}+\frac{1}{2}\right) &  & i\in[1, N_xN_y]\\
j=(n_\eta-1)N_x+n_\xi=\left(\frac{\eta}{\Delta y}-\frac{1}{2}\right)\frac{L_x}{\Delta x}+\left(\frac{\xi}{\Delta x}+\frac{1}{2}\right) &  & j\in[1, N_xN_y],
\end{array}
\label{ij}
\ee

\noi and displacements and surface forces become to  one-dimensional vectors made of  $N_xN_y$ elements each:

\be
\begin{array}{ccc}
u_z(x,y,0)  &\to&  u_z[i]\in\vec{u}_z\\
u_x(x,y,0)  &\to&  u_x[i]\in\vec{u}_x\\
u_z^0(x,y)  &\to&  u_z^0[i]\in\vec{u}_z^0\\
k_z(x,y)   &\to&  k_z[i]=\frac{c_z}{u_z^0[i]}\in \vec{k}_z \\
k_x(x,y)   &\to&  k_x[i]=c_xe^{-\frac{u_z[i]}{u_z^0[i]}} \in \vec{k}_x.
\end{array}
\label{u_vec}
\ee

\noi Correspondingly the Green's function on the contact plane ($z=\zeta=0$) is a $N_xN_y\times N_xN_y$ matrix:

\be
\begin{array}{ccc}
G(x,\xi;y,\eta;0,0)  &\to&  G_{ij}\in \hat{G}
\end{array}
\label{green_matrix}
\ee

\noi where each element can be calculated analitically thanks to Eqs.(\ref{green_2})-(\ref{green_3}), once $(x,y;\xi,\eta)$ $\to$ $(i,j)$ by means of Eq.(\ref{ij}). It remains to discretize  the contributions given  by the external foces $F_S$  and $F_N$ to Eq.(\ref{eq:x_solution_linear}) and  Eq.(\ref{eq:z_solution_linear}).
From Eqs.(\ref{G_x_top}) and (\ref{u_s_top}), it turns out that the term proportional to $F_S^{top}$ is constant across the entire contact interface, resulting in a constant vector whose components are 

\be
\begin{array}{ccc}
u_S^{top}(x,y,0) &\to& u_S^{top}[i]\in\vec{u}_S^{top}.
\end{array}
\label{f_s_top}
\ee

\noi On the other hand, the contribution coming from $F_S^{lat}$ always displays a non-trivial $x$-dependence, be the force applied with a shearing rod (Eq.(\ref{G_x_lat})) or uniformly on the sample side surface (Eq.(\ref{G_x_lat_simple})). After discretization of the contact plane,  this yields the   lateral shear contribution  to the  $i-$th component of $\vec{u}_x$ which arises from the discretization of the corresponding continuum component (Eq.(\ref{u_s_lat})):

\be
\begin{array}{ccc}
u_S^{lat}(x,y,0)  &\to&  u_S^{lat}[i]\in\vec{u}_S^{lat}.
\end{array}
\label{f_s_lat}
\ee

\noi Finally, from  Eq.(\ref{u_n}) we find that the contribution given  to the displacement $u_z$ by the loading force $F_N$ is constant over the whole interface, so that 

\be
\begin{array}{ccc}
u_N(x,y,0) &\to& u_N[i]\in\vec{u}_N.
\end{array}
\label{f_n}
\ee

\noi We can now  write the expressions (\ref{eq:x_solution_linear}) and  (\ref{eq:z_solution_linear}) for the discrete component $i$ of $\vec{u}_x$ and  $\vec{u}_z$

\be
u_x[i]=-\frac{(1+\nu)}{E}\sum\limits_{j=1}^{N_xN_y}\Delta x\Delta y
G_{ij}k_x[j]u_x[j]+u_S^{top}[i]+u_S^{lat}[i]+\langle u_x\rangle
\label{x_solution_dscrt}
\ee

\be
u_z[i]=\frac{(1+\nu)}{E}\sum\limits_{j=1}^{N_xN_y}\Delta x\Delta yG_{ij}k_z[j]\left\{u_z^0[j]-u_z[j]\right\}+u_N[i]
+\langle u_z\rangle
\label{z_solution_dscrt}
\ee

\noi where we made use of  Eqs.(\ref{stress_z}) and (\ref{stress_x}). Now, let us introduce the matrices $\hat{k}^x$ and $\hat{k}^z$, defined as

\be
k_{ij}^x=\left\{
\begin{array}{ccc}
\Delta x\Delta yk_x[j] &  & i=j\\
0   &  & otherwise
\end{array}
\right.
\label{k_x_matrix}
\ee

\be
k_{ij}^z=\left\{
\begin{array}{ccc}
\Delta x\Delta yk_z[j] &  & i=j\\
0   &  & otherwise.
\end{array}
\right.
\label{k_z_matrix}
\ee

\noi Thus expressions (\ref{stress_z}) and (\ref{stress_x}) can be cast in  vectorial form as

\be
\vec{u}_x=-\frac{(1+\nu)}{E}\hat{G}\hat{k}^x\vec{u}_x+\vec{u}_S^{top}+\vec{u}_S^{lat}+\vec{\langle u_x\rangle}
\label{x_solution_dscrt_vec}
\ee

\be
\vec{u}_z=\frac{(1+\nu)}{E}\left\{\hat{G}\hat{k}^z\vec{u}_z^0-\hat{G}\hat{k}^z\vec{u}_z\right\}+\vec{u}_N
+\vec{\langle u_z\rangle},
\label{z_solution_dscrt_vec}
\ee

\noi where we introduce the vectors $\vec{\langle u_x\rangle}$ and $\vec{\langle u_z\rangle}$ whose components are constant. Then  we can obtain the solutions by inversion:

\be
\vec{u}_x=\left[\hat{I}+\frac{(1+\nu)}{E}\hat{G}\hat{k}^x\right]^{-1}\left\{\vec{u}_S^{top}+\vec{u}_S^{lat}+\vec{\langle u_x\rangle}\right\}
\label{x_solution_vec}
\ee

\be
\vec{u}_z=\left[\hat{I}+\frac{(1+\nu)}{E}\hat{G}\hat{k}^z\right]^{-1}\left\{\frac{E}{(1+\nu)}\hat{G}\hat{k}^z\vec{u}_z^0+\vec{u}_N+\vec{\langle u_z\rangle}\right\},
\label{z_solution_vec}
\ee

\noi where $\hat{I}$ is the identity matrix. Finally we have to impose the constraints in Eq.(\ref{eq:additional_constraints}) that can now be written as

\be
\begin{array}{l}
\sum\limits_{i=1}^{N_xN_y}\Delta x\Delta yk_x[i]u_x[i] =F_S^{top}+F_S^{lat}\\
\sum\limits_{i=1}^{N_xN_y}\Delta x\Delta yk_z[i]\left\{u_z^0[i]-u_z[i]\right\}=-F_N.
\end{array}
\label{additional_constraints_dscrt}
\ee

\noi In order to fulfill Eq.(\ref{additional_constraints_dscrt}) we need the solution of Eqs.(\ref{x_solution_dscrt_vec}) and (\ref{z_solution_dscrt_vec}) for the $i$-th component of the displacement fields. Introducing the simplified expressions

\be
\hat{A}^x=\left[\hat{I}+\frac{(1+\nu)}{E}\hat{G}\hat{k}^x\right]^{-1},
\label{Ax}
\ee

\be
\hat{A}^z=\left[\hat{I}+\frac{(1+\nu)}{E}\hat{G}\hat{k}^z\right]^{-1}
\label{Az}
\ee

\noi and

\be
\frac{E}{(1+\nu)}\hat{G}\hat{k}^z\vec{u}_z^0 = \vec{v}_z^0,
\label{v_0}
\ee

\noi we obtain the final expressions (\ref{eq:x_solution_vec_i}), (\ref{eq:z_solution_vec_i}). Inserting Eqs.(\ref{eq:x_solution_vec_i}) and (\ref{eq:z_solution_vec_i}) in Eq.(\ref{additional_constraints_dscrt}) and making use of Eq.(\ref{k_x_matrix}),  the $x$ constraint can be written as

\be
\langle u_x\rangle\sum\limits_{i=1}^{N_xN_y}\sum\limits_{j=1}^{N_xN_y}A^x_{ij}k^x_{ii} =F_S^{top}+F_S^{lat}-\sum\limits_{l=1}^{N_xN_y}\sum\limits_{i=1}^{N_xN_y}\sum\limits_{j=1}^{N_xN_y}k^x_{li}A^x_{ij}\left\{u_S^{top}[j]+u_S^{lat}[j]\right\}
\label{additional_constraints_dscrt_x},
\ee

\noi from which

\be
\langle u_x\rangle =\frac{F_S^{top}+F_S^{lat}-\sum\limits_{l=1}^{N_xN_y}\sum\limits_{i=1}^{N_xN_y}\sum\limits_{j=1}^{N_xN_y}k^x_{li}A^x_{ij}\left\{u_S^{top}[j]+u_S^{lat}[j]\right\}}{\sum\limits_{i=1}^{N_xN_y}\sum\limits_{j=1}^{N_xN_y}A^x_{ij}k^x_{ii}}.
\label{average_u_x}
\ee

\noi In the same way we can handle the constraint along $z$: from Eq.(\ref{additional_constraints_dscrt}) we have

\be
\begin{array}{l}
\langle u_z\rangle\sum\limits_{i=1}^{N_xN_y}\sum\limits_{j=1}^{N_xN_y}A^x_{ij}k^z_{ii} =\\
F_N+\sum\limits_{i=1}^{N_xN_y}\sum\limits_{j=1}^{N_xN_y}k^z_{ij}[i]u_z^0[j]-\sum\limits_{i=1}^{N_xN_y}\sum\limits_{j=1}^{N_xN_y}\sum\limits_{l=1}^{N_xN_y}k^z_{li}A^x_{ij}\left\{v_z^0[j]+u_N[j]\right\}
\end{array}
\label{additional_constraints_dscrt_z}
\ee

\noi and

\be
\begin{array}{l}
\langle u_z\rangle =\\
\frac{F_N+\sum\limits_{i=1}^{N_xN_y}\sum\limits_{j=1}^{N_xN_y}k^z_{ij}[i]u_z^0[j]-\sum\limits_{i=1}^{N_xN_y}\sum\limits_{j=1}^{N_xN_y}\sum\limits_{l=1}^{N_xN_y}k^z_{li}A^x_{ij}\left\{v_z^0[j]+u_N[j]\right\}}{\sum\limits_{i=1}^{N_xN_y}\sum\limits_{j=1}^{N_xN_y}A^x_{ij}k^z_{ii}}.
\end{array}
\label{average_u_z}
\ee

\noi This completes the solution of the model: indeed inserting Eqs.(\ref{average_u_x}) and (\ref{average_u_z}) in Eqs.(\ref{x_solution_vec}) and (\ref{z_solution_vec}) respectively, we calculate the displacement fields at any point $(x,y)$ of the contact plane.

\section{Quasi-static dynamics \label{sec:quasi}}

In this section we discuss the quasi-static dynamics for the discrete elastic model. The adiabatic parameter is the shear force $F_S$ ($F_S^{lat}$, $F_S^{top}$ or both), and the dynamical protocol is enforced as  schematically shown in the flowchart in  Fig.\ref{fig:S11}. 

\noi First, we set the  value of the normal force $F_N$ and  choose a set of $u_z^0[i]$ ($u_z^0[i]$  are sampled from a Gaussian distribution, see appendix \ref{sec:calibration}). Second, we find the value of $\langle u_z\rangle$ according to Eq.(\ref{average_u_z}) and we plug it into Eq.(\ref{eq:z_solution_vec_i}): this gives the set of equilibrium displacements  $\vec{u}_z$, satisfying the second condition in Eq.(\ref{additional_constraints_dscrt}). It is possible that in some point $u_z[i]>u_z^0[i]$, which physically means that the slider $i-$th bottom plane element is not in contact with the underlying rough surface, i.e. the spring is \emph{detached} and the corresponding values of the interfacial stiffnesses $k_x[i]$ and $k_z[i]$ are set to 0, as well as the matrix elements $k^z_{ii}$ and $k^x_{ii}$. Third, a  small shear force $F_S$ is applied: we then calculate the equilibrium configuration along $x$ Eq.(\ref{eq:x_solution_vec_i}) using  Eq.(\ref{average_u_x}). 

Let us introduce the short notations for the internal stresses at the interface

\be
\begin{array}{l}
\left|\tau_{surf}(x,y)\right|=\left|\sigma_{xz}(x,y,0)\right|=\left|-\varphi^x_{surf}(x,y)\right|=k_x(x,y)\left|u_{x}(x,y,0)\right|\\
\left|\sigma_{surf}(x,y)\right|=\left|\sigma_{zz}(x,y,0)\right|=\left|-\varphi^z_{surf}(x,y)\right|=k_z(x,y)\left|u_{z}^0(x,y)-u_{z}(x,y,0)\right|.
\end{array}
\label{stresses_surf}
\ee

\noi After discretizing the contact plane in a mesh, i.e. $\tau_{surf}(x,y)\to\tau_{surf}[i]$ and $\sigma_{surf}(x,y)\to\sigma_{surf}[i]$ (see Eqs.(\ref{eq:tau_surf_dscrt}) and  (\ref{eq:sigma_surf_dscrt})), the local friction law (\ref{eq:local_rule}) requires that any site $i$ undergo the macroscopic laws of friction. In particular whenever the discrete condition (\ref{eq:local_rule_dscrt})

\be
\left|\tau_{surf}[i]\right|\geq \mu \left|\sigma_{surf}[i]\right|
\label{Amonton_dscrt}
\ee

\noi is satisfied, the corresponding spring is considered \emph{broken}: 
$k_x[i]=k_z[i]=0$, $k^z_{ii}=k^x_{ii}=0$. Althought the mathematical conditions for a spring to be broken or detached are the same, they are dynamically different as the following analysis is going to show.

Given the equilibrium configurations $\vec{u}_z$ and $\vec{u}_x$ for the initial values of $F_N$ and $F_S$, we start increasing the shear force on the slider. We want to determine the smallest next value of $F_S$ at which one of the \emph{attached} sites for which $k_x[i]\neq 0$ ($k_z[i]\neq 0$)  fulfills the condition in Eq.(\ref{Amonton_dscrt}). For such a purpose we first multiply the shear force by a factor $\lambda[i]>1$:

\be
\begin{array}{l}
F_S^{top}\to\lambda[i]F_S^{top}\\
u_S^{top}\to\lambda[i]u_S^{top}\\
F_S^{lat}\to\lambda[i]F_S^{top}\\
u_S^{lat}\to\lambda[i]u_S^{lat}.
\end{array}
\label{lambda_force}
\ee

\noi Then we insert the previous relations in Eq.(\ref{average_u_x}), achieving

\be
\langle u_x\rangle\to\lambda[i]\langle u_x\rangle.
\label{lambda_average_u_x}
\ee

\noi Plugging Eqs.(\ref{lambda_force}) and (\ref{lambda_average_u_x}) in (\ref{eq:x_solution_vec_i}),

\be
u_x[i]\to\lambda[i]u_x[i]
\label{lambda_x_solution_vec_i}.
\ee

\noi Now, the requirement Eq.(\ref{Amonton_dscrt}) translates to 

\be
\lambda[i]=\mu \frac{k_z[i]\left(u_{z}^0[i]-u_{z}[i]\right)}{k_x[i]\left|u_{x}[i]\right|},
\label{lambda_Amonton_dscrt}
\ee

\noi yielding the set of $\lambda[i]$ at which any $i$-th spring will be breaking.  Hence we take the smallest $\lambda[i]>1$ and we increase $F_S$ according to Eq.(\ref{lambda_force}): the corresponding site will be considered broken (putting $k_x[i]=k_z[i]=0$) and erased from the list of the \emph{available springs}, i.e. those attached, for which $k_z[i]\neq 0$ and $k_x[i]\neq 0$, and the detached ones ($k_x[i]=k_z[i]=0$, $u_{z}[i]>u_{z}^0[i]$ which not fulfilling Eq.(\ref{Amonton_dscrt})). With this new list of available springs, we proceed again with the equilibration along $z$ Eqs.(\ref{average_u_z})-(\ref{eq:z_solution_vec_i}). We notice that some of the detached sites for which previously $u_z[i]>u_z^0[i]$, could  now attach ($u_z[i]<u_z^0[i]$): for these we put $k_x[i]=c_xe^{-\frac{u_z[i]}{u_z^0[i]}}$ and $k_z[i]=\frac{c_z}{u_z^0[i]}$  according to (\ref{u_vec}); to the contrary, some of the formerly ``attached'' site could now detach: for these $k_x[i]=k_z[i]=0$. In either case, a new equilibrium along $z$ is needed. When the values of $\vec{k_x}$ and $\vec{k_z}$  in the list of available springs has not changed, we proceed further to the equilibration along $x$ Eqs.(\ref{average_u_x})-(\ref{eq:x_solution_vec_i}). After the equilibrium has been reached, it is still  possible that one of the attached sites could now fulfill the friction law Eq.(\ref{Amonton_dscrt}) and get broken; if  more than one site satisfy the rupture condition, we erase that with the largest elastic energy $\frac{1}{2}k_x[i]u_{x}[i]^2$. Hence, since the list of available springs has newly changed, we go back to the equilibration along $z$. It is important to notice that we do not increase further $F_S$ untill the list of available springs has not been modified by any of the possible events: attachment, detachment or rupture. When none of these occur, $F_S$ is further increased according to Eq.(\ref{lambda_Amonton_dscrt}) and the quasi-static protocol is repeated until all the  springs are broken.

\section{Model calibration \label{sec:calibration}}

In this section we describe the parameters that enter our model. They can be divided into three sets: \emph{material parameters}, \emph{sample parameters} and \emph{adjustable parameters}.

\begin{itemize}
\item \emph{Material parameters}. To this set belong the Young's modulus $E$, the Poisson's ratio $\nu$ and the friction coefficient $\mu$.

\noi In the case of PMMA $E=3.833 \times 10^9 N$, $\nu=0.38$, $\mu=0.5$.

\item \emph{Sample parameters}. To this set belong the macroscopic dimension of the sliders $L_x$, $L_y$ and $L_z$, the loading force $F_N$ and the parameters concerning the lateral shear $F_S^{lat}$: $h$ and $\Delta h$.

\item \emph{Adjustable parameters}. To this set belong the parameters that characterize the frictional properties of the slider-bottom plane interactions in our model. They are not directly inferred from the experimental setup, since they depend on to our main assumption that surface forces are purely elastic; however we can adjust them in order to reproduce the experimental features of Ref.\cite{rubinstein2006}. They are the mesh sizes $\Delta x$ and $\Delta y$ which set the degree of accuracy in the description of the contact plane slider-rough surface; $c_x$ and $c_z$ which will set the stregth of the linear interactions between the slider and the underlying surface along the $x$ and $z$ directions respectively; the  ``noise'' along the $z$ direction $\vec{u}_z^0$, i.e. the springs rest lengths: this is independent of the slider  properties but depends only on the surface  roughness. 
\end{itemize}

The following discussion will only concern the set of \emph{adjustable parameters}. The value of $\Delta x$ and $\Delta y$ are crucial for the computing time. As a matter of fact,  from Eq.(\ref{L_dscrt}), the smaller the mesh sizes, the higher will be the number of elements $N_x$ and $N_y$, increasing considerably the number of the vector and matrix  components appearing in our analysis. 

\noi The values of $\vec{u}_z^0$ will be set as follows. Firstly we notice that it corresponds to the average ``height'' that the underlying rough surface attains within the area $\Delta x\times\Delta y$ centered around the point $(x,y)$ (see Fig.\ref{fig:S1}). In general (see by instance Ref. \cite{persson2006})
 a rough surface is self-affine on distances $\ll\xi$, i.e. given two points $(x,y)$ and $(x',y')$ they are correlated whenever $\sqrt{(x-x')^2+(y-y')^2}\ll\xi$, where $\xi$ is the surface  correlation length. As a consequence of this, the noise $u_z^0(x,y)$ is a correlated Gaussian noise on scales $\ll\xi$, but is an uncorrelated Gaussian noise for distances  $\gg\xi$. Throughout our analysis we take  $\frac{\Delta x}{\xi}=\frac{\Delta y}{\xi}\gg 1$, since we assume $\xi\simeq 10\mu$ for PMMA and having chosen $\Delta x=\Delta y= 1 $mm. We can thus consider the components $u_z^0[i]$ as uncorrelated and Gaussian distributed according to 

\be
P(u_z^0[i])=\frac{e^{-\frac{(u_z^0[i]-\langle u_z^0\rangle)^2}{2\sigma_0^2}}}{\sqrt{2\pi\sigma_0^2}}\theta\left(u_z^0[i]\right).
\label{P_u_0}
\ee

\noi The average height $\langle u_z^0\rangle$ is  the root mean square surface roughness $\sqrt{w^2}$, while the distribution variance is $\sigma_0^2=w^2$, being $\sqrt{w^2}\simeq 1 \mu$ \cite{rubinstein2006bis}

\noi Finally we consider both constants $c_x$ and $c_z$. To set the the constant $c_z$ we refer to the experiment reported in Ref.\citep{rubinstein2006}
 where it was shown that the real contact area increases linearly with the applied load

\be
A_R=\int_0^{L_x}dx\int_0^{L_y}dy\, \rho(x,y)=\alpha F_N
\label{rubinstein}
\ee

\noi where the constant $\alpha$ can be determined from the experimental curves (see Fig.\ref{fig:1}(b)), and $\rho(x,y)$ represents the real contact area density. According to the assumption that each surface element behaves as a macroscopic object, we can assume that Eq.(\ref{rubinstein}) is valid locally as

\be
A_R(x,y)\propto \Delta x\Delta y e^{-\frac{u_z(x,y,0)}{u^0_z(x,y)}}\theta\left(u^0_z(x,y)-u_z(x,y,0)\right)
\label{rubinstein_local}
\ee

\noi thanks to Eq.(\ref{microscopical_load}). Now, we have to enforce a limit condition for the local real area of contact $A_R(x,y)$: first we want that the area is exactly 0 when $u^0_z(x,y)=u_z(x,y,0)$, hence

\be
A_R(x,y)=\gamma \Delta x\Delta y \left[e^{-\frac{u_z(x,y,0)}{u^0_z(x,y)}}-e^{-1}\right]\theta\left(u^0_z(x,y)-u_z(x,y,0)\right)
\label{rubinstein_local_1}
\ee

\noi with $\gamma>0$. Then we want that for $u_z(x,y,0)=0$ $A_R(x,y)=\Delta x\Delta y$, which gives $\gamma=\frac{e}{e-1}$

\be
A_R(x,y)=\frac{\Delta x\Delta y}{ \left[e-1\right]}\left[e^{-\frac{u_z(x,y,0)-u_z^0(x,y)}{u_z^0(x,y)}}-1\right]\theta\left(u^0_z(x,y)-u_z(x,y,0)\right)
\label{rubinstein_local_final}.
\ee

\noi In Movie S1 we show the adiabatic evolution of the real contact area for three different loading conditions. Relation (\ref{rubinstein}) can be cast in the following form after discretizing the contact plane

\be
\sum\limits_{i=1}^{N_xN_y}\frac{\Delta x\Delta y}{ \left[e-1\right]}\left[e^{-\frac{u_z[i]-u_z^0[i]}{u_z^0[i]}}-1\right]\theta(u_z^0[i]-u_z[i])=\alpha F_N.
\label{rubinstein_vec}
\ee

\noi By tuning the value of $c_z$, we change the equilibrium values $\vec{u}_z$ given by  Eqs.(\ref{eq:z_solution_vec_i})-(\ref{average_u_z}), in order to fulfill Eq.(\ref{rubinstein_vec}) (see Fig.\ref{fig:2}(a)). The best value that we found is $c_z= 1.65\times  10^8\frac{N}{m^2}$. In Fig.\ref{fig:S2} we show the variation of the real area of contact during the shearing process. Fig.\ref{fig:S3} shows  the same quantity for samples with different nominal area $A_0= L_x\times L_y$, undergoing the same loading $F_N$.

\noi Finally, we set $c_x$ by varying the stiffness of the springs along the $x$ direction for the optimal reproduction of the precursors quasi static dynamics  for a given sample, as shown in Fig.\ref{fig:2}(b). The best fit was achieved for $c_x=1.65\times  10^{12}\frac{N}{m^3}$.

\section{Displacement controlled shear \label{sec:displacement}}

The experiments reported in Ref. \cite{rubinstein2006} are performed by applying a lateral shear force 
through a spring with elastic constant $K_S=4\times 10^6$N/m. Hence the external force appearing in  Eq.(\ref{eq:x_solution}) is replaced by Eq. (\ref{eq:spring_shear_force}). We can thus write the solutions for the displacements at the slider bottom plane as

\be
\begin{array}{l}
u_x(x,y,0)=\langle u_x\rangle + \frac{(1+\nu)}{E}\left\{\int_0^{L_x}d\xi\int_0^{L_y}d\eta
G(x,\xi;y,\eta;0,0)\varphi_{surf}^x(\xi,\eta,0)+\right.\\
\left.+\frac{K_S\left(U_s-\langle u_x\rangle\right)}{L_y2\Delta h}
\int_0^{L_y}d\eta\int_{h-\Delta h}^{h+\Delta h}d\zeta G(x,0;y,\eta;0,\zeta)\right\},
\end{array}
\label{x_solution_spring}
\ee

\noi and adopt the shorthand notation

\be
K_S(x;h,\Delta h) = \frac{K_S}{L_y2\Delta h}
\int_0^{L_y}d\eta\int_{h-\Delta h}^{h+\Delta h}d\zeta G(x,0;y,\eta;0,\zeta)
\label{K_spring},
\ee

\noi where the value of the integral is furnished  in Eq.(\ref{G_x_lat}). Notice that in this case  we consider only the lateral contribution to the external shearing force, neglecting, for the moment, the force exerted on top of the slider. Once we apply the discretization scheme at the frictional interface, the displacements along $\hat{x}$ become

\be
u_x[i]=-\frac{(1+\nu)}{E}\sum\limits_{j=1}^{N_xN_y}\Delta x\Delta y
G_{ij}k_x[j]u_x[j]+K_S[i]\left(U_s-\langle u_x\rangle\right)+\langle u_x\rangle,
\label{x_solution_dscrt_spring}
\ee

\noi which in vectorial form read

\be
\vec{u}_x=-\frac{(1+\nu)}{E}\hat{G}\hat{k}^x\vec{u}_x+\vec{K}_S\left(U_s-\langle u_x\rangle\right)+\vec{\langle u_x\rangle}
\label{x_solution_dscrt_vec_spring}
\ee

\noi where we implicitely made use of Eq.(\ref{k_x_matrix}). Inverting Eq.(\ref{x_solution_dscrt_vec_spring}) we finally achieve

\be
\vec{u}_x=\left[\hat{I}+\frac{(1+\nu)}{E}\hat{G}\hat{k}^x\right]^{-1}\left\{\vec{K}_S\left(U_s-\langle u_x\rangle\right)+\vec{\langle u_x\rangle}\right\}
\label{x_solution_vec_spring},
\ee

\noi where we recall that $\vec{\langle u_x\rangle}$ is a vector with all components constant.

\noi The force balance in Eq.\ref{additional_constraints_dscrt} now takes te form

\be
\sum\limits_{i=1}^{N_xN_y}\Delta x\Delta yk_x[i]u_x[i] = K_S\left(U_s-\langle u_x\rangle\right)
\label{additional_constraints_dscrt_spring}.
\ee

\noi By inserting Eq.(\ref{x_solution_vec_spring}) into Eq.(\ref{additional_constraints_dscrt_spring}) and making use of the defintion Eq.(\ref{Ax}) we obtain, after straightforward passages,

\be
\langle u_x\rangle =U_S\frac{\left\{K_S-\sum\limits_{l=1}^{N_xN_y}\sum\limits_{i=1}^{N_xN_y}\sum\limits_{j=1}^{N_xN_y}k^x_{li}A^x_{ij}K_S[j]\right\}}{\left\{K_S-\sum\limits_{l=1}^{N_xN_y}\sum\limits_{i=1}^{N_xN_y}\sum\limits_{j=1}^{N_xN_y}k^x_{li}A^x_{ij}K_S[j]+\sum\limits_{i=1}^{N_xN_y}\sum\limits_{j=1}^{N_xN_y}A^x_{ij}k^x_{ii}\right\}}.
\label{average_u_x_spring}
\ee

\noi This complete the solution, indeed after Eq.(\ref{average_u_x_spring}) has been plugged into Eq.(\ref{x_solution_vec_spring}) the set of the displacements at the frictional interfaces can be obtained.

\section{Calculation of Coulomb stress \label{sec:calculation}}

The evaluation of the normal and tangential stresses on a generic point of the sample $(x,y,z)$ is given by the following expressions

\be
\begin{array}{l}
\left|\tau(x,y,z)\right|= \left|\sigma_{xz}(x,y,z)\right|=\frac{E}{(1+\nu)}\frac{\partial u_x}{\partial z}\\
\left|\sigma(x,y,z)\right|=\left|\sigma_{zz}(x,y,z)\right|=\frac{E}{(1+\nu)}\frac{\partial u_z}{\partial z}.
\end{array}
\label{stresses_P}
\ee

\noi We start our analysis from the shear internal stress $\tau(x,y,z)$. Inserting Eq.(\ref{eq:x_solution}) in Eq.(\ref{stresses_P}) immediately reveals that we need to calculate the derivative of the Green function at $(x,y,z)$:

\be
\begin{array}{l}
\tau(x,y,z)=\left\{\int_0^{L_x}d\xi\int_0^{L_y}d\eta
\frac{\partial G(x,\xi;y,\eta;z,0)}{\partial z}\varphi_{surf}^x(\xi,\eta,0)-\right.\\
-\frac{F_s^{top}}{L_xL_y}\int_0^{L_x}d\xi\int_0^{L_y}d\eta\frac{\partial G(x,\xi;y,\eta;z,L_z)}{\partial z}+\\
\left.+\int_0^{L_y}d\eta\int_{h-\Delta h}^{h+\Delta h}d\zeta \frac{\partial G(x,0;y,\eta;z,\zeta)}{\partial z}\frac{F_s^{lat}}{L_y2\Delta h}\right\}.
\end{array}
\label{tau_P_def}
\ee

\noi The first derivative appearing in the former expression reads

\be
\begin{array}{l}
\frac{\partial G(x,\xi;y,\eta;z,0)}{\partial z}=-\frac{8}{L_xL_yL_z^2}\left\{\sum\limits_{n=1}\frac{n\sin\left(\frac{n\pi z}{L_z}\right)}{4\left(\frac{n\pi}{L_z}\right)^2}+\right.\\
+
\sum\limits_{m=1}\sum\limits_{n=1}\frac{n\sin\left(\frac{n\pi z}{L_z}\right)\cos\left(\frac{m\pi y}{L_y}\right)\cos\left(\frac{m\pi \eta}{L_y}\right)}{2\left[\left(\frac{m\pi}{L_y}\right)^2+\left(\frac{n\pi}{L_z}\right)^2\right]}+\\
+\sum\limits_{n=1}\sum\limits_{p=1}\frac{n\sin\left(\frac{n\pi z}{L_z}\right)\cos\left(\frac{p\pi x}{L_x}\right)\cos\left(\frac{p\pi \xi}{L_x}\right)}{2\left[\left(\frac{n\pi}{L_z}\right)^2+\left(\frac{p\pi}{L_x}\right)^2\right]}+\\
\left.+\sum\limits_{m=1}\sum\limits_{n=1}\sum\limits_{p=1}\frac{n\sin\left(\frac{n\pi z}{L_z}\right)\cos\left(\frac{m\pi y}{L_y}\right)\cos\left(\frac{m\pi \eta}{L_y}\right)\cos\left(\frac{p\pi x}{L_x}\right)\cos\left(\frac{p\pi \xi}{L_x}\right)}{\left(\frac{n\pi}{L_z}\right)^2+\left(\frac{n\pi}{L_z}\right)^2+\left(\frac{m\pi}{L_y}\right)^2+\left(\frac{p\pi}{L_x}\right)^2}\right\}.
\end{array}
\label{derivative_G_A}
\ee

\noi The first sum in Eq.(\ref{derivative_G_A}) can be evaluated according to Ref. \cite{prudnikov1981} as:

\be
\frac{1}{4}\sum\limits_{n=1}\frac{n\sin\left(\frac{n\pi z}{L_z}\right)}{\left(\frac{n\pi}{L_z}\right)^2}=L_z\frac{(L_z-z)}{8\pi},
\label{derivative_G_A_1}
\ee

\noi and the second can be rewritten in the form

\be
\begin{array}{l}
\frac{1}{2}\sum\limits_{m=1}\sum\limits_{n=1}\frac{n\sin\left(\frac{n\pi z}{L_z}\right)\cos\left(\frac{m\pi y}{L_y}\right)\cos\left(\frac{m\pi \eta}{L_y}\right)}{\left(\frac{m\pi}{L_y}\right)^2+\left(\frac{n\pi}{L_z}\right)^2}=\\
\frac{L_z}{8\pi}\left\{\mathscr{G}(|y-\eta|,z;L_y,L_z)+\mathscr{G}(y+\eta,z;L_y,L_z)\right\}
\end{array}
\label{derivative_G_A_2}
\ee

\noi in terms of the function

\be
\mathscr{G}(a,b;L_a,L_b)=L_b\sum\limits_{m=1}\frac{\cos\left(\frac{m\pi a}{L_a}\right)\sinh\left(\frac{m\pi (L_b-b)}{L_a}\right)}{\sinh\left(\frac{m\pi L_b}{L_a}\right)}.
\label{g}
\ee

\noi The third term can be treated in the same way. The fourth element in the bracket of Eq.(\ref{derivative_G_A}) is easily evaluated:

\be
\begin{array}{l}
\sum\limits_{m=1}\sum\limits_{n=1}\sum\limits_{p=1}\frac{n\sin\left(\frac{n\pi z}{L_z}\right)\cos\left(\frac{m\pi y}{L_y}\right)\cos\left(\frac{m\pi \eta}{L_y}\right)\cos\left(\frac{p\pi x}{L_x}\right)\cos\left(\frac{p\pi \xi}{L_x}\right)}{\left(\frac{n\pi}{L_z}\right)^2+\left(\frac{n\pi}{L_z}\right)^2+\left(\frac{m\pi}{L_y}\right)^2+\left(\frac{p\pi}{L_x}\right)^2}=\\
\frac{L_z}{8\pi}\left\{\mathscr{Q}(|x-\xi|,|y-\eta|,z;L_x,L_y,L_z)+\mathscr{Q}(|x-\xi|,y+\eta,z;L_x,L_y,L_z)+\right.\\
+\mathscr{Q}(x+\xi,|y-\eta|,z;L_x,L_y,L_z)+\mathscr{Q}(x+\xi,y+\eta,z;L_x,L_y,L_z)+\\
\left.+\mathscr{Q}(x+\xi,y+\eta,z;L_x,L_y,L_z)\right\}
\end{array}
\label{derivative_G_A_4}
\ee

\noi where 

\be
\begin{array}{l}
\mathscr{Q}(a,b,c;L_a,L_b,L_c)=\\
L_z\sum\limits_{m=1}\sum\limits_{p=1}\frac{\cos\left(\frac{p\pi a}{L_a}\right)\cos\left(\frac{m\pi b}{L_b}\right)\sinh\left(\frac{\pi (L_c-c)}{L_aL_b}\sqrt{(pL_b)^2+(mL_a)^2}\right)}{\sinh\left(\frac{\pi L_c}{L_aL_b}\sqrt{(pL_b)^2+(mL_a)^2}\right)}.
\end{array}
\label{q}
\ee

\noi Hence from Eqs.(\ref{derivative_G_A_1}), (\ref{derivative_G_A_2}), (\ref{derivative_G_A_4}), Eq.(\ref{derivative_G_A}) transforms to

\be
\begin{array}{l}
\frac{\partial G(x,\xi;y,\eta;z,0)}{\partial z}=-\frac{1}{L_xL_yL_z}\left\{L_z-z+\mathscr{G}(|y-\eta|,z;L_y,L_z)+\right.\\
+\mathscr{G}(y+\eta,z;L_y,L_z)+\mathscr{G}(|x-\xi|,z;L_x,L_z)+\mathscr{G}(x+\xi,z;L_x,L_z)+\\
+\mathscr{Q}(|x-\xi|,|y-\eta|,z;L_x,L_y,L_z)+\mathscr{Q}(|x-\xi|,y+\eta,z;L_x,L_y,L_z)+\\
+\mathscr{Q}(x+\xi,|y-\eta|,z;L_x,L_y,L_z)+\mathscr{Q}(x+\xi,y+\eta,z;L_x,L_y,L_z)+\\
\left.+\mathscr{Q}(x+\xi,y+\eta,z;L_x,L_y,L_z)\right\}.
\end{array}
\label{derivative_G_A_final}
\ee

\noi The second term in Eq.(\ref{tau_P_def}) is straightforwardly evaluated
 (see \cite{prudnikov1981})

\be
\int_0^{L_x}d\xi\int_0^{L_y}d\eta
\frac{\partial G(x,\xi;y,\eta;z,L_z)}{\partial z}=\frac{z}{L_z}.
\label{derivative_G_B}
\ee

\noi Finally, the third term is

\be
\begin{array}{l}
\int_0^{L_y}d\eta\int_{h-\Delta h}^{h+\Delta h}d\zeta\frac{\partial G(x,0;y,\eta;z,\zeta)}{\partial z}\Big|_{x,y,z}=\\
-\frac{1}{2L_x}\Big\{\frac{(z-h-\Delta h)^2-(z+h+\Delta h)^2+(z-h+\Delta h)^2+(z+h-\Delta h)^2}{2L_z}-\\
-\left|z-h-\Delta h\right|+\left|z+h+\Delta h\right|-\left|z-h+\Delta h\right|+\left|z-h-\Delta h\right|\Big\}-\\
-\frac{2}{\pi}\sum\limits_{n=1}\frac{\cosh\left(\frac{n\pi}{L_z}\left(L_x-x\right)\right)}{n\sinh\left(\frac{n\pi L_x}{L_z}\right)}\Big\{\cos\left(\frac{n\pi}{L_z}\left|z-h-\Delta h\right|\right)-\cos\left(\frac{n\pi}{L_z}\left(z+h+\Delta h\right)\right)+\\
+\cos\left(\frac{n\pi}{L_z}\left|z-h+\Delta h\right|\right)-\cos\left(\frac{n\pi}{L_z}\left|z+h-\Delta h\right|\right)\Big\}.
\end{array}
\label{derivative_G_C}
\ee

\noi Inserting Eqs.(\ref{derivative_G_A_final}), (\ref{derivative_G_B}) and (\ref{derivative_G_C}) in Eq.(\ref{tau_P_def}) we obtain the expression for the shear stress at any point of the sample. 

The normal stress can be written as

\be
\begin{array}{l}
\sigma(x,y,z)=\left\{\int_0^{L_x}d\xi\int_0^{L_y}d\eta
\frac{\partial G(x,\xi;y,\eta;z,0)}{\partial z}\varphi_{surf}^z(\xi,\eta,0)\right.-\\
\left.-\frac{F_N}{L_xL_y}\int_0^{L_x}d\xi\int_0^{L_y}d\eta\frac{\partial G(x,\xi;y,\eta;z,L_z)}{\partial z}\right\}
\end{array}
\label{sigma_P_def}
\ee

\noi and can be handled in the same way as the shear stress,  plugging Eqs.(\ref{derivative_G_A_final}) and (\ref{derivative_G_B}) into Eq.(\ref{sigma_P_def}).

\noi 
In Fig.\ref{fig:S10} we show the quasi-static evolution of the normal and shear stress at the reference  plane $z_P=2$mm. We notice that in the limit $z_P\to 0$, Eq.(\ref{tau_P_def}) and Eq.(\ref{sigma_P_def}) transform to \ref{stresses_surf}. Finally we can define the Coulomb stress at any  point in the sample

\be
\tau_C(x,y,z)=\left|\tau(x,y,z)\right|-\mu\left|\sigma(x,y,z)\right|.
\label{coulomb_stress}
\ee

\noi  When we perform the mesh discretization at the contact plane, we have to map the variables on the plane according to Eq.(\ref{ij}), i.e. $(x,y)\to i$. Hence the derivative of the Green's function Eq.(\ref{derivative_G_A_final}) transforms to the following matrix

\be
\frac{\partial G(x,\xi;y,\eta;z,0)}{\partial z}\to\frac{\partial \hat{G}_{ij}}{\partial z}
\label{derivative_G_A_final_matrix}.
\ee

\noi The second term in the shear stress (Eq.(\ref{tau_P_def})) is constant as it results from Eq.(\ref{derivative_G_B}), then we can write

\be
-\frac{F_s^{top}}{L_xL_y}\int_0^{L_x}d\xi\int_0^{L_y}d\eta\frac{\partial G(x,\xi;y,\eta;z,L_z)}{\partial z}\to \tau^{top}_{z}[i_P].
\label{derivative_G_B_dscrt}
\ee

\noi The same can be done for the third term, although from Eq.(\ref{derivative_G_C}) it is apparent its dependence on $x$:

\be
\frac{F_s^{lat}}{L_y2\Delta h}\int_0^{L_y}d\eta\int_{h-\Delta h}^{h+\Delta h}d\zeta \frac{\partial G(x,0;y,\eta;z,\zeta)}{\partial z}
\to \tau^{lat}_{z}[i].
\label{derivative_G_C_dscrt}
\ee

\noi The discrete form of the shear stress can be cast as 

\be
\tau_{z}[i]=-\Delta x\Delta y\sum\limits_{j=1}^{N_xN_y}\frac{\partial \hat{G}_{ij}}{\partial z}\left(k_x[j]u_x[j]+\tau^{top}_{z}[i]+\tau^{lat}_{z}[i]\right)
\label{tau_P_def_dcrt}
\ee

\noi and in the vectorial form

\be
\vec{\tau}_{z}=-\frac{\partial \hat{G}}{\partial z}\left(\hat{k}^x\vec{u}_x+\vec{\tau}^{top}_{z}+\vec{\tau}^{lat}_{z}\right)
\label{tau_P_def_dcrt_vec}
\ee

\noi thanks to Eq.(\ref{k_x_matrix}). Applying the same analysis to the normal stress Eq.(\ref{sigma_P_def}), we have for the second term the same as Eq.(\ref{derivative_G_B_dscrt}), i.e. 

\be
-\frac{F_N}{L_xL_y}\int_0^{L_x}d\xi\int_0^{L_y}d\eta\frac{\partial G(x,\xi;y,\eta;z,L_z)}{\partial z}\to \sigma^{N}_{z}[i],
\label{derivative_G_D_dscrt}
\ee

\noi and finally

\be
\sigma_{z}[i]=\Delta x\Delta y\sum\limits_{j=1}^{N_xN_y}\frac{\partial \hat{G}_{ij}}{\partial z}\left(k_z[j]\left(u_z^0[j]-u_z[j]\right)+\sigma^{N}_{z}[i]\right)
\label{sigma_P_def_dcrt}
\ee

\noi which in the vectorial form transforms to

\be
\vec{\sigma}_{z}=\frac{\partial \hat{G}}{\partial z}\left(\hat{k}^z\left(\vec{u}_z^0-\vec{u}_z\right)+\vec{\sigma}^{N}_{z}\right)
\label{sigma_P_def_dcrt_vec}
\ee

\noi  thanks to (\ref{k_z_matrix}). Hence the discrete expression of the Coulomb stress is given by

\be
\vec{\tau}_{z}^C=\left|\vec{\tau}_{z}\right|-\mu\left|\vec{\sigma}_{z}\right|
\label{coulomb_stress_dscrt}
\ee

\noi and its quasi-static evolution is shown in Movie S2.

\section{Geometrical dependence of front precursors \label{geometrical}}

We report here some general consideration on the role of geometry in the shape of precursors.
Plugging the continuous expression of Eq.(\ref{eq:x_solution_vec_i}) into the first of Eqs.(\ref{stresses_surf}), we obtain that  the shear stress at the interface can be written as

\be
\tau_{surf}(x,y)=k_x(x,y)\int_0^{L_x}d\xi\int_0^{L_y} d\eta\,A^x(x,y;\xi,\eta)\left[u_S^{top}(\xi,\eta)+u_S^{lat}(\xi,\eta)+\langle u_x\rangle\right],
\label{eq:tau_surf_cont}
\ee

\noi or, in its discrete form, as 

\be
\tau_{surf}[i]=k_x[i]\sum\limits_{j=1}^{N_xN_y}A^x_{ij}\left[u_S^{top}[j]+u_S^{lat}[j]+\langle u_x\rangle\right].
\label{eq:tau_surf_dscrt}
\ee

\noi $A^x(x,y;\xi,\eta)$ is the continuous version of the matrix $A^x_{ij}$ introduced in Eq.(\ref{eq:x_solution_vec_i}), connected to the inverse of the Green's operator and to $k_x(x,y)$ (see  Eq.(\ref{Ax})). According to Eq.(\ref{eq:tau_surf_cont}), $\tau_{surf}(x,y)$ can be  decomposed into three contributions stemming from  (i) forces applied to the top surface $F_S^{top}$,
(ii) forces applied on the trailing edge $F_S^{lat}$ and (iii) the average volume shift (see respectively Eqs.(\ref{u_s_top}), (\ref{u_s_lat}) and Eq.(\ref{average_u_x})).  Among these, only the second term  is responsible for the stress gradient at the interface. Indeed, when  loading is exerted on the top plate, $u_S^{lat}=0$ and $u_S^{top}$ is uniform across the entire slider bottom plane as it turns out from Eqs.(\ref{G_x_top}) and (\ref{u_s_top}). Moreover $\langle u_x\rangle$ is constant for any site, being the avergae displacement.

\noi The   normal stress at the interface, on the other hand,  takes the following continuous and discrete expression respectively

\be
\sigma_{surf}(x,y)=k_z(x,y)\left\{\int_0^{L_x}d\xi\int_0^{L_y} d\eta\,A^z(x,y;\xi,\eta)\left[v_0(\xi,\eta)+u_N(\xi,\eta)+\langle u_z\rangle\right]-u_z^0(x,y)\right\},
\label{eq:sigma_surf_cont}
\ee

\be
\sigma_{surf}[i]=k_z[i]\left\{\sum\limits_{j=1}^{N_xN_y}A^x_{ij}\left[v_0[j]+u_N[j]+\langle u_z\rangle\right]-u_z^0[j]\right\}.
\label{eq:sigma_surf_dscrt}
\ee

\noi $A^z(x,y;\xi,\eta)$ and $v_0(x,y)$ represent the continuous expressions of  Eqs.(\ref{Az}) and  (\ref{v_0}), $u_N(x,y)$ is introduced in Eq.(\ref{u_n}) and the average shift expression $\langle u_z\rangle$ is defined in its discrete form in Eq.(\ref{average_u_z}). The three terms in Eq.(\ref{eq:sigma_surf_cont}) do not exhibit any apparent gradient on the slider bottom plane, since they depend on the material bottom surface   heterogeneity ($v_0(x,y)$), which in average provides an uniform contribution; on the normal applied load $F_N$, which is again even and uniform throughout the frictional interface (the edge effects are playing indeed a minor role); and on $\langle u_z\rangle$ which is by definition uniform. 

\noi With the discrete expressions for the surface shear and normal stresses at hand, we can write the local slip condition (\ref{eq:local_rule}) as

\be
\tau^C_{surf}[i]=\left|\tau_{surf}[i]\right|-\mu\left|\sigma_{surf}[i]\right|.
\label{eq:local_rule_dscrt}
\ee

\noi  Based on the former analysis, when shear is applied uniformly  from above, i.e. $F_S\equiv F_S^{top}$, our model suggests that 
both shear ($\tau_{surf}$) and normal ($\sigma_{surf}$) component of the Coulomb stress are uniform on the contact surface and therefore no precursors
are expected. This is clearly seen in Fig.\ref{fig:3}(d) and corroborates the finding of one-dimensional friction models \cite{capozza2012}. However, scalar model does just furnish  an average picture of the internal stress behaviors when shear is applied uniformly on the top, as it does not account, by instance, for the elastic Poisson expansion, nor for  torque,  responsibles for the stresses heterogeneities observed  at the sample edges \cite{bendavid2010}.

 When the shear force is applied on the trailing edge, i.e. $F_S\equiv F_S^{lat}$, only the second term proportional to $u_S^{lat}$ survives in Eqs.(\ref{eq:tau_surf_cont}), (\ref{eq:tau_surf_dscrt}). Therefore the shear stress $\tau_{surf}$ displays an evident gradient across the contact interface, leading to detachment of the regions where this gradient is more pronounced and,  eventually, to precursors activity. In particular, when loading  uniformly
from the trailing edge ($\Delta h=L_z/2$), the corresponding contribution given by $u_S^{lat}$ to $\tau_{surf}$  is

\begin{equation}
\tau_{ii}(x,y)=\frac{(1+\nu)}{E}\frac{L_x}{L_yL_z}F_s^{lat}k_x(x,y)\left[\frac{x^2}{2L_x^2}-\frac{x}{L_x}+\frac{1}{3}\right],
\label{u_x_lat_uniform}
\end{equation}

\noindent and  the interface shear stress dependence on the sample geometry is entirely encapsulated in the ratio $R=\frac{L_x}{L_yL_z}$ (see Eqs.(\ref{G_x_lat_simple}), (\ref{u_s_lat})). Since we argue that the precursors envelope curves reflect the symmetries appering in the Coulomb stress $\tau_C(x,y,0)$,    the rescaled precursor profile $\ell/L_x$ should only depend on  $R$, in complete agreement with the numerics reported in Fig. \ref{fig:6}(a),(b),(c).

 If $F_S^{lat}$  is applied  by a rod at height $z=h$, a  direct evaluation of the corresponding of $u_S^{lat}$ contribution to the expressions (\ref{eq:tau_surf_cont}), (\ref{eq:tau_surf_dscrt})
yields a non-trivial dependence on  $L_x,L_y,L_z,h$ and $\Delta h$, as it is immediate to see by referring to Eqs.(\ref{G_x_lat}) and Eqs.(\ref{u_s_lat}).
As a result, no universal scaling on the sample dimensions and rod parameters  ($h$, $\Delta h$) can be found in this case,
as already illustrated in Figs. \ref{fig:5}(b), \ref{fig:S4}, \ref{fig:S5}, \ref{fig:S6}, \ref{fig:S7}, \ref{fig:S8}, \ref{fig:S9}.

 When the shear is applied from the slider top and trailing edge simultaneously, both terms $u_S^{lat}$ and $u_S^{top}$  in Eq.(\ref{eq:tau_surf_cont}) are competing: when the first dominates, no precursor nucleation should be seen and the slider detaches as a whole when the frictional force is reached. This is indeed what is shown in Fig.\ref{fig:6}(d). Yet, conclusions about uniform top shearing are to be taken with a grain of salt  as explained before, indeed they are valid only on average and for very large samples, since no edges nonuniformities are encompassed within the scalar elasticity theory.

Finally,  the role of the contact plane disorder heterogeneity  on the interfacial stresses appears  in Eqs.(\ref{eq:tau_surf_cont}), (\ref{eq:sigma_surf_cont})  through $A^x(x,y;\xi,\eta)$, $A^z(x,y;\xi,\eta)$, $k_x(x,y)$, $k_z(x,y)$ and $v_0(x,y)$. For the relative narrow distribution of the heights $u_0^z(x,y)$ characterizing the rough substrate here considered, the contribution of the interfacial disorder to the precursor nucleation is irrelevant and uniform in average.

\section{The finite element model \label{sec:FEM}}
The system has been initially simulated by means of a three-dimensional model based on the finite element method  (FEM).  In order to compare this model with experiments, we  chose parameters that correspond to the PMMA samples employed in Refs. \cite{rubinstein2004,rubinstein2006,rubinstein2007,bendavid2010,bendavid2011}. In particular, we make  use of the sample geometry and dimensions used in those experiments and of the known elastic constants (Young's modulus $E$ and Poisson's ratio $\nu$) of PMMA. In addition, the  slider bottom corners are rounded by a radius of 2\,mm to avoid both stress singularities at the edges and frustrated Poisson expansion \cite{rubinstein2006, bendavid2010, radiguet2013}.

For the simulations we used the commercial FEM software COMSOL. We approximated the geometry and displacements by use of quadratic shape functions and chose the size and local refinement of the tetrahedral finite element mesh such that further refinement would give no appreciable gain in accuracy. Resulting displacements within elements were interpolated using the polynomials of the shape functions,  stresses and strains were interpolated using gradients of the shape functions.

\subsection{Model calibration and stress profiles}

Our model encompasses the setting of only two parameters, i.e. the asperity spring stiffnesses $k_x$ and $k_z$, as the material elastic constants have been chosen to reproduce the PMMA properties. To tune the values of $k_x$ and $k_z$ our benchmark is the experiment of Ref.\cite{bendavid2010}, where the internal normal and shear stresses, $\sigma(x,y,z)$ and $\tau(x,y,z)$ respectively, were calculated at the points $(x,y=L_y/2,z=z_p)$ where $z_p$ is a plane placed 2mm above the bottom surface. The slider dimensions were $L_x=200mm$, $L_y=6mm$ and $L_z=100mm$ while no shearing force was applied on the PMMA block. We have reproduced the experimental setup by means of our FEM model, and adjusted the values of $k_x$ and $k_z$ until the shear and normal stress profiles respectively were showing a satisfactory agreement: the results are displayed in Fig.\ref{fig:S13}. In particular, it is possible to see how our 3D FEM model not only furnishes an accurate quantitative agreement with the experimental curves (dashed lines), but  accounts also for the high nonuniformity exhibited by normal and shear stress when no shearing is present. Indeed $\sigma(x,L_y/2,z_p)$ clearly displays divergences at the bottom edges due to the  interface pinnning, as well as $\tau(x,L_y/2,z_p)$. In this regard, in  the experiments reported in \cite{bendavid2010,bendavid2011},  shear stress nonuniformity deriving from differential Poisson expansion frustrated at the interface, was reduced by leaving the block edges free to expand. In our model, both normal and shear stress nonuniformities could be controlled thanks to the rounding of the block's bottom corners, providing a remarkable agreement of the stress profiles at $F_S=0$.

The situation considerably changes when $F_S\neq 0$. Experimentally, normal stress $\sigma(x,L_y/2,z_p)$ shows stronger nonuniformities localized to regions near the block's edges, as the effect of the torque induced by the application of $F_S$(see Fig.\ref{fig:S14}(a) and (c), dashed lines). This effect is much more pronounced when shear is applied at the sample's trailing edge ($F_S=F_S^{lat}$), by means of a 4mm-sized rod placed at  height $h=6$mm \cite{bendavid2010} (Fig.\ref{fig:S14}(a)). To partially avoid this byproduct, a controlled gradient in  $F_N$ was applied  by introducing a slight rigid tilt to the PMMA slider. This rigid tilt of the block was notably enhanced when shear was applied on the side.
We could not reproduce such a tilt in our FEM model, since this loading controlled gradient leads to a highly non-linear system response, causing severe stability problems during the FEM solution. Furthermore, also the shear stress  $\tau(x,L_y/2,z_p)$ is considerably altered when $F_S\neq 0$ (see Fig.\ref{fig:S14}(b) and (d), dashed lines). As in the case of $\sigma$, $\tau$ nonuniformities appear more marked while shear is impressed from aside (Fig.\ref{fig:S14}(b), dashed line) instead of uniformly on the block's top (Fig.\ref{fig:S14}(d), dashed line).  

\noi   While  trying to reproduce the internal stress profiles for $F_S\neq 0$, our FEM model fails. Indeed none of the simulated stresses, achieved for different values of the shearing force and different  modes of shearing,  exhibits a satisfactory agreement with the experimental curves, although qualitatively the observed trends are respected (see Fig\ref{fig:S14}). We believe that  our model failure is ultimately  ascribable  to the impossibility of transposing numerically the tilting block expedient employed in the experiment. As a matter of fact, when $F_S$ is applied uniformly on top of the slider, the  agreement between experimental and numerical stress profiles appear to be slightly better than when shear is imposed from the edge.

\bibliography{friction}


\pagebreak


\vspace{15cm}

\makeatletter 
\renewcommand{\thefigure}{S\@arabic\c@figure} 
\makeatother
\setcounter{figure}{0}

\begin{figure}[ht]
\centering
\includegraphics[width=0.8\textwidth]{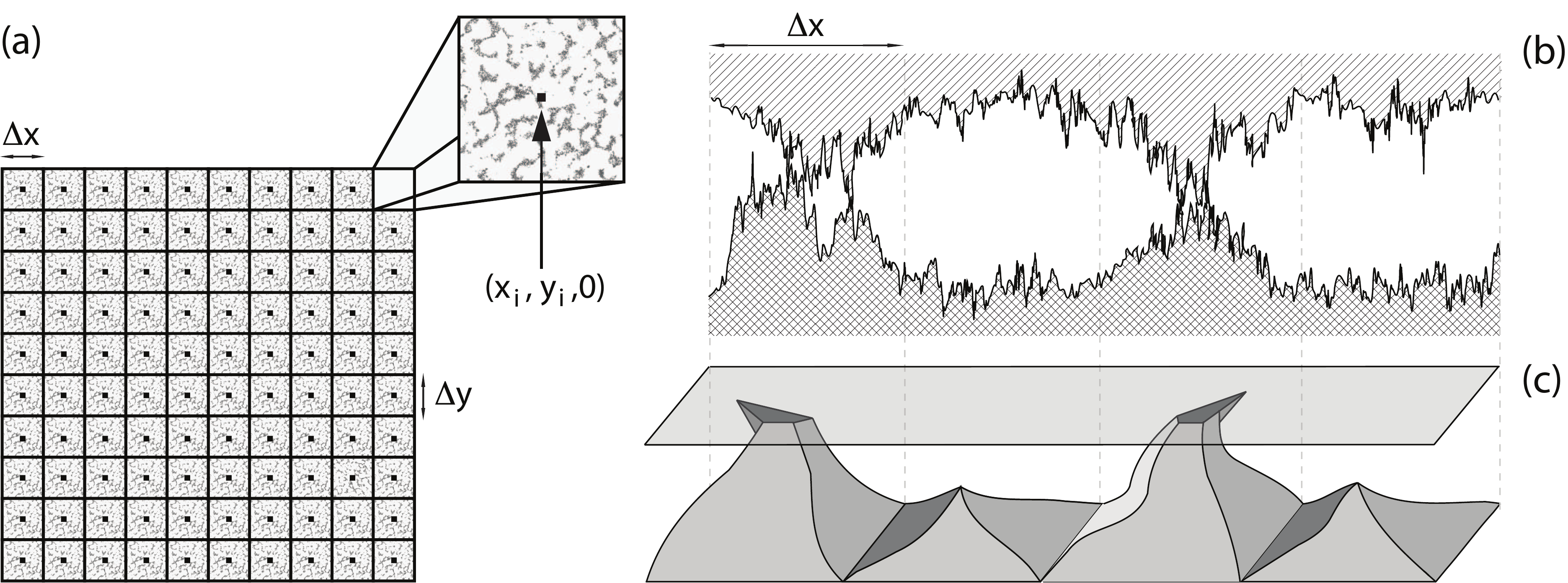}
\caption{{\bf Discretization}. Mesh of the slider-rough surface contact plane (a). The central point of each mesh  element is representative of the microscopical roughness-induced forces within  the interface portion $\Delta x\times\Delta y$ (b). Panel (c): coarse-grained  asperity-like interaction of the substrate with the upper slider. The elastic picture emerging in panel (c) is mathematically translated in sthe pring-like interaction as shown in Fig.\ref{fig:1} (bottom panel): the slider lais on a carpet of springs $u_z^0[i]$. }
\label{fig:S1}
\end{figure}

\clearpage
\begin{figure}[ht]
\centerline{\includegraphics[width=0.8\textwidth]{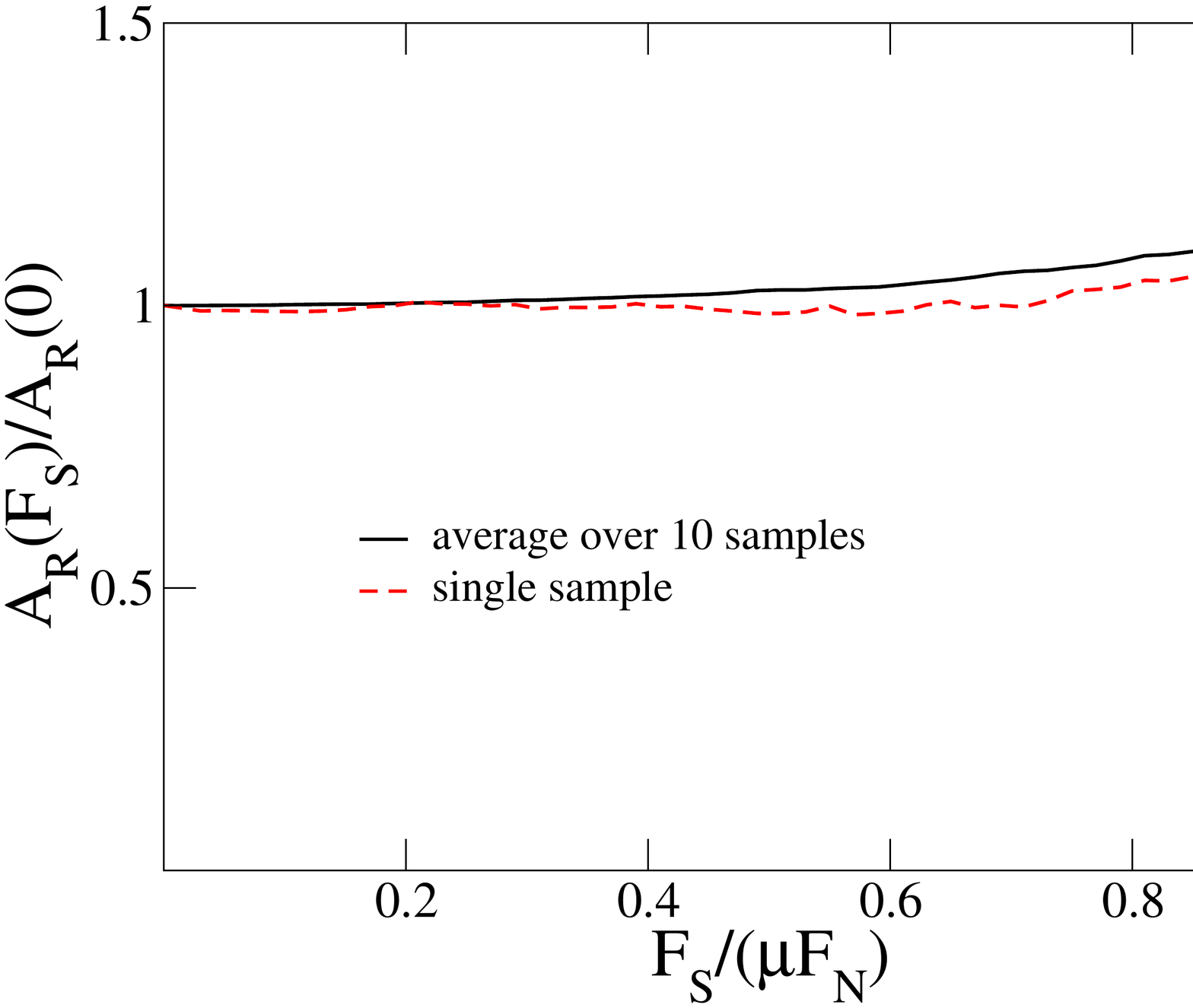}}
\caption{{\bf Area of real contact during slip}. Variation of the real area of contact (normalized to the initial value) as a function of the shearig $F_S$. Sample parameters are $L_x=100$mm, $L_y=7$mm, $L_z=75$mm, $F_N=4$kN. Uniform top shear is applied, i.e. $F_S\equiv F_S^{top}$ }
\label{fig:S2}
\end{figure}

\begin{figure}[ht]
\centerline{\includegraphics[width=0.8\textwidth]{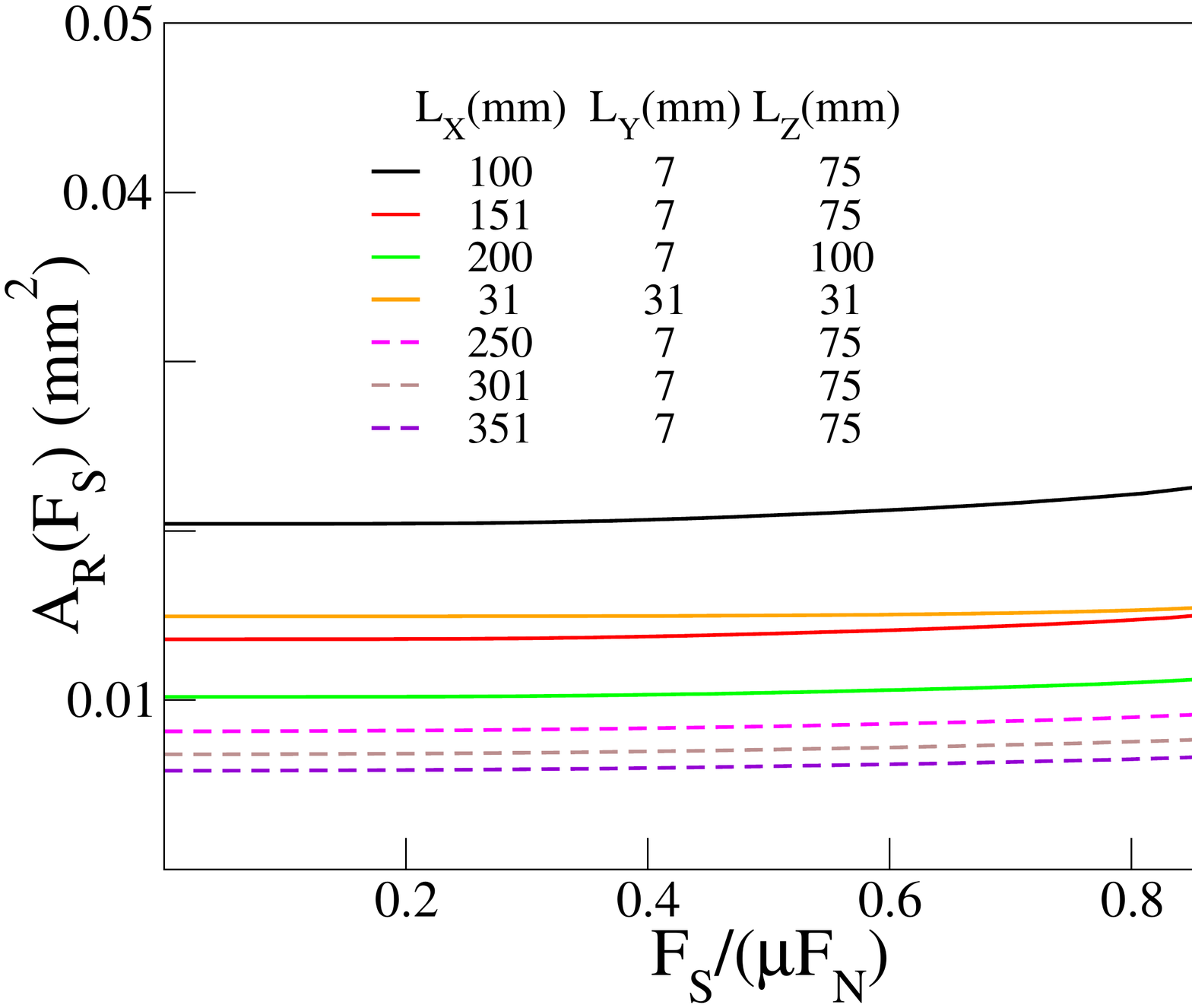}}
\caption{{\bf First and second Amontons' laws}. Real area of contact $A_R$ during the shearing process for various samples, $F_N=4$kN. Althoug nominal area $A_0=L_x\times L_y$ are very different for the shown samples,  $A_R$ appears to be only dependent on the load $F_N$ and independent from the nominal area of contact, i.e. $A_R=\alpha F_N $ \cite{greenwood1966} (this can be also seen in Fig.\ref{fig:2}(a)). Small variations of $A_R$ are persistent in different samples as shown in the figure, they are due  mainly to the mesh discretization. Solid lines correspond to a uniform lateral shearing from the sample  trailing edge, dashed lines correspond to a lateral shearing with a rod at height $h=6$mm ($\Delta h=2$mm). Plots are averaged over 10 independent realizations. Since the sliding happens \emph{always} at $\frac{F_S}{\mu F_N}=1$ the first Amontons' law is satisfied. Moreover, since $A_R$ solely depends on $F_N$, the frictional force $F_S$ at which the sliding takes place is only dependent on $A_R$ and not on $A_0$, what goes under the name of Amontons' second law.}
\label{fig:S3}
\end{figure}

\begin{figure}[ht]
\centering
\includegraphics[width=0.8\textwidth]{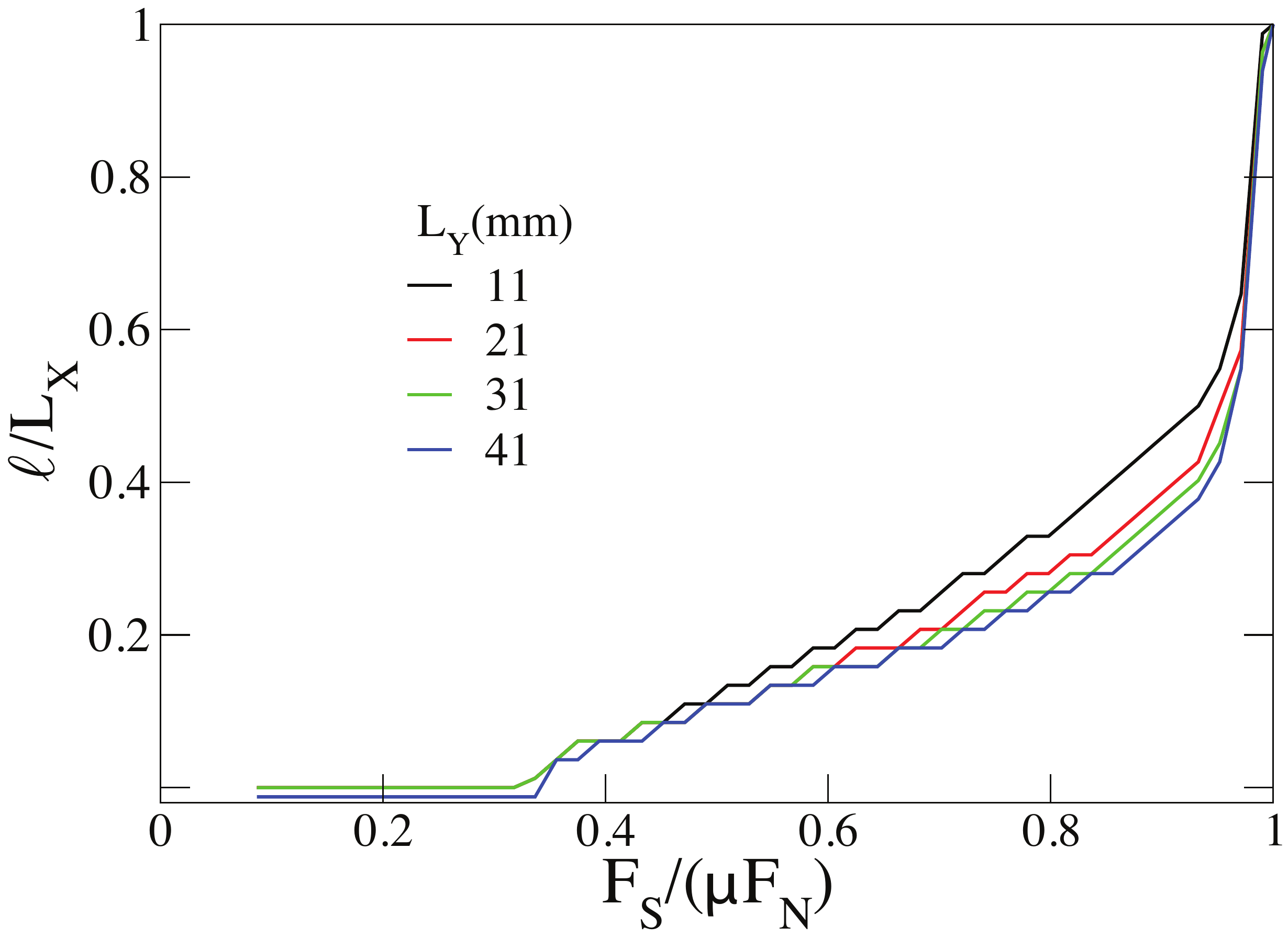}
\caption{{\bf The dependence of slip  precursors on the sample size $L_y$}. Precursor  quasi-static dynamics for different $L_y$. The sample is sheared with a rod  at the height $h=6$mm  ($\Delta h=2$mm), other sample parameters are $F_N=4$kN, $L_x=41$mm, $L_z=132$mm. Plots are averaged over 10 independent realizations.
\label{fig:S4}}
\end{figure}

\begin{figure}[ht]
\centering
\includegraphics[width=0.8\textwidth]{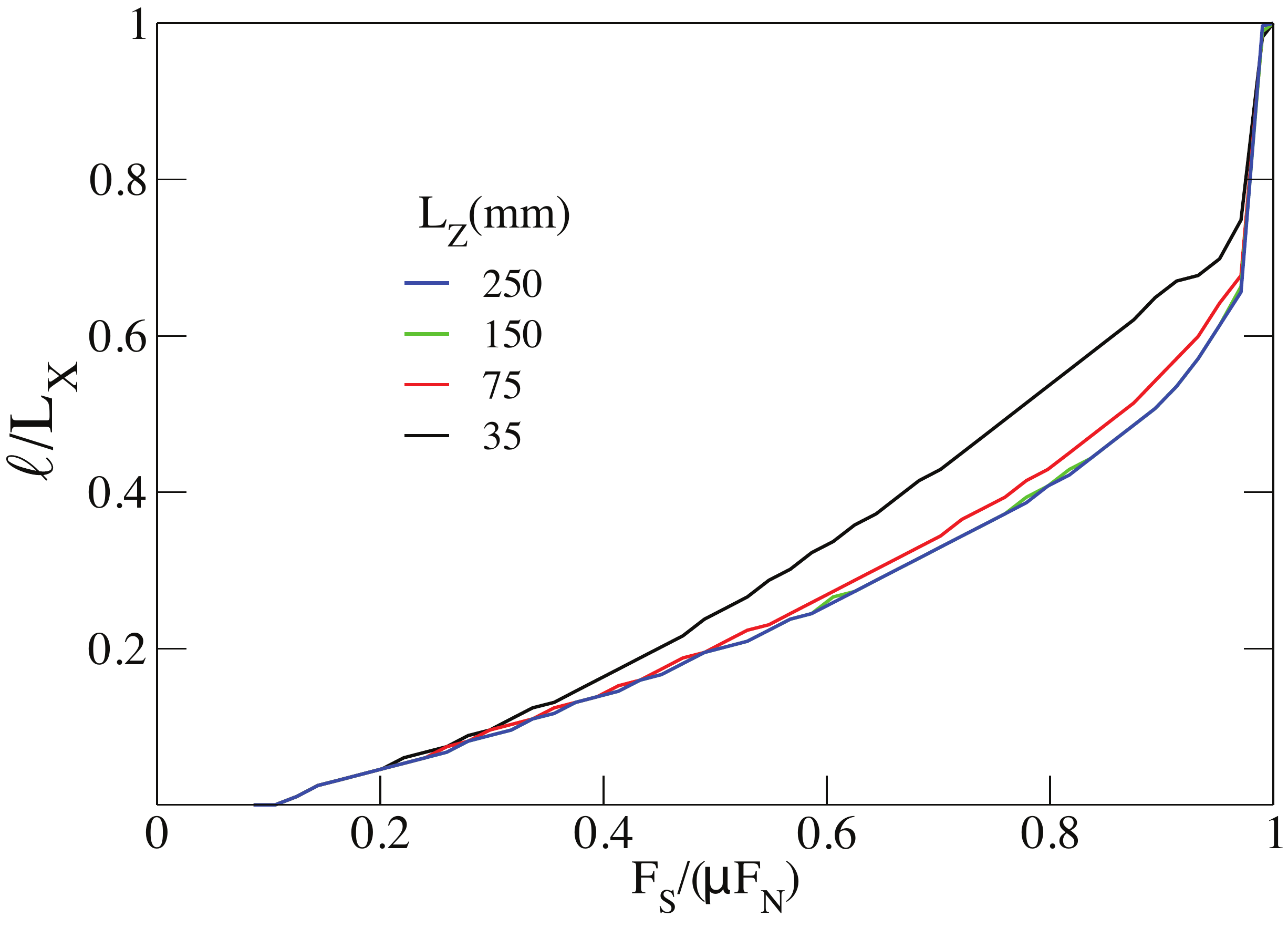}
\caption{{\bf The dependence of slip  precursors on the sample size $L_z$}. Precursor  quasi-static dynamics for different $L_z$. The sample is sheared with a rod  at the height $h=6$mm  ($\Delta h=2$mm), other sample parameters are $F_N=4$kN, $L_x=141$mm, $L_y=7$mm. Plots are averaged over 10 independent realizations.
\label{fig:S5}}
\end{figure}

\begin{figure}[ht]
\centering
\includegraphics[width=0.8\textwidth]{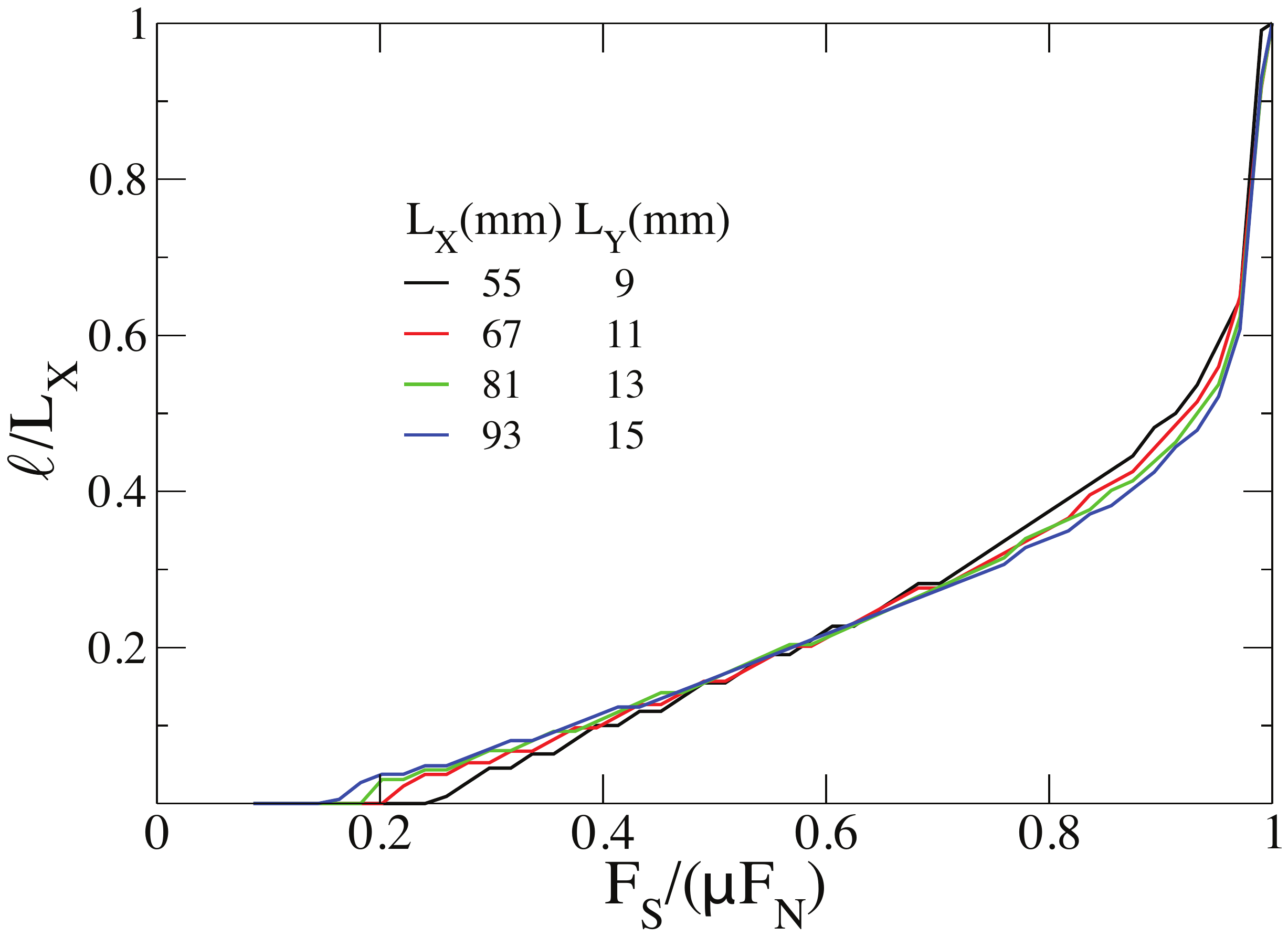}
\caption{{\bf The dependence of slip  precursors on the sample aspect ratio $L_x/L_y$}. Precursor  quasi-static dynamics for different  $L_x$ and $L_y$ but same aspect ratio $\frac{L_x}{L_y}\simeq 6$. The sample is sheared with a rod  at the height $h=6$mm  ($\Delta h=2$mm), other sample parameters are $F_N=4$kN,  $L_z=100$mm. Plots are averaged over 10 independent realizations.
\label{fig:S6}}
\end{figure}

\begin{figure}[ht]
\centering
\includegraphics[width=0.8\textwidth]{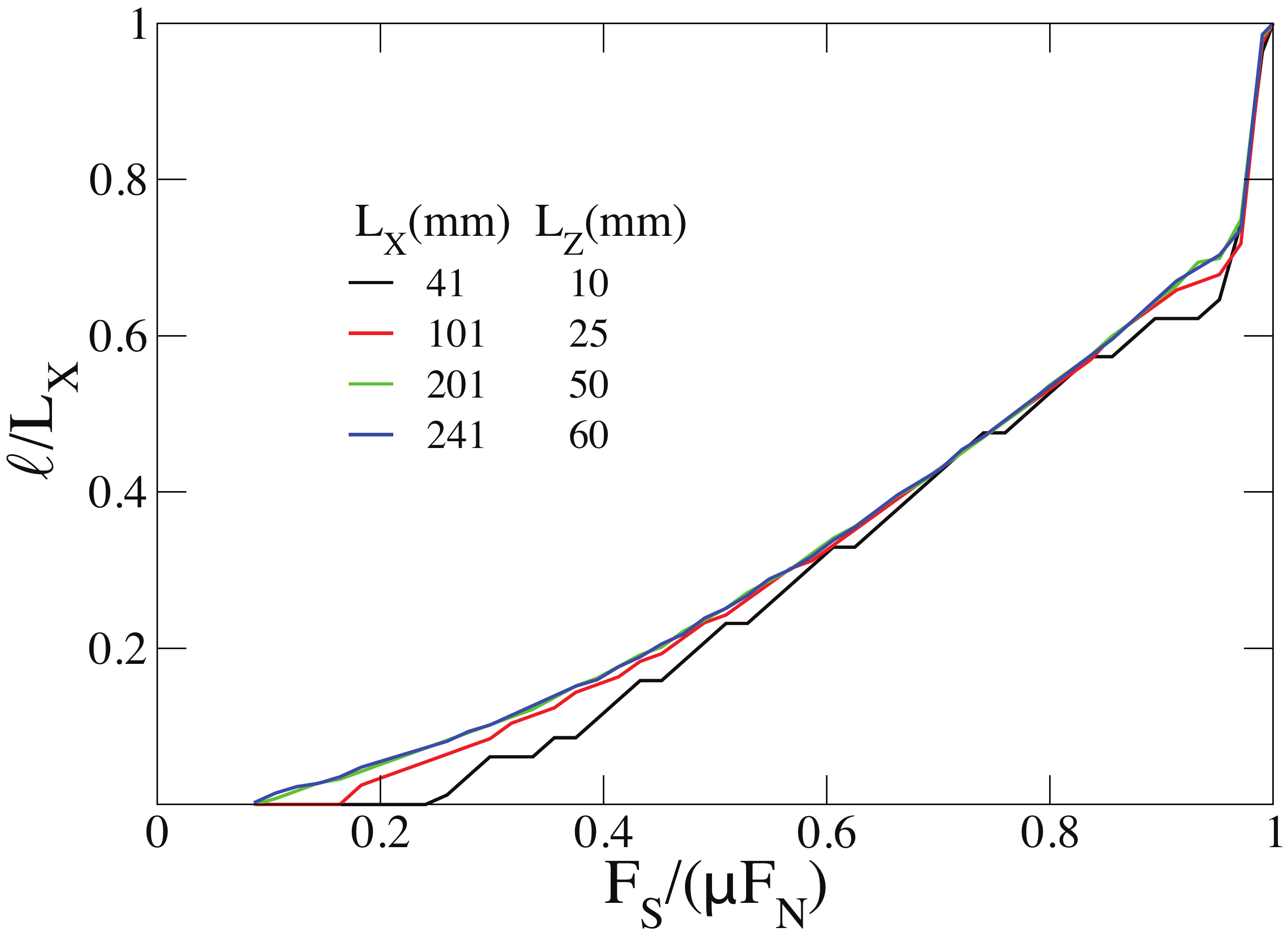}
\caption{{\bf The dependence of slip  precursors on the sample aspect ratio $L_x/L_z$}. Precursor  quasi-static dynamics for different  $L_x$ and $L_z$ but same aspect ratio $\frac{L_x}{L_z}\simeq 4$. The sample is sheared with a rod  at the height $h=6$mm  ($\Delta h=2$mm), other sample parameters are $F_N=4$kN,  $L_y=7$mm. Plots are averaged over 10 independent realizations.
\label{fig:S7}}
\end{figure}

\begin{figure}[ht]
\centering
\includegraphics[width=0.8\textwidth]{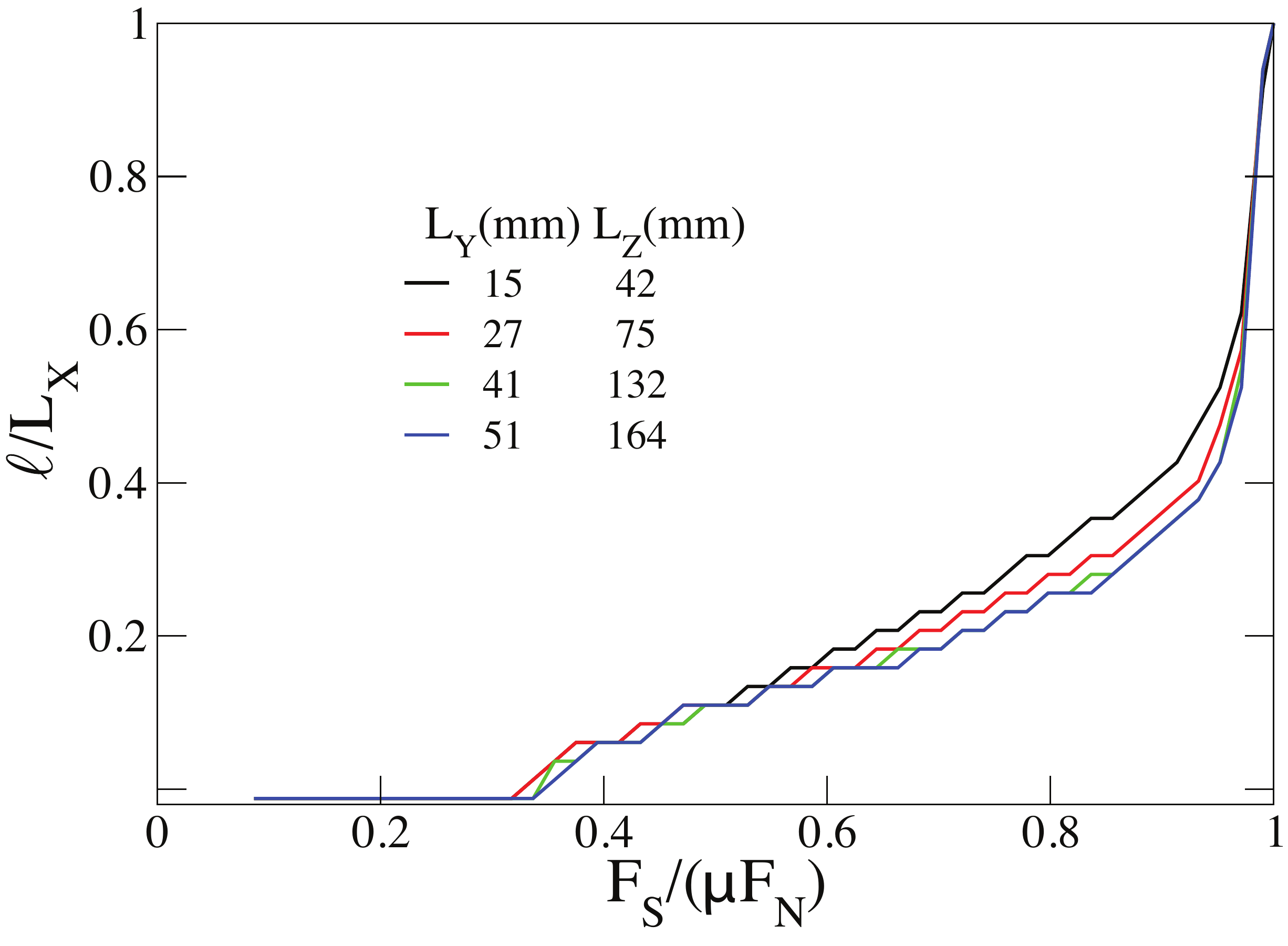}
\caption{{\bf The dependence of slip  precursors on the sample aspect ratio $L_y/L_z$}.Precursor  quasi-static dynamics for different  $L_y$ and $L_z$ but same aspect ratio $\frac{L_y}{L_z}\simeq 0.36$. The sample is sheared with a rod  at the height $h=6$mm  ($\Delta h=2$mm), other sample parameters are $F_N=4$kN,  $L_x=141$mm. Plots are averaged over 10 independent realizations.
\label{fig:S8}}
\end{figure}

\begin{figure}[ht]
\centering
\includegraphics[width=0.8\textwidth]{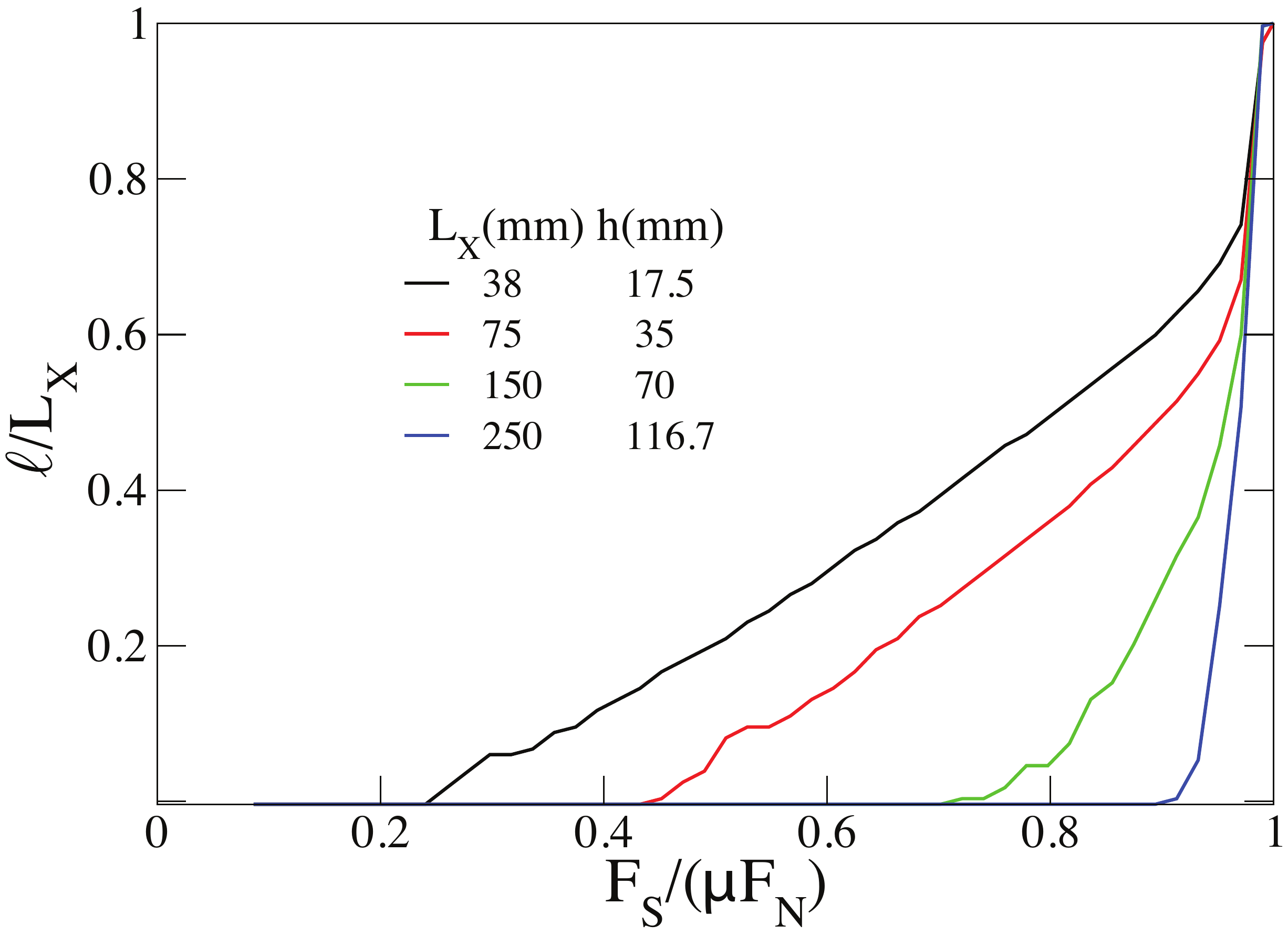}
\caption{{\bf The dependence of slip  precursors on the ratio $h/L_z$}. Precursor  quasi-static dynamics when the sample undergoes a lateral shearing with a rod at the height $h=6$mm  ($\Delta h=2$mm). Plots show different profiles obtained by varying $h$ and $L_z$ but keeping constant the  ratio $\frac{h}{L_z}\simeq 0.46$. Other sample parameters are $F_N=4$kN,  $L_x=141$mm,  $L_y=7$mm. Plots are averaged over 10 independent realizations.
\label{fig:S9}}
\end{figure}

\begin{figure}[ht]
\centering
\includegraphics[width=0.5\textwidth]{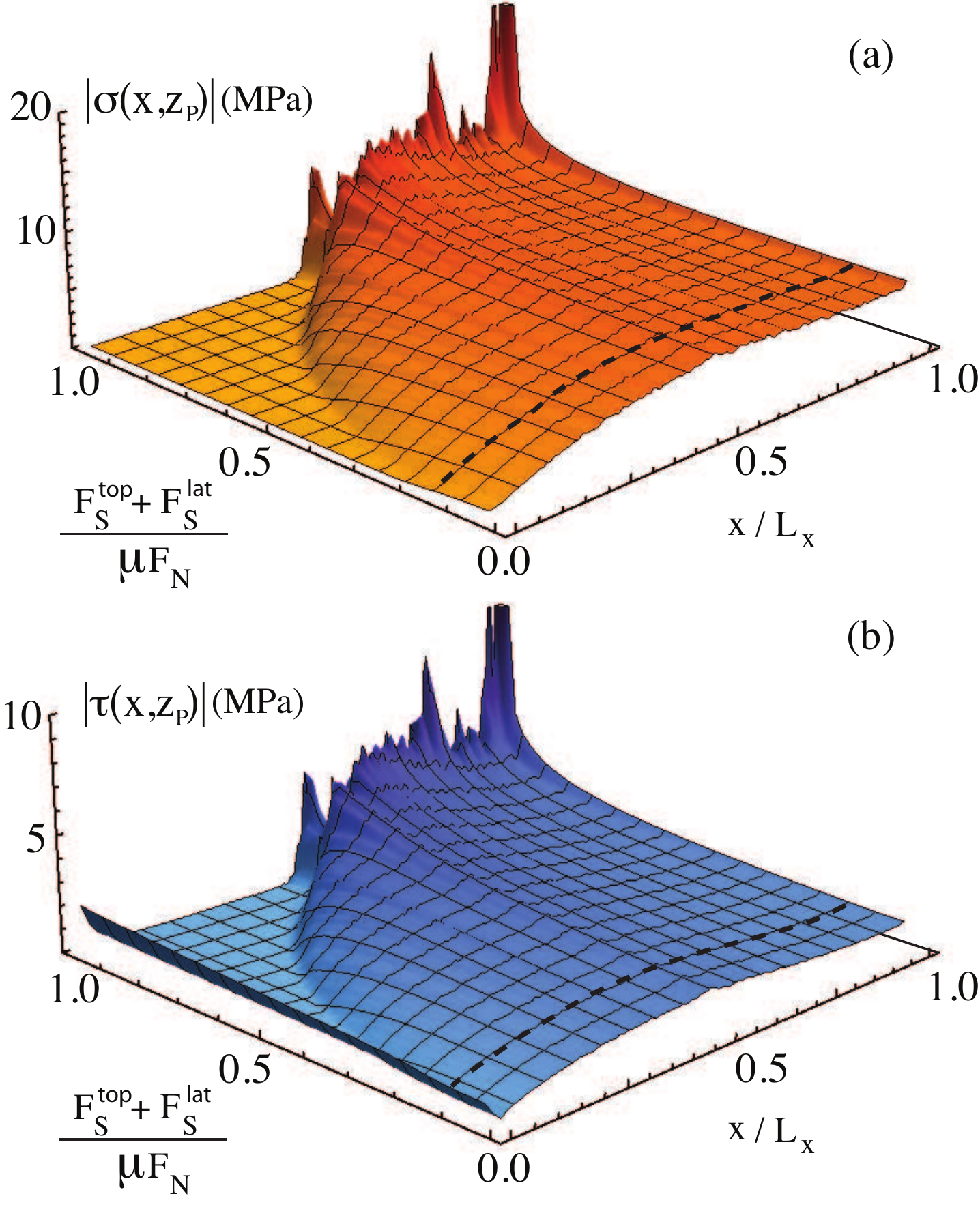}
\caption{{\bf Shear and normal stresses at $z_P=2$mm. Quasi-static evolution}.Average internal stresses calculated at the reference plane $z_P=2$mm. (a) Normal stress obtained from Eq.\ref{sigma_P_def} (Eq.\ref{sigma_P_def_dcrt} in its discrete form) and averaged along the $y$ direction, i.e. $|\sigma(x,z_P)|$ during a shearing protocol which involves both an edge and top pulling. The dashed black line corresponds to the $F_S$ value prior to precursor nucleation: the corresponding shape of $|\sigma(x,z_P)|$ is shown in Fig.\ref{fig:7}(a) (red curve). (b) Internal average shear stress obtained from Eqs.\ref{tau_P_def}-\ref{tau_P_def_dcrt} and averaged along the $y$ direction, i.e. $|\tau(x,z_P)|$. The dashed black line corresponds to the $F_S$ value prior to precursor nucleation: the corresponding shape of $|\tau(x,z_P)|$ is shown in Fig.\ref{fig:7}(a) (blue curve). $F_N=6.25$kN, $L_x=200$mm, $L_y=7$mm, $L_z=100$mm, $h=6$mm ($\Delta h=2$mm). Plots are averaged over 10 independent realizations.
\label{fig:S10}}
\end{figure}

\begin{figure}[ht]
\centerline{\includegraphics[width=0.5\textwidth]{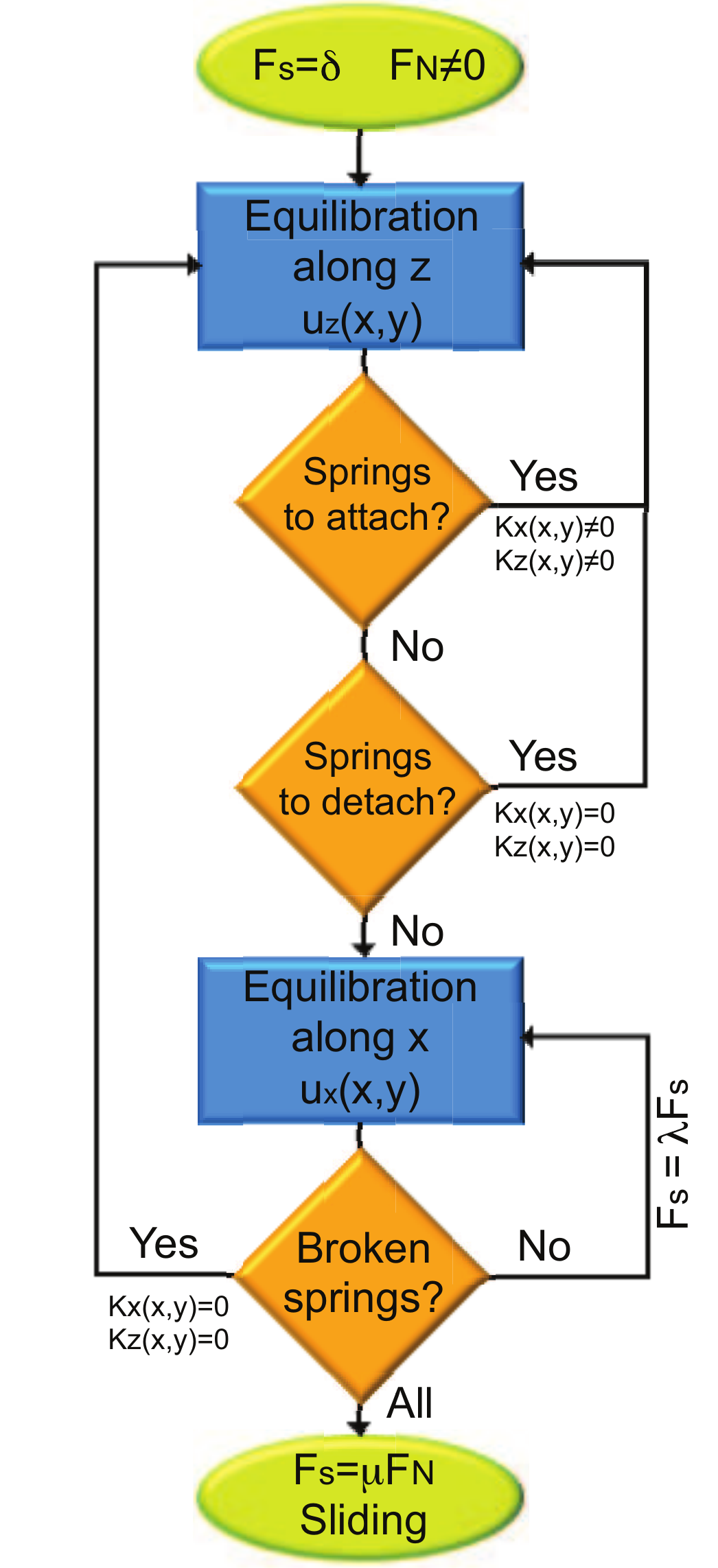}}
\caption{{\bf Scalar model algorithm}. Flowchart of the quasi-static dynamical protocol of the scalar model.}
\label{fig:S11}
\end{figure}

\begin{figure}[ht]
\centering
\includegraphics[width=0.8\textwidth]{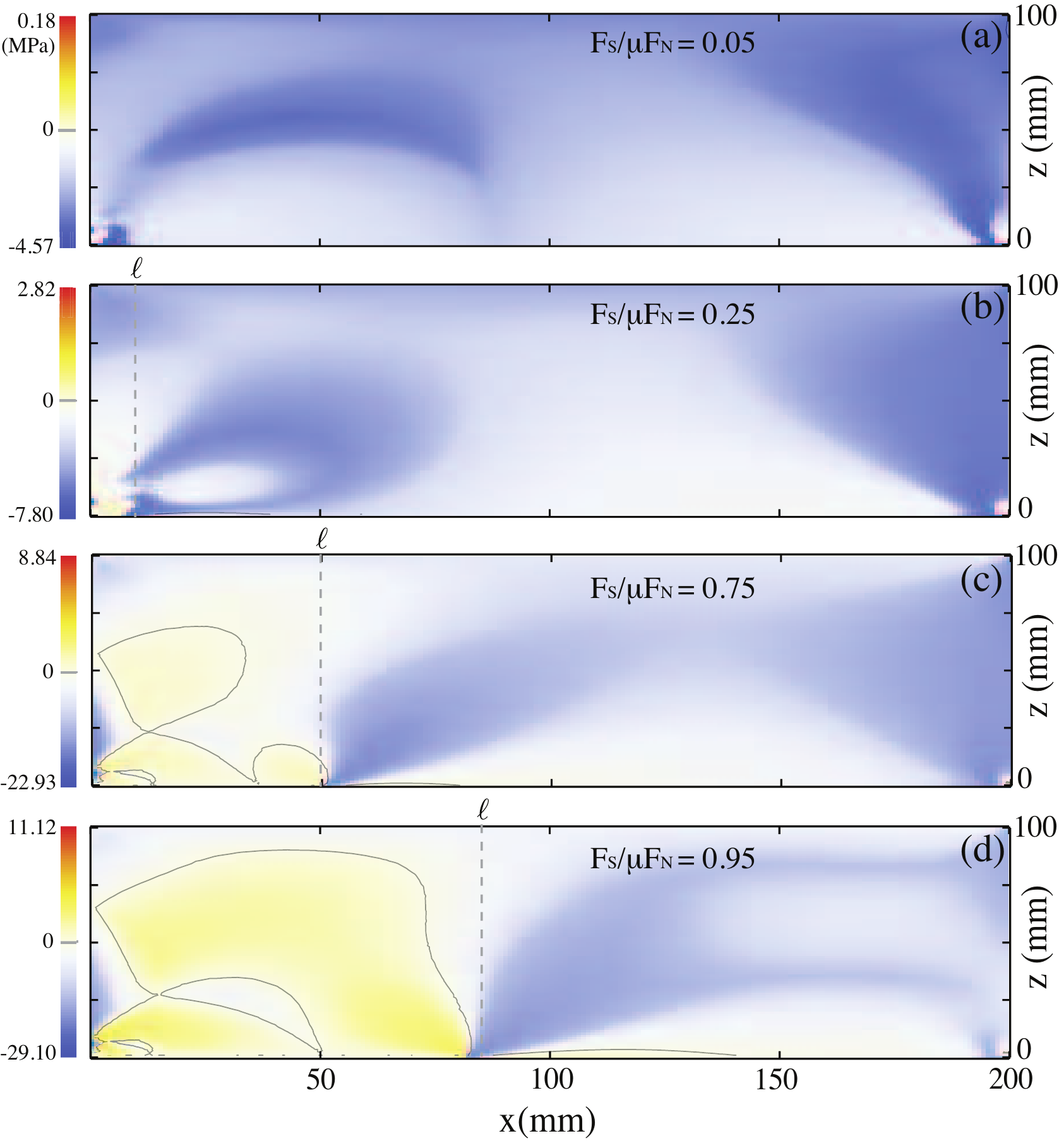}
\caption{{\bf Coulomb stress quasi-static evolution in FEM}. (a)-(d) Quasi-static evolution of the Coulomb stress (calculated at $y=L_y/2$) along the slab $x-z$ plane. Color code indicates regions where $\tau_C>0$ (yellow-red) from those for which $\tau_C<0$  (blu), grey solid lines correspond to the set of points fulfilling $\tau_C=0$. The loading conditions are the same as Fig.\ref{fig:8} and the area of the detached contact surface is $\ell\times L_y$, where $\ell$ is the corresponding precursor size achieved in the scalar model.
 \label{fig:S12}}
\end{figure}

\begin{figure}[ht]
\centering
\includegraphics[width=0.5\textwidth]{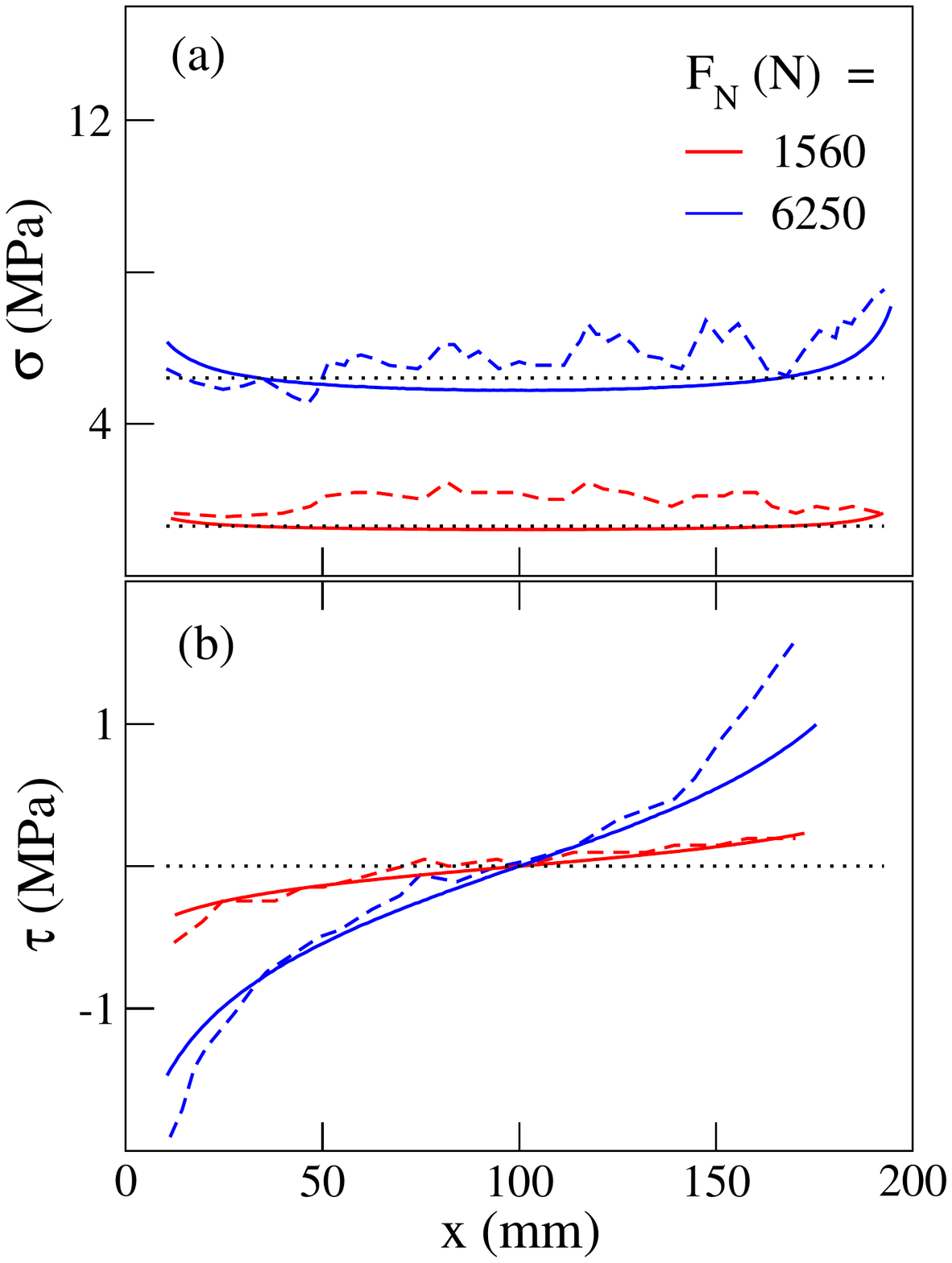} 
\caption{ {\bf FEM model: calibration and stress profiles}. (a)-(b) Calibration of the model by comparison with (a) normal and (b) shear stress profiles of the experiment reported in \cite{bendavid2010} (dashed lines), calculated on the set of points $(x,y=L_y/2,z=z_p=2mm)$. No shearing force is applied, and $L_x=200$mm, $L_y=6$mm, $L_z=100$mm. Solid lines represent the outcomes of FEM simulations after setting the interfacial asperities stiffnesses  to $k_x=10^5$MPa/m and $k_z=10^6$MPa/m. The model shows good agreement with the experiments, for both values of the applied normal force $F_N$. Dotted black lines symbolize normal and shear internal stresses  calculated by the scalar model: the stresses appear to be uniform ($\sigma=F_N/(L_xL_y)$, $\tau=0$) since no Poisson expansion and corner divergences are encompassed (interfacial rough fluctuations are averaged out).
\label{fig:S13}}
\end{figure}

\begin{figure}[ht]
\centerline{\includegraphics[width=0.8\textwidth]{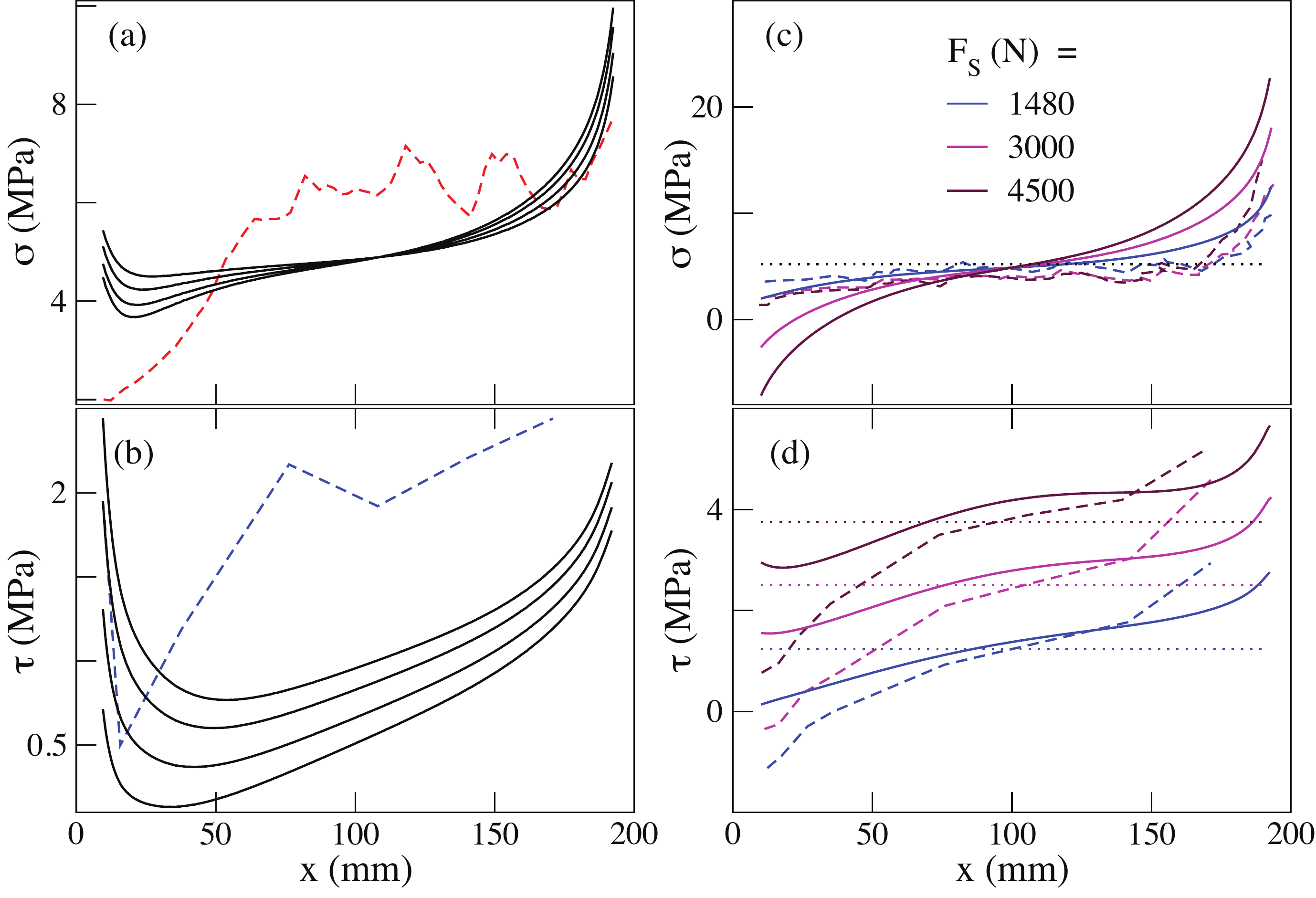}}
\caption{{\bf FEM shearing: internal stress profiles }.  Normal (a) and shear (b) internal stresses calculated on the set of points $(x,y=L_y/2,z=z_p=2mm)$, when the shear force is applied both at the trailing edge  and uniformly on top of the PMMA sample ($F_s=F_s^{top}+F_S^{lat}$). Sample dimensions are $L_x=200$mm, $L_y=6$mm, $L_z=100$mm and the normal load is $F_N=6250$N. The experimental curves (dashed lines) cannot be reproduced by the FEM model, whose stress profiles are represented by black solid lines  for values of $F_s=F_s^{top}+F_s^{lat}=1500$N, $1300$N, $1040$N  and $800$N from top to down ($F_s^{top}=F_s^{lat}$). (c) and (d) show the normal  and shear internal stresses when shear force is applied uniformly on top of the sample, i.e. $F_S=F_S^{top}$: solid lines represent the outcomes of FEM simulations and dashed lines stand for the experimental curves (extracted from Fig.1B-C of Ref.\cite{bendavid2010}).  Although the FEM model captures the trend, the agreement between numerics and experiments is not satisfactory (although slightly better than lateral shearing, in panels (a) and (b)).  Dotted lines represent the normal and shear stress profiles calculated according to the scalar model. In this case the stresses appear uniform as a consequence of the decoupling of the displacements $u_x$ and $u_z$ (scalar elasticity approximaton).
\label{fig:S14}}
\end{figure}

\end{document}